\documentclass[usenatbib,usegraphicx]{mn2e}
\usepackage{amssymb,amsmath}
\usepackage{times}
\usepackage{morefloats}
\usepackage{lscape}
\usepackage{color}
\usepackage{longtable}
\usepackage[totalwidth=17.8cm, totalheight=24.0cm]{geometry} 
\def\aj{AJ}			
\def\apj{ApJ}			
\def\apjl{ApJ}			
\def\apjs{ApJS}			
\def\aap{A\&A}			
\def\aaps{A\&AS}		
\def\mnras{MNRAS}		
\def\pasp{PASP}			

\def\MLsunV{(${\rm M_{\odot}/L_{\odot}}$)$_V$}
\def\MLsun36{(${\rm M_{\odot}/L_{\odot}}$)$_{\rm [3.6]}$}
\def\kms{$\mbox{km s}^{-1}$}
\def\kmsM{km\,s$^{-1}$\,Mpc$^{-1}$}    

\newcommand{\sauron}{{\texttt {SAURON}}}

\newcommand{\se}{$\sigma_{\rm e}$}
\newcommand{\Le}{$L_{\rm e}$}
\newcommand{\re}{$R_{\rm e}$}
\newcommand{\rev}{$R_{\rm e,V}$}
\newcommand{\reIR}{$R_{\rm e,[3.6]}$}
\newcommand{\lR}{$\lambda_{\rm R}$}
\newcommand{\mue}{$\langle\mu_{\rm e}\rangle$}
\newcommand{\muev}{$\langle\mu_{\rm e,V}\rangle$}
\newcommand{\hbeta}{\hbox{H$\beta$}}
\newcommand{\hbetao}{\hbox{H$\beta_{\rm o}$}}
\newcommand{\ml}{$M/L$}
\newcommand{\steml}{$\gamma_\ast$}
\newcommand{\stemlV}{$\gamma_{\ast,V}$}
\newcommand{\stemlIR}{$\gamma_{\ast,[3.6]}$}

\title[The SAURON project - XIX.]
{The SAURON Project - XIX. Optical and near-infrared scaling relations of 
nearby elliptical, lenticular and Sa galaxies}

\author[Falc\'on-Barroso et al.]
{J.\,Falc{\'o}n-Barroso,$^{1,2,3}$\thanks{Email: jfalcon@iac.es} 
G.\,van de Ven,$^{4,5}$ 
R.\,F.\,Peletier,$^{6}$ 
M.\,Bureau,$^{7}$ 
H.\,Jeong,$^{8,9}$ 
R.\,Bacon,$^{10}$ 
\newauthor 
M.\,Cappellari,$^{7}$ 
R.\,L.\,Davies,$^{7}$ 
P.\,T.\,de Zeeuw,$^{11,12}$ 
E.\,Emsellem,$^{11,10}$ 
D.\,Krajnovi{\'c},$^{11}$ 
\newauthor
H.\,Kuntschner,$^{13}$ 
R.\,M.\,McDermid,$^{14}$ 
M.\,Sarzi,$^{15}$ 
K.\,L.\,Shapiro,$^{16,17}$
\newauthor
R.\,C.\,E.\,van den Bosch,$^{4}$ 
G.\,van der Wolk,$^{6}$
A.\,Weijmans,$^{18}$\thanks{Dunlap Fellow} and 
S.\,Yi,$^{19}$\\
$^{1}$Instituto de Astrof\'isica de Canarias, V\'ia L\'actea s/n, La Laguna, Tenerife, Spain\\
$^{2}$Departamento de Astrof\'isica, Universidad de La Laguna, E-38205 La Laguna, Tenerife, Spain\\
$^{3}$European Space and Technology Centre, Keplerlaan 1, 2200~AG Noordwijk, The Netherlands\\
$^{4}$Max-Planck Institute for Astronomy, K\"{o}nigstuhl 17, 69117 Heidelberg, Germany\\
$^{5}$Institute for Advanced Study, Einstein Drive, Princeton, NJ 08540, USA\\
$^{6}$Kapteyn Astronomical Institute, University of Groningen, Postbus 800, 9700 AV Groningen, The Netherlands\\
$^{7}$Sub-Dept. of Astrophysics, Dept. of Physics, University of Oxford, Denys Wilkinson Building, Keble Road, Oxford, OX1 3RH, UK\\
$^{8}$Korea Astronomy and Space Science Institute, Daejeon 305-348, Korea\\
$^{9}$Yonsei University Observatory, Seoul 120-749, Korea\\
$^{10}$Universit\'e Lyon 1, Observatoire de Lyon, Centre de Recherche Astrophysique de Lyon \\
~~~~~~and Ecole Normale Sup\'erieure de Lyon, 9 avenue Charles Andr\'e, F-69230 Saint-Genis Laval, France\\
$^{11}$European Southern Observatory, Karl-Schwarzschild-Str. 2, 85748 Garching, Germany\\
$^{12}$Sterrewacht Leiden, Leiden University, Postbus 9513, 2300 RA Leiden, The Netherlands\\
$^{13}$Space Telescope European Coordinating Facility, European Southern  Observatory, Karl-Schwarzschild-Str.~2, 85748 Garching, Germany\\
$^{14}$Gemini Observatory, Northern Operations Centre, 670 N. A`ohoku Place, Hilo, HI 96720, USA\\
$^{15}$Centre for Astrophysics Research, University of Hertfordshire, Hatfield, Herts AL1 9AB, UK\\
$^{16}$Department of Astronomy, University of California-Berkeley, Berkeley, CA 94720, USA\\
$^{17}$Aerospace Research Laboratories, Northrop Grumman Aerospace Systems, Redondo Beach, CA 90278, USA\\
$^{18}$Dunlap Institute for Astronomy \& Astrophysics, University of Toronto, 50 St. George Street, Toronto, ON M5S 3H4, Canada\\
$^{19}$Department of Astronomy, Yonsei University, Seoul 120-749, Korea}
\begin{document}
\maketitle

\clearpage
\begin{abstract}

We present ground-based MDM $V$-band and \textit{Spitzer}/IRAC 3.6$\mu$m-band
photometric observations of the 72 representative galaxies of the \sauron\  
Survey. Galaxies in our sample probe the elliptical E, lenticular S0 and spiral
Sa populations in the nearby Universe, both in field and cluster environments. 
We perform aperture photometry to derive homogeneous structural quantities. In 
combination with the \sauron\ stellar velocity dispersion measured within an 
effective radius (\se), this allows us to explore the location of our galaxies 
in the colour-magnitude, colour-\se, Kormendy, Faber-Jackson and Fundamental 
Plane scaling relations. 
We investigate the dependence of these relations on our recent kinematical 
classification of early-type galaxies (i.e. Slow/Fast Rotators) and the stellar 
populations. 
Slow Rotator and Fast Rotator E/S0 galaxies do not populate distinct locations
in the scaling relations, although Slow Rotators display a smaller intrinsic
scatter. 
We find that Sa galaxies deviate from the colour-magnitude and colour-\se\
relations due to the presence of dust, while the E/S0 galaxies define tight
relations. Surprisingly, extremely young objects do not display the bluest
$(V-[3.6])$ colours in our sample, as is usually the case in optical colours.
This can be understood in the context of the large contribution of TP-AGB stars
to the infrared, even for young populations, resulting in a very tight
$(V-[3.6])-$\se\ relation that in turn allows us to define a strong correlation
between metallicity and \se. 
Many Sa galaxies appear to follow the Fundamental Plane defined by E/S0 
galaxies. Galaxies that appear offset from the relations correspond mostly to 
objects with extremely young populations, with signs of on-going, extended star 
formation.
We correct for this effect in the Fundamental Plane, by replacing luminosity
with stellar mass using an estimate of the stellar mass-to-light ratio, so that
all galaxies are part of a tight, single relation. The new estimated
coefficients are consistent in both photometric bands and suggest that
differences in stellar populations account for about half of the observed tilt
with respect to the virial prediction.
After these corrections, the Slow Rotator family shows almost no intrinsic
scatter around the best-fit Fundamental Plane.
The use of a velocity dispersion within a small aperture (e.g. \re/8) in the
Fundamental Plane results in an increase of around 15\% in the intrinsic scatter
and an average 10\% decrease of the tilt away from the virial
relation.\looseness-2
\end{abstract}


\section{INTRODUCTION}
\label{sec:intro}

Galaxies are fundamental building blocks of our universe, and our knowledge of
their distribution, structure and dynamics is closely tied to our general
understanding of structure growth. So-called scaling relations, that is
correlations between well-defined and easily measurable galaxy properties, have
always been central to our understanding of nearby galaxies. With high redshift
studies now routine, scaling relations are more useful than ever, allowing us to
probe the evolution of galaxy populations over a large range of lookback times
\citep[e.g.][]{bell04,conselice05,ziegler05,saglia10}. 

The colour-magnitude relation (CMR) was already recognised in the sixties and
seventies \citep{devac61,sandage72,vs77} and has served as an important
benchmark for theories of galaxy formation and evolution since
\citep[e.g.][]{bower92b,bell04,bernardi05}. The main drivers are thought to be
galaxy metallicity, which causes more metal rich galaxies to be redder, and age,
causing younger galaxies to be bluer. Galaxies devoid of star formation
are thought to populate the red sequence, while star-forming galaxies lie in the
blue cloud \citep[e.g.][]{baldry04}. The dichotomy in the distribution of 
galaxies in this relation has opened a very productive avenue of research to
unravel the epoch of galaxy assembly 
\citep[e.g.][]{delucia04,andreon06,arnouts07}. 

Since its discovery \citep{dd87,dressler87}, the Fundamental Plane (FP) has been
one of the most studied relations in the literature. Given its tightness, like
many other scaling relations the FP was quickly envisaged as a distance
estimator as well as a correlation to understand how galaxies form and evolve
(e.g.\ \citealt{saglia93, jfk96, pahre98, kelson00, bernardi03b, vdwel04,
holden05, macarthur09}). It is widely recognised that the FP is a manifestation
of the virial theorem for self-gravitating systems averaged over space and time
with physical quantities total mass, velocity dispersion, and gravitational
radius replaced by the observables mean effective surface brightness (\mue),
effective (half-light) radius (\re), and stellar velocity dispersion ($\sigma$).
Since velocity dispersion and surface brightness are distance-independent
quantities, contrary to effective radius, it is common to express the FP as
$\log($\re$)\,=\,\alpha\log(\sigma)+\beta$\mue$+\gamma$, to separate
distance-errors from others. If galaxies were homologous with constant total
mass-to-light ratios, the FP would be equivalent to the virial plane and be
infinitely thin, with slopes $\alpha=2$ and $\beta=0.4$ (in the notation used
here). By studying the intrinsic scatter around the FP, one can study how galaxy
properties differ within the observed sample.

Some projections of the FP, known earlier in time, have also been widely
studied. \citet{kormendy77} found that the surface brightness density (i.e. mean
surface brightness) of a galaxy changes as a function of its size. This relation
is usually known as the Kormendy relation (hereafter KR). The correlation is
such that larger galaxies have lower surface brightness densities, compared to
their smaller counterparts. The size-luminosity relation (SLR) is widely used to
establish the size evolution of galaxies as a function of redshift
\citep[e.g.][]{trujillo06,dokkum08}. Finally, the last projection of the FP that
we consider in this paper is the Faber-Jackson relation (\citealt{fj76};
hereafter FJR), which relates the luminosity of a galaxy to its stellar velocity
dispersion.

The highest quality and best understood scaling relations in the 
optical/near-infrared are the ones for early-type galaxies. This is because
observationally they are much simpler than spirals, with less complicated star
formation histories, and less extinction by dust, and thus tighter scaling 
relations \citep[e.g.][]{ls10}. It is for that reason that often the two groups 
are treated separately in physically similar relations: the prime example being 
the relations between the central stellar velocity dispersion and absolute 
luminosity of elliptical galaxies (FJR), and the rotation velocity and 
absolute luminosity of disc galaxies \citep{tf77}. In an attempt to unify 
properties of these two groups, spiral galaxies are usually studied in terms of 
their bulge and disc properties. The resemblance of bulges to ellipticals has 
lead to their inclusion in the scaling relations of early-type systems 
\citep[e.g.][]{bender92,khos00,fb02}, although they often reveal a much larger 
scatter and show, on average, an offset with respect to the relations of 
early-type galaxies. While this might not be surprising due to the mentioned 
effects of (younger) stellar populations and dust, part of the reason might 
also be the (often far from trivial) decoupling of the bulge from the disc.
 
These scaling relations exist for galaxy parameters at various radii. While
photometric quantities inside one effective radius are easy to measure, the
inherent limitations of traditional (single-aperture or long-slit) spectrographs
have restricted the measurement of the stellar velocity dispersion to the
central regions of galaxies. For instance, papers based on the Sloan Digital Sky
Survey (SDSS) data \citep[e.g.][]{bernardi03a,graves10} use velocity dispersions
that have been determined from central galaxy apertures, corrected to effective
velocity dispersions using standard aperture corrections. In this paper we
follow the approach of \citet[][hereafter Paper IV]{cappellari06} and make use
of the panoramic capabilities of \sauron\ to measure velocity dispersions in
circular apertures going out to an effective radius (\se), and present scaling
relations for which all the parameters are measured within the same aperture.
This method offers the interesting possibility of presenting the spiral Sa
galaxies in the same relations as early-type E/S0 galaxies. When measuring \se\
from the integrated galaxy spectrum galaxy broadening can be caused by intrinsic
velocity dispersions, or by galaxy rotation. Despite this uncertainty, \se\ will
still be a measure of the mass in a galaxy inside \re. In addition, these
velocity dispersions will not be affected by the presence of central discs,
which often show low velocity dispersions \citep[e.g.][]{fb03}. Since \re\ for
most of our Sa galaxies is much larger than the radius inside which the galaxy
bulge dominates, the scaling relations will give us information about both the
bulges and the inner discs of the Sa galaxies. This paper tends to investigate
both issues by combining photometry with integral-field spectroscopy for a
representative sample of E to Sa galaxies, treated in a consistent manner with a
homogeneous database and methods. The importance of this last point should not
be overlooked, as supposedly standard parameters can vary greatly when measured
by different groups. An example of this is provided by the measurement of
nuclear cusp slopes
\cite[e.g.][]{ferrarese94,byun96,gebhardt96,carollo97,rest01}. 

With these goals in mind, we have carried out an optical spectroscopic survey of
$72$ representative nearby E/S0 galaxies and Sa galaxies to one \re, using the
custom-designed panoramic integral-field spectrograph \sauron\ mounted on the
William Herschel Telescope, La Palma \citep[][hereafter Paper I]{bacon01}. The
\sauron\ representative sample was chosen to populate uniformly
$M_B$--$\,\epsilon$ planes, equally divided between cluster and field objects
\citep[][hereafter Paper II]{dezeeuw02}. The work in this paper builds on
previous results of our survey on scaling relations in other wavelength domains
(Paper IV; \citealt{jeong09}, hereafter Paper XIII). The reader is referred to
other papers of the \sauron\ survey for results on the stellar kinematics
\citep{emsellem04} and kinematic classification \citep{emsellem07,cappellari07}
of early-type galaxies and their stellar populations
\citep{kuntschner06,shapiro10,kuntschner10} and on the kinematics \citep{fb06}
and population of early spirals \citep{peletier07}. Hereafter, we will refer to
them as Paper III, IX, X, VI, XV, XVII, VII and XI respectively.\looseness-2

We present in this paper homogeneous ground-based $V$-band and \textit{Spitzer} 
3.6$\mu$m-band imaging observations of the $24$ elliptical E, $24$ lenticular S0 
and 24 spiral Sa galaxies of the \sauron\ representative sample. Aperture
photometry (growth curve analysis) is carried out to homogeneously derive a
number of characteristic quantities to which the more complex \sauron\
integral-field observations are compared. We introduce the sample selection,
biases and completeness in \S~\ref{sec:sample}. The observations and basic data
reduction are presented in \S~\ref{sec:obs_data}. We describe the aperture
photometry and determination of the spectroscopic quantities in
\S~\ref{sec:aper_phot} and \S~\ref{sec:sauron}. Bivariate scaling relations
are shown in \S~\ref{sec:scaling}, while the Fundamental Plane relation is
specifically addressed in \S~\ref{sec:FP}. We summarise our results and conclude
briefly in \S~\ref{sec:conclusions}. Description of the stellar population
synthesis models and methods used to derive stellar mass-to-light ratios for our
galaxies are presented in Appendices~\ref{app:models} and \ref{app:ML}.
Scaling relations showing the dependencies with kinematic substructure and
environment are shown in Appendix~\ref{app:relations}. Tables with the measured
quantities are presented in Appendix~\ref{app:photab}.\looseness-2

Throughout the paper we adopt the \textit{WMAP} (Wilkinson Microwave Anisotropy
Probe) cosmological parameters for the Hubble constant, the matter density and
the cosmological constant, of respectively $H_0=73$\,\kmsM, $\Omega_M=0.24$ and
$\Omega_L=0.76$ \citep{spergel07}, although these parameters only have a small
effect on the physical scales of the galaxies due to their proximity.

\begin{figure*}
\begin{center}
\includegraphics[angle=90,width=0.99\linewidth]{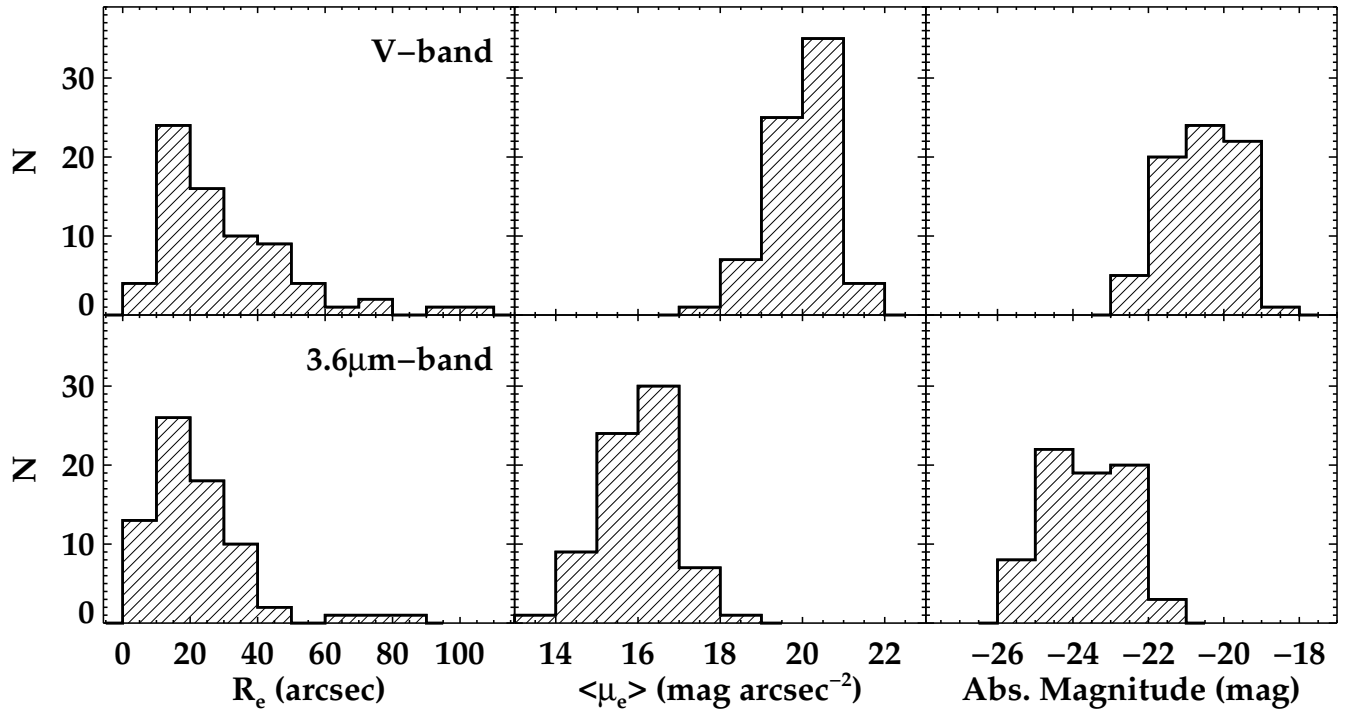}
\caption{Photometric characterisation of the \sauron\ survey. Top row: $V$-band
effective radius (\re), mean effective surface brightness (\mue) and absolute
total magnitudes. Bottom row: same quantities in the 3.6$\mu$m band from the
\textit{Spitzer}/IRAC dataset. The sample is defined such that galaxies uniformly
populate the desired ($B$-band) absolute magnitude range and that our
integral-field observations reach typically as far as one \re\ (see 
\citealt{dezeeuw02}).}
\label{fig:phot}
\end{center}
\end{figure*}

\section{SAMPLE SELECTION, BIASES AND COMPLETENESS}
\label{sec:sample}
The \sauron\ sample is designed to be representative of the population of
early-type galaxies in the nearby Universe. By construction, E, S0 and Sa
galaxies were selected in equal numbers (24 in each group) to populate uniformly
the absolute magnitude versus ellipticity diagram. Within each morphological
class, galaxies were chosen to sample the field and cluster environments equally
(12 in the field and 12 in clusters). The sample therefore consists of 72
galaxies. The basis for the sample selection was the Lyon/Meudon Extragalactic
Database (LEDA; \citealt{paturel97}). 

Besides the astrophysically-motivated criteria, the instrument specifications
impose further constraints on the sample selection:
$6^{\circ}<\delta<64^{\circ}$ to avoid instrument flexure; $cz\leq3000$ \kms\ to
ensure that all the lines of interest are in the observed spectral range;
M$_B\leq-18$ mag to ensure that all central velocity dispersions (above 75 \kms)
can be measured; $\mid b\mid\geq15^\circ$ to avoid crowded fields and large
Galactic extinctions. These restrictions make the \sauron\ set a representative 
but incomplete luminosity- and volume-limited sample of galaxies (see Paper II).

In Figure~\ref{fig:phot} we show some of the main properties of the \sauron\
sample in the $V$- and $3.6\mu$m-bands. The photometric quantities have been
derived as outlined in Section~\ref{sec:aper_phot}. The field-of-view (FoV) for 
a single pointing of the \sauron\ integral-field unit (IFU) is $33\times41$
arcsec$^2$, and therefore, as shown in the figure, covers up to one effective
(half-light) radius \re\  for most of our galaxies. Larger galaxies were usually
mosaiced with several \sauron\ pointings to reach the effective radius. As
intended in our sample selection, our galaxies uniformly populate the desired
absolute magnitude range above the selection cut. 

While Figure~\ref{fig:phot} illustrates the limits in size, mean surface
brightness and luminosity of our sample, it still lacks information about
potential biases (other than the luminosity) and the level of completeness of
our sample. Preliminary checks with larger, more complete samples of early-type
galaxies (\citealt{bernardi03a}, hereafter B03; \citealt{labarbera10};
\citealt{cappellari10}) reveal that our \sauron\ representative sample
covers rather well the parameter space defined by these three photometric
indicators for \mue$\lesssim$21 mag arcsec$^{-2}$ at $V$-band. We illustrate this
using the $g$-band catalogue of B03 as a comparison sample in
\S\ref{sec:scaling} and \S\ref{sec:FP}. This dataset consists of around 9000
early-type galaxies up to redshift 0.3 with stellar velocity dispersions above
70 \kms.

\section{OBSERVATIONS AND DATA REDUCTION}
\label{sec:obs_data}

\subsection{MDM dataset}
\label{sec:mdm} Part of the imaging data were obtained at the 1.3m McGraw-Hill
Telescope of the MDM Observatory located on Kitt Peak, Arizona, over 5 observing
runs totalling $40$ nights: 2003 March 25--April 6, 2003 October 27--November 2,
2004 February 18--25, 2005 April 11--17 and 2005 November 2--6. The entire
\sauron\ representative sample was observed (Paper II). The thin
$2048\times2048$~pix$^2$ backside-illuminated SITE `echelle' CCD was used, and
an additional $128$ column virtual overscan region was simultaneously obtained
with every frame. In direct imaging mode ($f/7.5$), this yields pixels of
$0\farcs508\times0\farcs508$ and a $17\farcm4\times17\farcm4$ field-of-view,
ensuring proper sampling of the seeing and plenty of sky around the targets for
sky subtraction. The seeing was typically $1\farcs8$ to $2\farcs6$ and no
observation with a seeing above $3\farcs0$ was used. The readout noise and gain
were typically $3.0$~e$^-$ and $5.7$~e$^-$~ADU$^{-1}$. The {\it Hubble Space
Telescope} (HST) F555W and F814W filters were used, similar to the Cousins $V$
and $I$ optical filters. Long non-photometric exposures were obtained during
most nights. To reach a sufficient depth and allow correction of CCD defects
when combining the images, our stated goal was to acquire at least three long
offset exposures in each filter for every object. Exposure times for individual
long exposures were typically $400$~s in F555W, although we adjusted both the
exposure times and the number of exposures based on weather conditions. To
internally calibrate the photometry, we also acquired a short ($100$~s) exposure
of every object during the few truly photometric nights.

\subsubsection{Data reduction}
\label{sec:mdmred}
The data reduction of the MDM images was carried out using standard procedures
in {\small IRAF}\footnote{{\small IRAF} is distributed by National Optical
Astronomy Observatories, which is operated by the Association of Universities
for Research in Astronomy, Inc., under cooperative agreement with the National
Science Foundation, USA.}. The bias was subtracted in two steps. First, the
average of the overscan region of each frame was subtracted from each column.
Second, a `master' bias was subtracted from every frame. Since the bias was
found to be stable, we used a run-wide average of (overscan-subtracted) bias
frames obtained at the beginning and end of each night. Dark current was found
to be non-negligible and was subtracted using a `master' dark, again resulting
from a run-wide average of dark frames obtained at night during bad weather
conditions. All galaxy and standard star exposures were then divided by a
`master' flatfield frame, resulting from an average of twilight frames obtained
in each filter. Surprisingly, the flatfields were found to vary from night to
night by up to $2$ per cent, so a night average was used whenever possible. This
is not a major limitation, however, as we are mostly interested in
azimuthally-averaged quantities (see \S~\ref{sec:aper_phot}). All images of a
given galaxy in each filter were registered using the {\it match} routine by
Michael Richmond (available at http://spiff.rit.edu/match/), based on the
algorithm by \citet{valdes95}. Star lists were obtained using {\it SExtractor}
\citep{ba96}. Position uncertainties in the registered images were typically
smaller than a tenth of a pixel. Individual images were then sky subtracted and
combined using a sigma clipping algorithm and proper scaling. Independently of
the photometric calibration, the short exposures are essential in the central
region of many objects, where long exposures are often saturated. The combined
images were flux calibrated using the photometric transformations determined,
night by night, in the way explained below. We estimate the limiting surface
brightness of our $V$-band data ($\approx$25 mag arcsec$^{-2}$) as the surface 
brightness level 1$\sigma$ above the uncertainties in the determination of the 
sky.\looseness-2

\subsubsection{F814W images}
\label{sec:halo}
During the reduction of the data a few important issues were identified in the
F814W images. First, the images suffered from fringing at the $1$--$2$ per cent
level. Since no nighttime exposure of blank fields was obtained, we attempted to
devise an alternative correction from standard star exposures, and also from the
galaxies' exposures themselves. Standard star images, however, proved to have
too low signal-to-noise ratios $S/N$, while median-combined galaxy exposures
failed to remove all the galaxy signal, as most of our targets cover a
substantial fraction of the field-of-view. We therefore could not correct for
the fringing. Second, and most importantly, stars in the F814W long-exposure
images showed a faint, but extended halo around them. This was mostly noticeable
in the saturated stars. This problem, commonly termed 'red halo point spread
function (PSF)', is due to the use of thinned CCDs together with other effects
related to the aging of the telescope coating (see \citealt{michard02} for an
in-depth study on the issue). These cause the PSF in the F814W filter to extend
to well over 100 arcsec (see also \citealt{wu05} and \citealt{dejong08}).
Although one could in principle devise a correction, it would be rather
uncertain. Since this issue will affect any measurement with this filter, we
deemed the F814W dataset unreliable and discarded it from our analysis.

The effective radii presented in previous papers of the \sauron\ series (Papers
IV, VI, IX, X, \citealt{mcdermid06,scott09}) were computed from combined F814W
HST and MDM images before we identified this issue. Nevertheless, the values
calculated there appear to be 1\% smaller than the ones measured here in the
$V$-band (thus consistent with the presence of colour gradients) and with a
scatter implying a small error of 8\%. Note that this issue has no effect on any
of the conclusions in those papers and has a very small effect on the quantities
derived from our integral-field data, which have been updated in Paper XV and
subsequent papers of the \sauron\  series.

\subsubsection{Flux calibration}
\label{sec:mdmflux}
During photometric nights, in addition to galaxies, we also performed repeated
observations of stellar fields including standard stars from \citet{landolt92},
covering a range of apparent magnitudes and colours; for many fields, the
observations were repeated at the beginning, in the middle and at the end of the
night, in order to also calibrate the dependence on airmass. In total, we
acquired 84 stellar fields, with several (3 to $\approx$ 10) standard stars in
each one. The aim was to calibrate the data using photometric solutions of the
following form:
\begin{eqnarray}
\label{eq:phot}
m_{\rm V,std} & = & m_{\rm F555W,ins}\,+\,z_{\rm F555W}\,+\,k_{\rm F555W}A\,+\\
\nonumber & & c_{\rm F555W}\,(m_{\rm V,std}\,-\,m_{\rm I,std}),
\end{eqnarray}

\noindent where $m_{\rm std}$ are the standard magnitudes from
\citet{landolt92}, $m_{\rm ins}$ is the instrumental magnitude, $z$ the
photometric zero-point, $k$ the atmospheric extinction coefficient, $A$ the
airmass and $c$ the colour correction coefficient. In practice, for each of our
stellar fields, we identified the standard stars with the help of the maps
published by \citet{landolt92}, and for each star measured the magnitude which
enters in equation~\ref{eq:phot} as $m_{\rm ins}$. This was done by means  of
standard {\small IRAF} tasks, within the {\it noao.digiphot.daophot} and {\it
noao.digiphot.photcal} packages. The sky background was evaluated taking the
mode of the intensity in an annular region around each star, situated 3-4 times
the full-width half maximum (FWHM) of the stellar profile away from its peak;
the star was sky-subtracted and the computed magnitude corrected for aperture
effects. All the stars affected by scattered light or saturated, and those for
which the fits performed in {\small IRAF} did not converge, were removed. We
took all the remaining stars with measured instrumental magnitudes and solved
equation \ref{eq:phot} for $z$, $k$ and $c$. We then grouped the stars according
to the night in which they had been observed and solved again equation
\ref{eq:phot} on a 'per-night' basis, keeping the colour term $c$ -which is very
close to zero- fixed to the `global' value determined at the previous step and
fitting for the airmass term $k$ and the zero point $z$ only. Given that the
standard magnitudes of the reference Landolt stars are in the $V$-band (Johnson)
filter, the photometry of our images is based on that filter. During the flux
calibration, we have therefore converted our images from the HST F555W to the
$V$-band (Johnson) filter. 

The internal accuracy of our flux calibration is around 0.03 magnitudes. In
order to investigate systematic departures of our calibration, we compared our
measurements to apparent magnitudes measured with the same aperture on HST/WFPC2
PC1 images. The set of galaxies used in the comparison is that published by
\citet{lauer05} and for which sky values are reported. The choice of aperture
was arbitrarily set to \re/2, with the constraint of it being larger than
5$\arcsec$ to avoid uncertainties derived from the different PSFs and smaller than
15$\arcsec$ to be fully included within the PC1 chip. The difference between our
$V$-band magnitudes and those of Lauer's HST/F555W imaging is better than 0.05
magnitudes rms, with a small systematic offset. The MDM flux calibration
predicts slightly brighter ($\leq$0.05 mag) galaxies than HST. This shift is,
however, within the expected $V-$~F555W zero-point transformation for different
late-type stellar templates (i.e. it conforms to the dominant old population in
our galaxies)\footnote{see Table 5.2 in the HST/WFPC2 handbook for zero-point
transformations (http://www.stsci.edu/hst/wfpc2).}.

\begin{figure}
\begin{center}
\includegraphics[angle=0,width=0.99\linewidth]{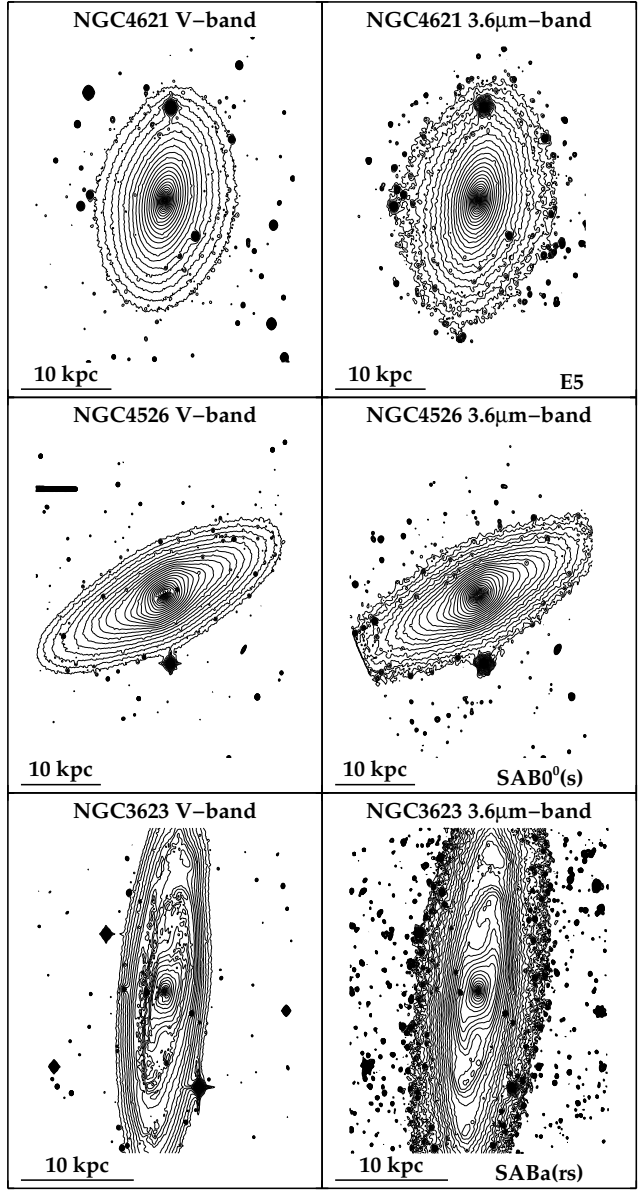}
\caption{Contours of the surface brightness of three galaxies in the \sauron\ 
sample (in 0.25 mag arcsec$^{-2}$ steps) in both the $V$- and $3.6\mu$m bands.
The largest isophotes are set to reach the limiting surface brightness of the
survey in each band ($\approx$25 and $\approx$21 mag arcsec$^{-2}$
respectively). The orientation of the images is such that north is up and east
is left. The Hubble morphological type is indicated in the bottom right corner.
The field-of-view of the observations is much larger than displayed here.}
\label{fig:contours}
\end{center}
\end{figure}

\subsection{Spitzer/IRAC $3.6\mu$m dataset}
\label{sec:spitzer}
In an attempt to extend our analysis of the scaling relations to longer
wavebands than the optical, and to alleviate the loss of the F814W MDM
observations, we decided to use the homogeneous IRAC $3.6\mu$m imaging dataset
provided by the \textit{Spitzer} telescope. This dataset has the great advantage
of being less sensitive to the presence of dust and provides a better tracer for
the underlying, dominant, predominantly old stellar mass component of galaxies.

We retrieved the InfraRed Array Camera (IRAC) images of our sample galaxies at
$3.6\mu$m through the Spitzer Science Center (SSC) archive. These archival
images cover a significant fraction of the \sauron\ sample, and were
acquired in the context of several programs. The remaining objects were observed
as part of the specific program 50630 (PI: G. van der Wolk) during Cycle 5,
meant to complete observations of the \sauron\ sample in both IRAC and
MIPS (Multiband Imaging Photometer for Spitzer) bands. We used the BCD images,
and mosaiced them together using the MOPEX software. This avoided the artificial
point sources sometimes present in the centre of the PBCD images. Details about
the reduction are given in the IRAC instrument
Handbook\footnote{http://ssc.spitzer.caltech.edu/irac/dh/}. The output mosaics
were then sky subtracted in the standard way. As indicated in the handbook, a
zero-point of 280.9 Jy was assumed for flux calibration of the data into the
Vega system. Further details regarding the data reduction can be found in Paper
XV and in van der Wolk et al. (in preparation).

After the data reduction and flux calibration processes, we estimate the
limiting surface brightness of our $3.6\mu$m-band data to be $\approx$21 mag
arcsec$^{-2}$. In order to illustrate the quality of our imaging, in
Figure~\ref{fig:contours} we show isophotal contours down to the limiting
surface brightness of a few galaxies in our sample. The consistency in the
photometric depth of both datasets ensures that our measured parameters truly
reflect the potential structural changes as a function of waveband, and are not
affected by poor imaging.

\subsection{Distances}
\label{sec:distances}

We have made a comprehensive effort to collect the best available distance
estimates in the literature for all galaxies in our sample. We have assigned
distance estimates adopting the following priority order in the methods used:

\begin{enumerate}
 \item For 42 galaxies, the distances were obtained with the surface brightness
fluctuation (SBF) method by \cite{Mei2007} for the ACS Virgo Cluster Survey,
\cite{Cantiello2007} using archival ACS imaging, and \cite{Tonry2001} using
ground-based imaging. We subtract 0.06 mag from the latter distance moduli
(i.e., a 2.7\% decrease in distance) to convert to the same zero point as the
HST Key Project Cepheid distances \citep{Freedman2001}.

 \item For 1 galaxy (NGC\,5308), the distance is derived using Supernovae type
Ia luminosities from \cite{Reindl2005}, subtracting 0.43 mag from the distance
moduli to convert to $H_0=73$\,\kmsM.
 
 \item For 10 galaxies within 12$^\circ$ of the Virgo cluster center (defined by
the galaxy M\,87) and with heliocentric velocities $<2500$\,\kms, we followed
\cite{Crook2007} and assigned the Virgo cluster distance modulus of
$31.092$\,mag \citep{Mei2007} and an error of $0.082$\,mag due to the depth of
the cluster.
 
 \item For 2 galaxies (NGC\,2273 and NGC\,5448), the distances are based on
"Look-Alike" galaxies \citep{paturel84} from \cite{Terry2002}, calibrated with
the HST Key Project Cepheid distances.
 
 \item For 4 galaxies, the distances are based on the correlation between galaxy
luminosity and linewidth (Tully-Fisher relation) from \cite{Tully2008},
calibrated with the HST Key Project Cepheid distances.
 
 \item For 1 galaxy (NGC\,5198), the distance is based on the $D_n$--$\sigma$
\citep[see ][]{dressler87} relation from \cite{Willick1997}, adopting
$H_0=73$\,\kmsM\ and an error of $0.40$\,mag in the distance modulus.
 
 \item  For the remaining 12 galaxies, the distances are based on their observed
heliocentric radial velocities given by
NED\footnote{http://nedwww.ipac.caltech.edu/}, using the velocity field model of
\cite{Mould2000} with the terms for the influence of the Virgo Cluster and the
Great Attractor.
 
\end{enumerate}

The methods (i)--(iii) typically yield errors in the distances of $\lesssim10$\%,
while methods (iv)--(vi) are expected to be accurate to better than
$\lesssim20$\%. Comparing these reliable distance estimates for 60/72 galaxies
in our sample with the distance estimates based on the observed redshifts
using method (vii), we find a (biweight) dispersion of $\approx23$\%. Taking
into account the average 7\% error in the accurate distance estimates, we thus 
adopt for the redshift distances a typical error of $\approx22$\%, i.e. 
$0.487$\,mag in the redshift distance modulus. Tables~\ref{tab:photpars_ESO} and
\ref{tab:photpars_Sa} list the adopted distances as well as the sources of the 
distance estimates. 

\section{PHOTOMETRIC QUANTITIES}
\label{sec:aper_phot}
One of the main goals of this project is the measurement of homogeneous
photometric quantities. As in Paper IV, we have opted for simple, yet frequently
used, methods to derive these values. This has the advantage of being well
reproducible and of allowing comparison with a wide range of studies
\citep[e.g.][]{burstein87,jfk92} . The values measured here are based on
aperture photometry. For each galaxy, in both the $V$-band and the
$3.6\mu$m-band, we extracted radial profiles on circular apertures using the
{\it mge\_fit\_sectors} IDL\footnote{http://www.ittvis.com/} package of
\citet{cappellari02}. Foreground stars and nearby objects were masked using the
{\it SExtractor} lists generated for the registration of the images (see
\S~\ref{sec:mdmred}). The profiles were flux calibrated and corrected for
Galactic extinction using the A$_{\rm V}$ and A$_{{\rm 3.6}\mu{\rm m}}$ values
from NED, which are based on \textit{COBE}, \textit{IRAS} and the
Leiden-Dwingeloo HI emission maps as discussed by \citet{sfd98}. We have made no
attempt to correct the observed fluxes or luminosities for internal extinction.

Throughout the figures of this paper, circles denote E/S0 galaxies and diamonds
Sa galaxies. Filled symbols indicate galaxies with good distance estimates,
whereas open symbols denote those with only recession velocity determinations. 
In blue we highlight Fast Rotator galaxies, in red Slow Rotator galaxies (see
\S\ref{sec:kinparams}) and in green Sa galaxies. The special case of NGC\,4550,
a galaxy known to consist of two counter-rotating stellar discs of similar mass
(\citealt{rix92}; Paper X), is marked in yellow.

\subsection{Effective Radii, mean effective surface brightnesses and absolute magnitudes}
\label{sec:r14_fit}

We have determined the effective radii and mean effective surface brightnesses
for our sample galaxies by fitting our aperture photometry profiles with
$R^{1/n}$ \citep{sersic68} growth curves of the form
\begin{eqnarray}
L(<R) & = & 2\pi \int_0^R I(R') \, R'  \, dR'\\
\nonumber & = & 2\pi \, n \, I_{\rm e} R_{\rm e}^{2} \, \frac{{\rm e}^{b_n}}{(b_n)^{2n}} 
	\, \gamma\left[2n,b_n(R/R_{\rm e})^{1/n}\right],
\end{eqnarray}
with $\gamma$ the incomplete gamma function, \re\ the effective radius, $I_{\rm
e}$ the effective surface brightness (at \re),  $n$ the S\'ersic index
describing the curvature of the radial profile, and we adopt \citep[][]{cb99}
\begin{equation}
b_n=2\,n-\frac{1}{3} + \frac{4}{405\,n} + \frac{46}{25515\,n^2},
\end{equation}
which is an approximation to better than $10^{-4}$ for  $n>0.36$. 

\begin{figure}
\begin{center}
\includegraphics[angle=0,width=0.99\linewidth]{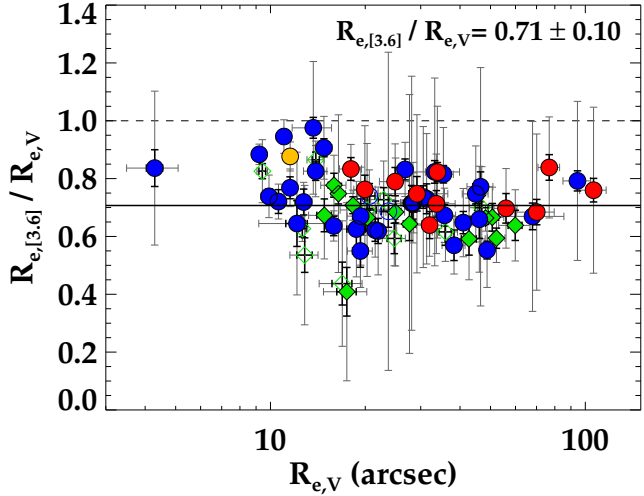}
\caption{Ratio between \rev\ and \reIR\ for our sample of galaxies. The solid
thick line shows the average ratio (\reIR/\rev). The dashed line marks the
one-to-one relation. Circles denote E/S0 galaxies, diamonds Sa galaxies. Filled
symbols indicate galaxies with good distance estimates, open symbols those with
only recession velocity determinations. In blue we highlight Fast Rotators, in
red Slow Rotators and in green the Sa galaxies. The special case of NGC\,4550,
with two similarly-massive counter-rotating disc-like components, is marked in
yellow. Black/gray error bars denote the minimum/maximum uncertainties in our
analysis (see \S\ref{sec:r14_fit})\looseness-2}
\label{fig:recomp}
\end{center}
\end{figure}

When fitting the growth curve profiles, generally the inner $\approx10\arcsec$
as well as regions outside 1$\sigma$ above the sky level have been ignored. The
former avoids potential complications due to the point spread function and the
latter reduces the uncertainties associated with imperfect sky subtraction. 
We have used the integrated S\'ersic profile only to extrapolate the
outermost part of the growth curve to infinity and estimate the total galaxy
luminosity. After this, we have determined \re\ from the radius where the growth
curve profiles are equal to half this total luminosity, i.e., equal to \Le$=
L$($<$\re). In other words, we have not adopted the \re\ values from the
S\'ersic fit, even though they turned out to be very similar to those from the
growth curve profiles after obtaining \Le\ from the S\'ersic fit.
While the surface brightness profiles of galaxies are not perfectly described by
a $R^{1/4}$ law \citep[e.g.][]{ccd93, graham96,macarthur03}, the growth curves
for the vast majority of the objects in our sample, typically early-type
galaxies, were well represented at large radii by $n=4$. However, the growth
curves of galaxies displaying extended discs and intermediate to edge-on
configurations (i.e. mainly spirals, but also some lenticulars) were often
poorly described by a de Vaucouleurs law, hence we fitted a S\'ersic law with
$n<4$ instead. Overall, the adopted S\'ersic indices $n$ were the same in both
the $V$-band and the $3.6\mu$m-band. The adopted S\'ersic $n$ values together
with all the other photometric quantities for the sample galaxies are listed in
Tables~\ref{tab:photpars_ESO} and \ref{tab:photpars_Sa}. 
The approach of using an $R^{1/4}$ growth curve to extrapolate the
galaxy luminosity to infinity was the same adopted by classic studies
\citep[e.g.][]{burstein87,jfk95} and by previous papers of our survey on scaling
relations (e.g. Paper IV). This allows for a direct comparison of our numbers with
theirs.
In addition to the measurements of \Le\ and \re,  we have computed their
uncertainties via MonteCarlo realisations. Besides including the uncertainty in
the background sky subtraction, we included correlations among the pixels in the
images in two different ways, providing lower and upper limits to the
uncertainties. First, we assumed that the dominant source of error is Poisson
noise and that the pixels are un-correlated when we include the errors in the
sky, except for scales smaller than the FWHM of the PSF. For those scales we
defined a correlation length (FWHM/pixel scale) which we set to 2.0 after some
tests. The choice of this correlation length does not significantly change the
output uncertainties for values below 6. Second, we assumed that the pixels are
fully correlated and that this is significantly higher than the Poisson noise
(which is also included). We adopted as the uncertainty the one produced by the
first method throughout the paper. The second estimate is a good test to assess
the maximum error one could expect in the worse possible situation. For
reference we show the different error measurements for \re\ in
Fig.\ref{fig:recomp}.

The total apparent magnitude and corresponding uncertainty follows from \Le,
simply as $m=-2.5\log(2L_{\rm e})$.
The mean effective surface brightness was computed by dividing \Le, the total
luminosity measured within one effective radius, by the area of the aperture,
$A_{\rm e}=\pi$\re$^2$, and expressing it in magnitudes, \mue$=-2.5\log(L_{\rm
e}/A_{\rm e})$. Since the effective luminosity and radius are
correlated, the uncertainty in the mean effective surface brightness was derived
after first computing the \mue\ values for each of the corresponding Monte Carlo
realisations of \Le\ and \re.

As already shown by other authors
\citep[e.g.][]{pahre99,bernardi03a,macarthur03,labarbera04}, \re\ values in the
infrared appear to be, in general, smaller than those in optical bands. This is
mostly due to the fact that galaxies tend to be bluer in the outer parts and
therefore emit less light at these longer wavelengths \citep{peletier90}. In
Figure~\ref{fig:recomp} we show the relation we find between the independently
measured \rev\ and \reIR\ values; \reIR\ is on average 29\% smaller than \rev.
As a results of this, and as shown in \S~\ref{sec:scaling} and
\ref{sec:FP}, differences are thus found in the scaling relations for the two
photometric bands.

\begin{figure*}
\begin{center}
\includegraphics[angle=90,width=0.99\linewidth]{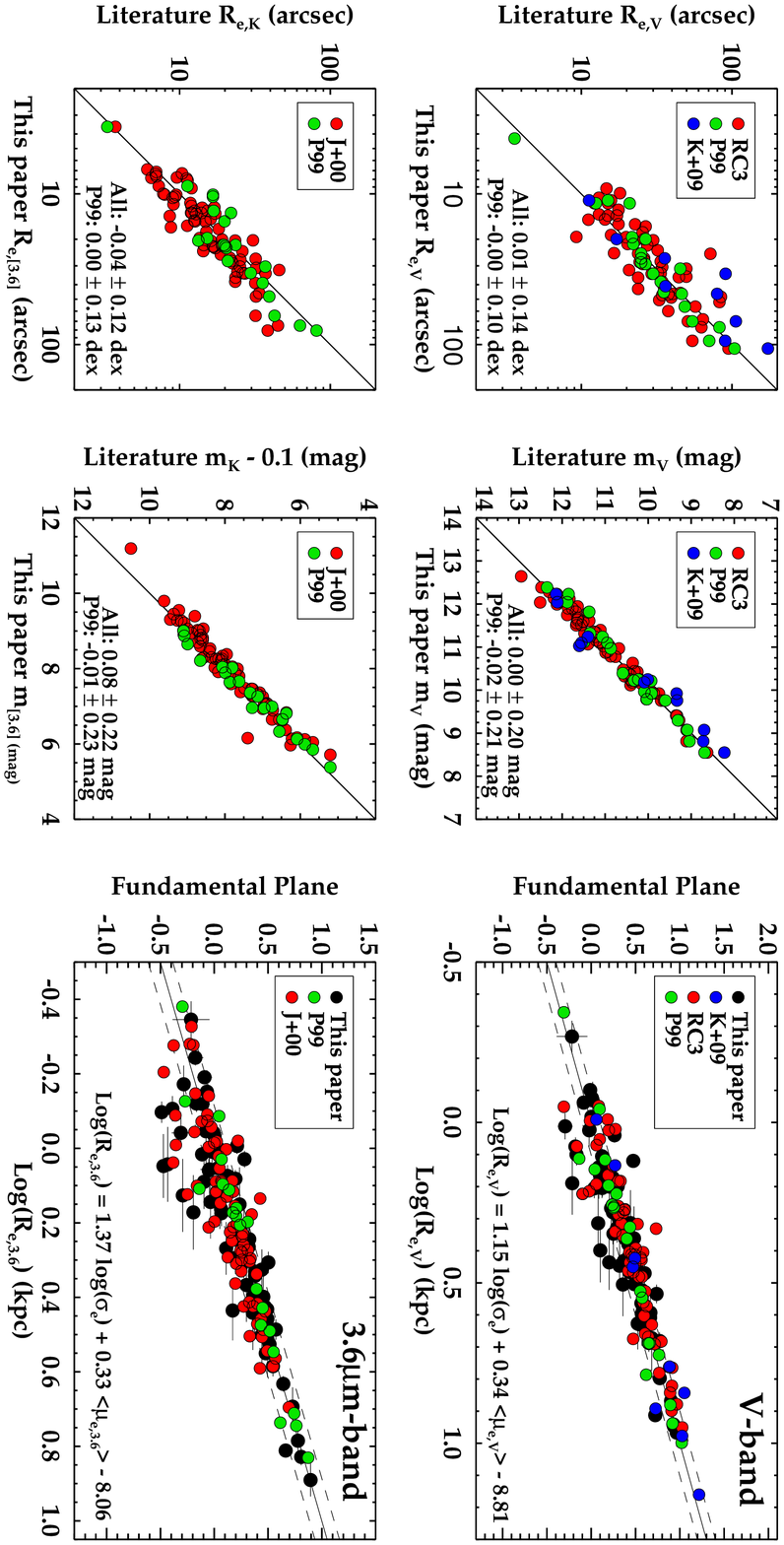}
\caption{Comparison of our aperture photometry with the literature. Four leftmost 
panels: \re\ and apparent magnitudes from this paper are compared to those in the 
literature (RC3, \citealt{rc3}; P99, \citealt{pahre99}; J+00, \citealt{jarrett00}; K+09,
\citealt{kormendy09}) for $V$- and $3.6\mu$m-bands. We subtracted 0.1 mag to the 
literature $K$-band values (i.e. the mean $K-$[3.6] colour for old stellar 
populations) to compensate for the colour difference. Mean offset and scatter 
are indicated for all sources (All) and P99 alone (as it is the only sample 
available in the two bands for the two quantities). Two rightmost panels: the 
best Fundamental Plane relations derived in \S~\ref{sec:FPfit} showing our data 
as well as the literature data for our sample of galaxies (see details in 
\S~\ref{sec:litcomp}). Dashed lines in the Fundamental Plane relations 
mark the $1\sigma$ uncertainty.}
\label{fig:litcomp}
\end{center}
\end{figure*}

\subsection{Literature comparison}
\label{sec:litcomp}

In order to test the reliability of our measurements, we have compared
them with published values in the literature. This exercise can reveal large
differences among sources, mainly depending on the depth of the data,
photometric band and most importantly the methodology used to derive the
relevant quantities. In Figure~\ref{fig:litcomp} we show the comparison of our
estimated values with a few references in the literature \citep{rc3, pahre99,
jarrett00,kormendy09}. When necessary, the literature major-axis \re\ estimates
have been transformed to geometric-mean radius to compare with our values
obtained from circular apertures. As shown in the figure, the agreement with the
different sources is generally good with a typical scatter of about 0.14 dex in
\re\ and 0.2 mag in apparent magnitude. The most notorious difference in our
sample is that of NGC\,4486 (M87), where our measured value in the $V$-band of
$106\arcsec$, contrasts with other values in the literature ($95\arcsec$,
\citealt{rc3}; $104\arcsec$, \citealt{pahre99}; $171\arcsec$,
\citealt{ferrarese94}; $194\arcsec$, \citealt{kormendy09}).

In addition, we show the best-fitting FP relations presented in
\S~\ref{sec:FPfit}, including data from the same sources. We carry out this
comparison to assess whether different methods to estimate the structural
parameters can affect our relations. In order to minimize uncertainties, we have
re-derived \se\ for each source's \re\ value. In addition, if not provided, our
estimated distances were used to convert \re\ from arcsec to kiloparsecs. In the
case of \citet{kormendy09}, we do not use their tabulated mean surface
brightnesses, but instead compute them ourselves from the total luminosity and
effective radii they provide in their Table~1 (columns 9 and 17). This is to
mimic as much as possible our procedure to compute the photometric quantities.
The figure shows that even though the structural parameters individually might
vary among the different sources, in combination they agree well within the
observed scatter, and thus our best-fitting parameters should not be biased in
any particular way.

\subsection{$V-[3.6]$ colours}
\label{sec:colors}

Colour measurements have been widely used in the past to extract first order 
information on the stellar content of galaxies and constraint different 
formation scenarios \citep[e.g.][]{white80,carlberg84}. Here we determine the 
effective colour
\begin{equation}
(V-[3.6])_{\rm e} = -2.5\log(L_{V}(<R_{{\rm e},V})/L_{\rm [3.6]}(<R_{{\rm e},V})) + {\rm const.}
\end{equation}
\noindent measured within a \rev\ aperture. The choice of \rev\ was preferred
over \reIR\ to match the aperture used to extract our \sauron\ spectroscopic
quantities (see \S~\ref{sec:sauron}). Aperture corrections, as devised in the
IRAC Instrument Handbook, have been taken into account when deriving the
colours. We will use the information from these parameters, together with
absorption line-strength indices, to establish the stellar population properties
of our galaxies in different regions of the scaling relations presented here.

Central colours (e.g. within \rev /8) were not derived due to the complexity in
matching the MDM and \textit{Spitzer} PSFs. Furthermore, $V-[3.6]$
colour gradients were also discarded given the presence of dust in many of our
galaxies, which introduces features in the colour profiles that cannot be
accounted for with a single linear relation. Nevertheless, an in-depth analysis
of colour profiles in the \sauron\ sample, using the
\textit{Spitzer}/IRAC $3.6$ and $4.5\mu$m bands, will be the subject of study in
Peletier et al. (2011, in preparation).\looseness-2

\section{SAURON QUANTITIES}
\label{sec:sauron}

In addition to the aperture photometry extracted from the MDM and
\textit{Spitzer}/IRAC images, we have determined a number of quantities from our
\sauron\ integral-field data that are key for the analysis presented in this
paper. These are parameters describing the richness in dynamical substructures
and the stellar content of the galaxies in our sample. They have been computed
following the procedures detailed in previous papers of the \sauron\ survey.
Here we will briefly summarise the main aspects and refer the reader to the
relevant papers for a full description of the methods employed. For convenience,
the final set of spectroscopic quantities is listed in
Tables~\ref{tab:saupars_ESO} and ~\ref{tab:saupars_Sa}.\looseness-2

\subsection{Kinematic quantities}
\label{sec:kinparams}

The stellar kinematics of our sample galaxies have been extracted following the
procedure outlined in Papers IX and X. Briefly, we used pPXF \citep{capem04} to
fit a linear combination of stellar templates from the MILES library
\citep{sanchez06} and derive the best mean velocity $V$ and velocity dispersion
$\sigma$ for each spectrum in our datacubes. In this paper we are mostly
interested in extracting the true first two velocity moments of the
line-of-sight velocity distribution (LOSVD), and therefore we deliberately do
not fit the higher order moments ($h_3$, $h_4$). We use the extracted $V$ and
$\sigma$ maps to compute \lR, a parameter that measures the specific angular
momentum within \rev\ and that has led to the new kinematical classification of
galaxies presented in Paper IX. Throughout this paper we will identify as
\textit{Slow Rotators} (hereafter SR) those galaxies with \lR$\leq0.1$, and as
\textit{Fast Rotators} (hereafter FR) the rest (as all the papers in the
\sauron\ series since Paper IX). We note that for this sample, the
notation is consistent with that based on the improved criterion defined in
\citet{emsellem11} for the galaxies in the ATLAS$^{\rm 3D}$
sample\footnote{http://purl.org/atlas3d} \citep{cappellari10}.

In order to plot some of the scaling relations in \S~\ref{sec:scaling}, we have
measured the mean stellar velocity dispersion within an \rev\ aperture. For that
purpose, we summed up all the spectra available within such a \textit{circular}
aperture and then computed $\sigma$ following the same procedure described
above. Whenever the aperture was not fully contained within the \sauron\ FoV, we
used the velocity dispersion calibration of Paper IV (equation~1) to correct our
values. For some aspects in \S~\ref{sec:FPscatter}, we have also measured the
mean stellar velocity dispersion within an \rev/8 aperture ($\sigma_{\rm{e},8}$). 
As in other papers in the survey (see Papers IV and XVII), we adopt a random 
error of 5\% for our measured velocity dispersion values.

Finally, we make use of the results in \citet[][ hereafter Paper XII]{davor08}
to describe the level of kinematic substructure present in our maps (e.g. inner
discs, kinematically decoupled cores, kinematic twists).

\subsection{Stellar population quantities}
\label{sec:stepopparams}

As well as stellar kinematic quantities, we have also measured line-strength
indices within \rev. In this paper, in order to be fully consistent with the
procedures employed to derive the stellar kinematics and to minimize the
uncertainties in the absolute calibration of the line-strength indices, we have
opted to measure the indices in the recently defined \textit{Line Index System}
(LIS) LIS-14.0\,\AA\ \citep[][hereafter VAZ10]{vaz10}. This method has the
advantage of circumventing the use of the so-called Lick fitting functions for
the model predictions, which requires the determination of often uncertain
offsets to account for differences in the flux calibration between models and
observations. The only required step to bring our flux-calibrated data to the
LIS-14.0\,\AA\ system is to convolve the aperture spectra to a total FWHM of
14.0\,\AA. The choice of LIS-14.0\,\AA\ over other proposed systems (e.g.
LIS-5.0\,\AA\ or LIS-8.4\,\AA) is imposed by the galaxy with the largest \se\ in
our sample (i.e. NGC\,4486).

\begin{figure}
\begin{center}
\includegraphics[angle=0,width=1\linewidth]{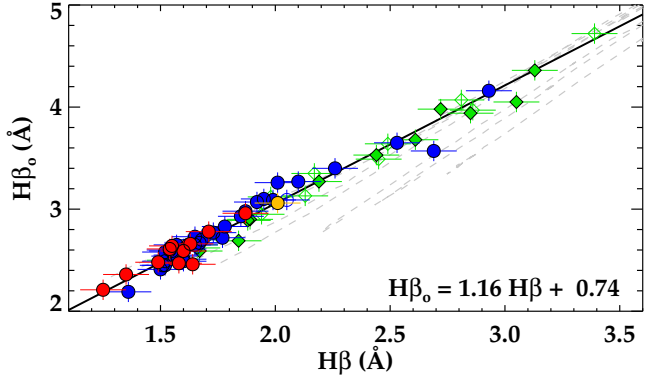}
\end{center}
\caption{Relation between the \hbeta\ and \hbetao\ line-strength indices for our 
sample galaxies. Colours and symbols as in Fig.\ref{fig:recomp}. The dashed gray 
lines mark the predictions of the \citet{vaz10} models for different metallicities.
([M/H]=[$-2.32,-1.71,-1.31,-0.71,-0.40,0.0,+0.2$]). Our sample appears to closely 
follow the predictions at high metallicities.}
\label{fig:hbetao}
\end{figure}

Besides the standard Lick indices \citep{worthey94} that can be measured within
our wavelength range (i.e. \hbeta, Fe5015, Mg$b$), we have also measured the
\hbetao\ index presented in \citet{cv09}. This new index is similar to the
classical \hbeta\ index, but it has the advantage of being less sensitive to
metallicity. For convenience, we show the relation between the two indices for
our galaxies in Fig.~\ref{fig:hbetao}. We however warn the reader that this
relation is necessarily biased by our sample selection. Since we are using
stellar population models with solar abundance ratios, we use the
[MgFe50]$^\prime$
index\footnote{[MgFe50]$^\prime$=0.5$\times$[0.69$\times$Mg$b$+Fe5015]} to
minimise the effects of $\alpha$-elements over abundances (see Paper XVII). For
galaxies with \rev\ larger than the \sauron\ coverage, we applied the
line-strength aperture corrections devised in Paper VI. As established in Paper
XVII, typical random errors for our measured values are 0.1\,\AA, with
systematic uncertainties being 0.06, 0.15 and 0.08\,\AA\ for \hbeta, Fe5015 and
Mg$b$ respectively. We assume the same aperture corrections and errors as
\hbeta\ for the \hbetao\ index.

\begin{figure}
\begin{center}
\includegraphics[angle=0,width=1.\linewidth]{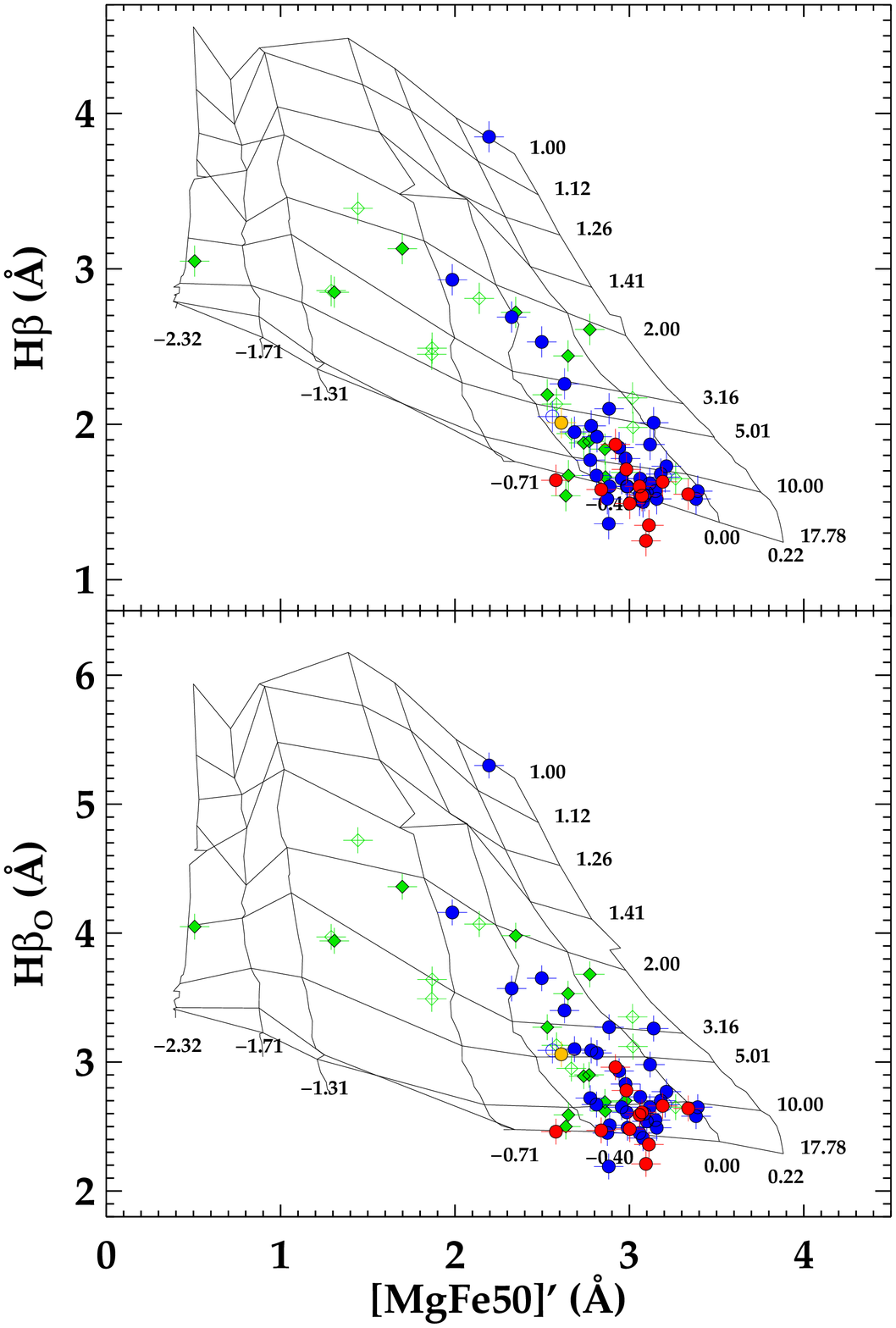}
\end{center}
\caption{Line-strength index relations for the \sauron\ sample galaxies in the
LIS-14.0\,\AA\ system \citep{vaz10}. Colours and symbols as in
Fig.~\ref{fig:recomp}. The line-strength indices presented have been measured
within a \rev\ aperture. Grids show model predictions for constant ages
(horizontal lines with labels in Gyr) and metallicities (vertical lines labelled
with [M/H] with respect to solar metallicity). The top panel shows the classical
\hbeta\ Lick index \citep{worthey94} versus [MgFe50]$^\prime$. The bottom panel
shows the same diagram, but with the the recently defined \hbetao\ index
\citep{cv09} instead of \hbeta.}
\label{fig:sauls}
\end{figure}

Figure~\ref{fig:sauls} shows the location of our galaxies in the \hbeta\ and
\hbetao\ vs [MgFe50]$^\prime$ diagrams in the LIS-14.0\,\AA\ system. The figure
illustrates the main reasons for adopting the \hbetao\ index over the
traditional Lick \hbeta\ index: (1) the \hbetao\ is more insensitive to
metallicity than \hbeta\ down to [M/H]$\approx-0.71$, which makes the diagram
more orthogonal, and (2) the vast majority of our galaxies fall within the
model predictions, which is crucial for a proper estimation of the stellar
mass-to-light ratios (\steml).

Throughout this paper we investigate the potential dependencies on age via the
line-strength index \hbetao. This is to be able to include the Sa galaxies in
the same diagrams. While the use of ages, metallicities and abundance ratios is
in general desired, estimates of these parameters from a single-stellar
population (SSP) analysis, as was done for the 48 E and S0 galaxies in Paper
XVII, are not recommended given the more continuous star-formation activity they
have experienced (see Paper XI for more details on these and other caveats). The
use of the \hbetao\ index will instead provide a robust first order indication
of the presence of young stars in our galaxies.

As is required in some of the relations we are presenting in this paper, we have
estimated the \steml\ of our sample galaxies in both the $V$- and
$3.6\mu$m-bands. We dedicate Appendices~\ref{app:models} and~\ref{app:ML} to the
description of the set of models and the methods used to derive these values.

\section{SCALING RELATIONS}
\label{sec:scaling}

In this section we show some of the classic scaling relations for the \sauron\
sample of early-type galaxies. Although much work has been devoted to these
relations in the literature, mostly separating galaxies by their morphological
classification, here we will focus on how deviations from scaling relations
depend on the kinematic information and stellar populations from our
integral-field data. Since we found no significant correlation with the level of
kinematic substructure or environment in any of the considered scaling relations
(demonstrated in Appendix~\ref{app:relations}), we focus on differences between
the SR E/S0, FR E/S0, and Sa galaxies.

In the following sections we derive linear fits of the form $y = \alpha \, x +
\beta$  to all relations, except the Fundamental Plane in \S~\ref{sec:FP}. We
started from the \textsc{fitexy}\footnote{Based on a similar routine by
\citet{press92}.} routine taken from the IDL Astro-Library \citep{landsman93}
which fits a straight line to data with errors in both directions, which we
extended to include possible correlations between the errors in both directions.
To find the straight line that best fits a set of $N$ data points $x_j$ and
$y_j$, with symmetric errors $\Delta x_j$ and $\Delta y_j$ and covariance
$\mathrm{Cov}(x_j,y_j)$, the routine minimizes
\begin{equation}
   \label{eq:chi2}
   \chi^2 = \sum_{j=1}^N \frac{(y_j-\alpha \, x_j-\beta)^2}
   {\Delta_{\mathrm{obs},j}^2 + \Delta_\mathrm{int}^2},
\end{equation}
where the combined observational error is given by
\begin{equation}
   \label{eq:obserror}
   \Delta_{\mathrm{obs},j}^2 = (\Delta y_j)^2 + \alpha^2 (\Delta x_j)^2 
   	- 2 \, \alpha \, \mathrm{Cov}(x_j,y_j),
\end{equation}
and $\Delta_\mathrm{int}$ is the intrinsic scatter, which is increased until the
value of $\chi^2$ per degrees of freedom is unity. Next, finding the changes in
$\alpha$ and $\beta$ needed to increase $\chi^2$ by unity, yields the
(1-$\sigma$) uncertainties on the best-fit parameters. The values of $x_j$ are
normalized by subtracting the corresponding observational quantities per galaxy
by the median value of all galaxies. This choice for the reference value (or
pivot point) minimizes the uncertainty in the fitted slope $\alpha$ and its
correlation with the intercept $\beta$. The details and benefits of this
approach are described in \citet{tremaine02}.

In deriving the errors $\Delta x_j$ and $\Delta y_j$, we include the
uncertainties in all contributing quantities, i.e., the errors in the distances,
the photometric quantities, the kinematic quantities (stellar velocity
dispersion) and the stellar population quantities. Correlations in the
photometric quantities are taking into account via our Monte Carlo realizations
of \S~\ref{sec:r14_fit}); for example, when deriving the error in \mue$=-2.5
\log \left[ L_{\rm e}/(\pi R_{\rm e}^2)\right]$, we first compute from all
realizations of $L_{\rm e}$ and \re\ the corresponding \mue\ values, and then
derive the error as the (biweight) standard deviation. Similar Monte Carlo
simulations of all quantities are used to calculate the covariance
$\mathrm{Cov}(x_j,y_j)$, which can be significant especially when the same
quantities are used in both $x_j$ and $y_j$, such as \re\ in the Kormendy
relation and the distance in the size-Luminosity relation.

\begin{figure*}
\begin{center}
\includegraphics[angle=0,width=0.99\linewidth]{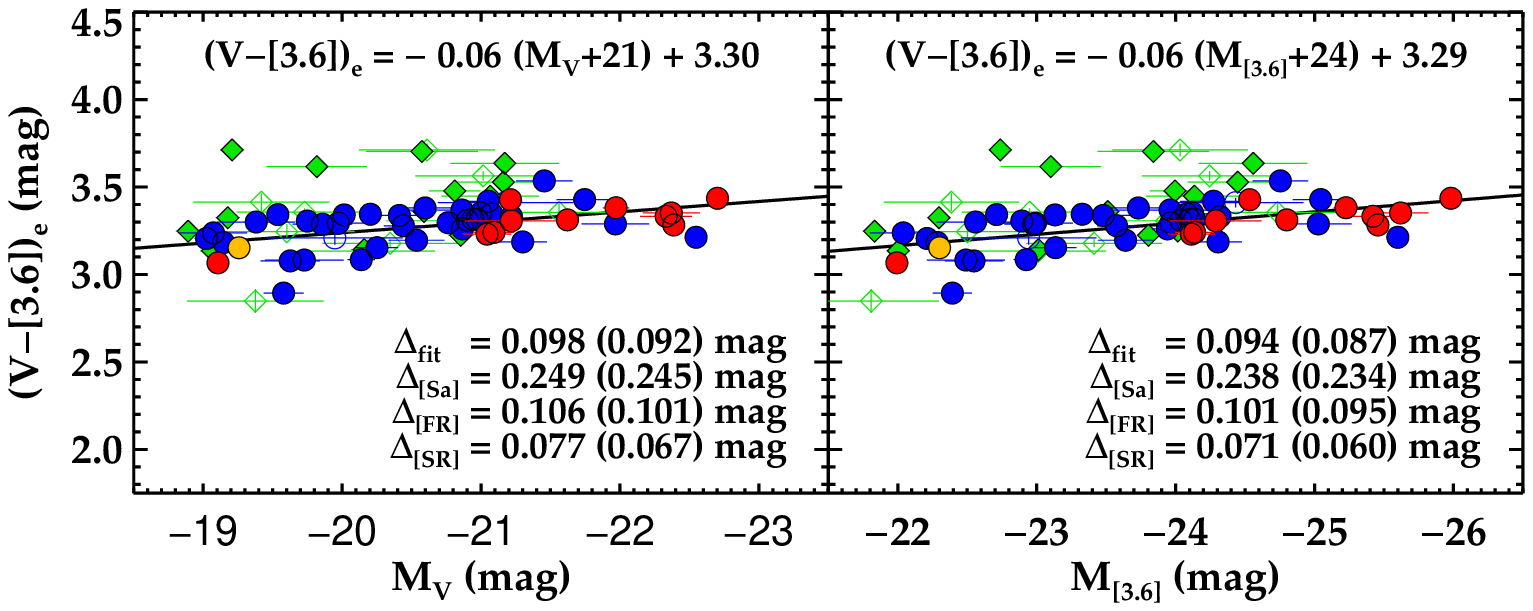}
\end{center}
\vspace{0.25cm}
\begin{center}
\includegraphics[angle=0,width=0.99\linewidth]{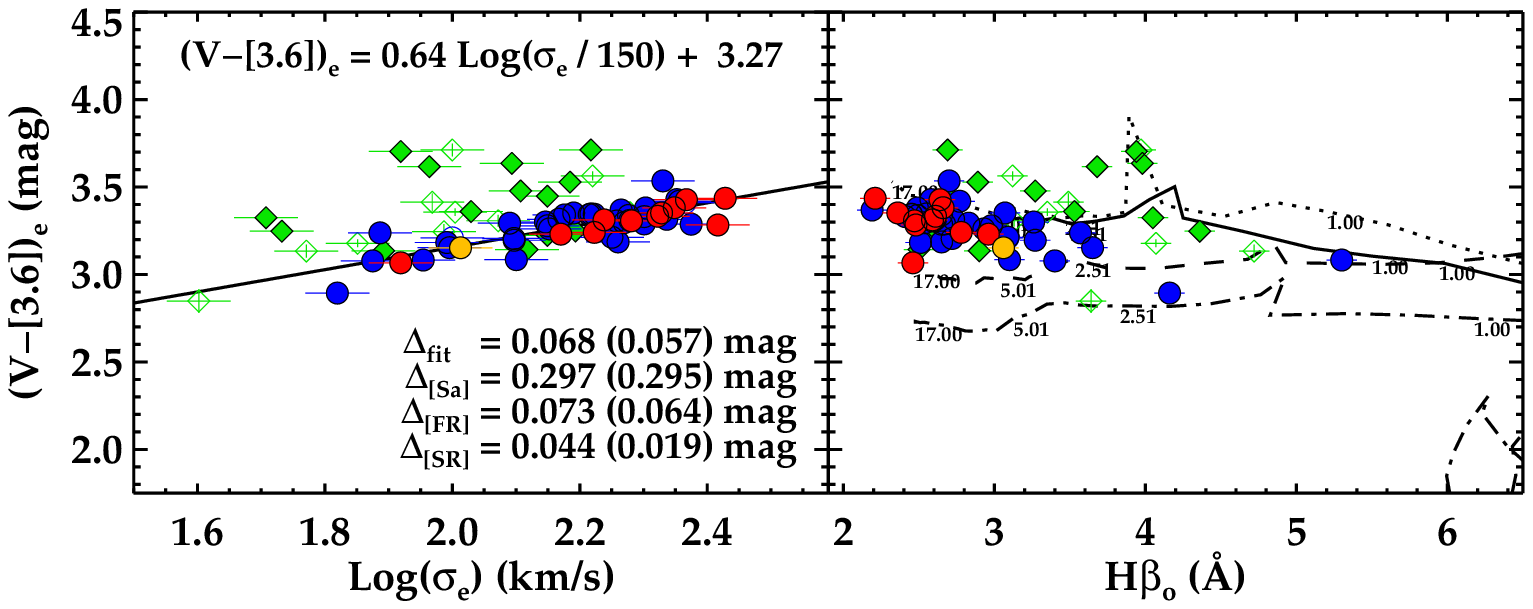}
\end{center}
\caption{Colour-magnitude, colour-\se, and colour-\hbetao\ relations for the
\sauron\ sample of galaxies. The best-fit linear relations to the E/S0 galaxies
with good distance determinations are indicated in each panel and are shown with
solid black lines. $\Delta_{\rm fit}$ indicates the observed and intrinsic
scatters (the latter in parentheses) for the galaxies used to obtain the fit,
while $\Delta_{[\dots]}$ represent the measured values for the different
families around the best-fit relations. Colours and symbols as in
Figure~\ref{fig:recomp}. The lines in the bottom-right panel correspond to the
\citet{marigo08} model colour predictions combined with MILES \citep{vaz10}
Lick/IDS line-strength indices for different metallicities (see \S~\ref{sec:cmd}
for details): [M/H]=[$-0.71$ (dotted-dashed), $-0.40$ (dashed), $0.00$ (solid),
$+0.20$ (dotted)]. For reference, some ages (in Gyr) are indicated along the
different metallicity tracks. Peaks seen in the colour tracks are due to an
enhancement in the production rate of AGB stars, that causes a transient red
phase in the integrated SSP colours \citep[see][]{gb98}.}
\label{fig:CMR}
\end{figure*}

Unless mentioned otherwise, for consistency and hence for easy comparison with
most publications on scaling relations for early-type galaxies, only E/S0
galaxies with reliable distance estimates are included in the fits (i.e. 46
objects). The resulting best-fit relations are written in the figures
themselves, while the best-fit parameters and corresponding errors are given in
Table~\ref{tab:params}. The values indicated by $\Delta_\mathrm{fit}$ represent
the scatter around the best-fit relation in the vertical direction, i.e., the
(biweight) standard deviation of $y_j - \alpha \, x_j - \beta$, for the galaxies
$j$ used in the fit. The value in parentheses is the intrinsic scatter, after
subtracting, in quadrature, from $\Delta_\mathrm{fit}$ the (biweight) mean of
the combined observational errors $\Delta_\mathrm{obs,j}$, which, as indicated
in equation~(\ref{eq:obserror}), include potential covariances between the
variables. The intrinsic scatter values and corresponding error estimates can be
found in Table~\ref{tab:params}. We confirmed that well within these errors, the
intrinsic scatter values are the same as $\Delta_\mathrm{int}$ in
equation~\ref{eq:chi2} when the value of $\chi^2$ per degrees of freedom is
unity. The other quoted scatter values $\Delta_{[\dots]}$ are based on
\emph{all} galaxies within the family/families indicated by the index in square
brackets, i.e., also the galaxies with distances based on their recession
velocities.

\begin{figure*}
\begin{center}
\includegraphics[angle=0,height=11cm]{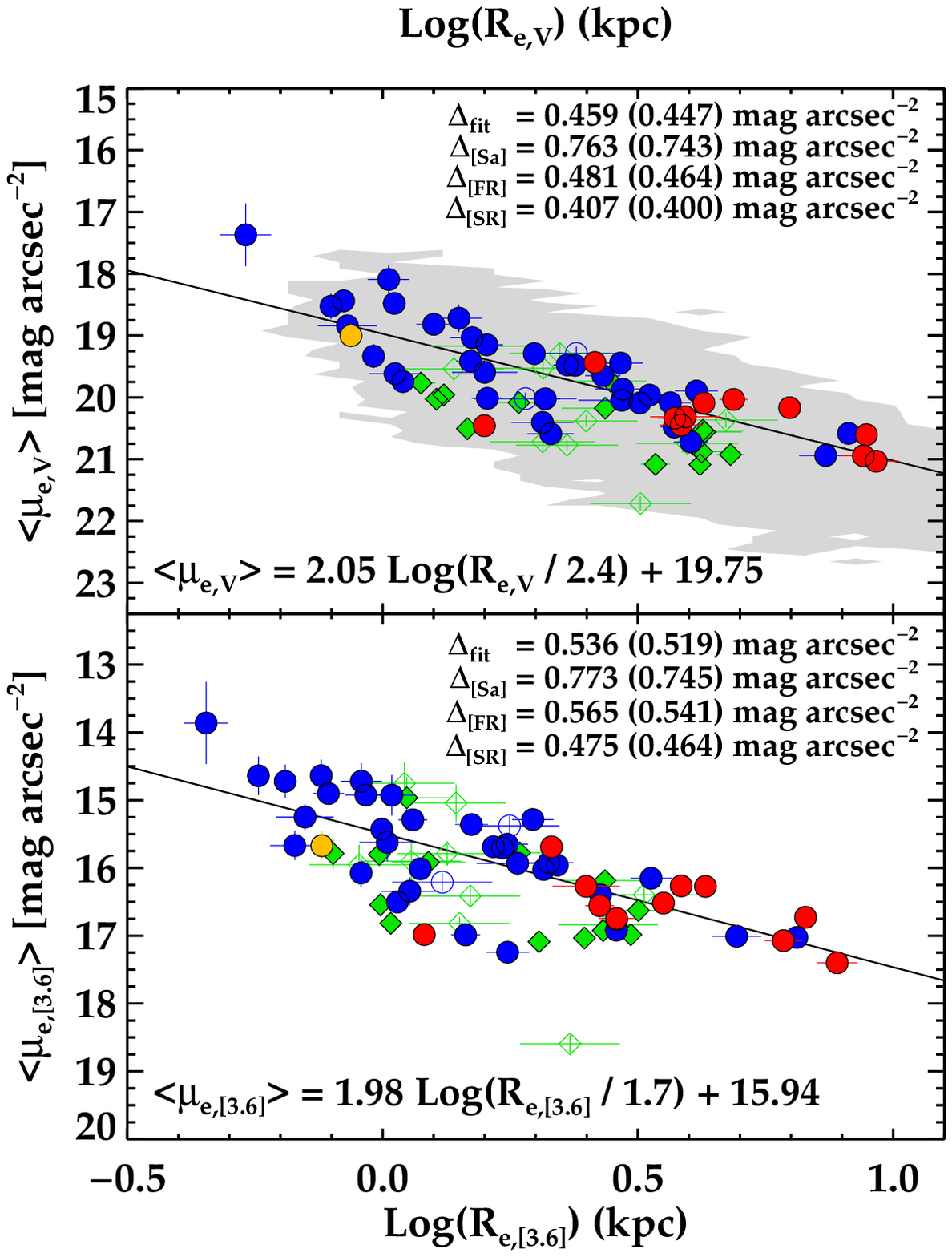}
\hspace{0.7cm}
\includegraphics[angle=0,height=11cm]{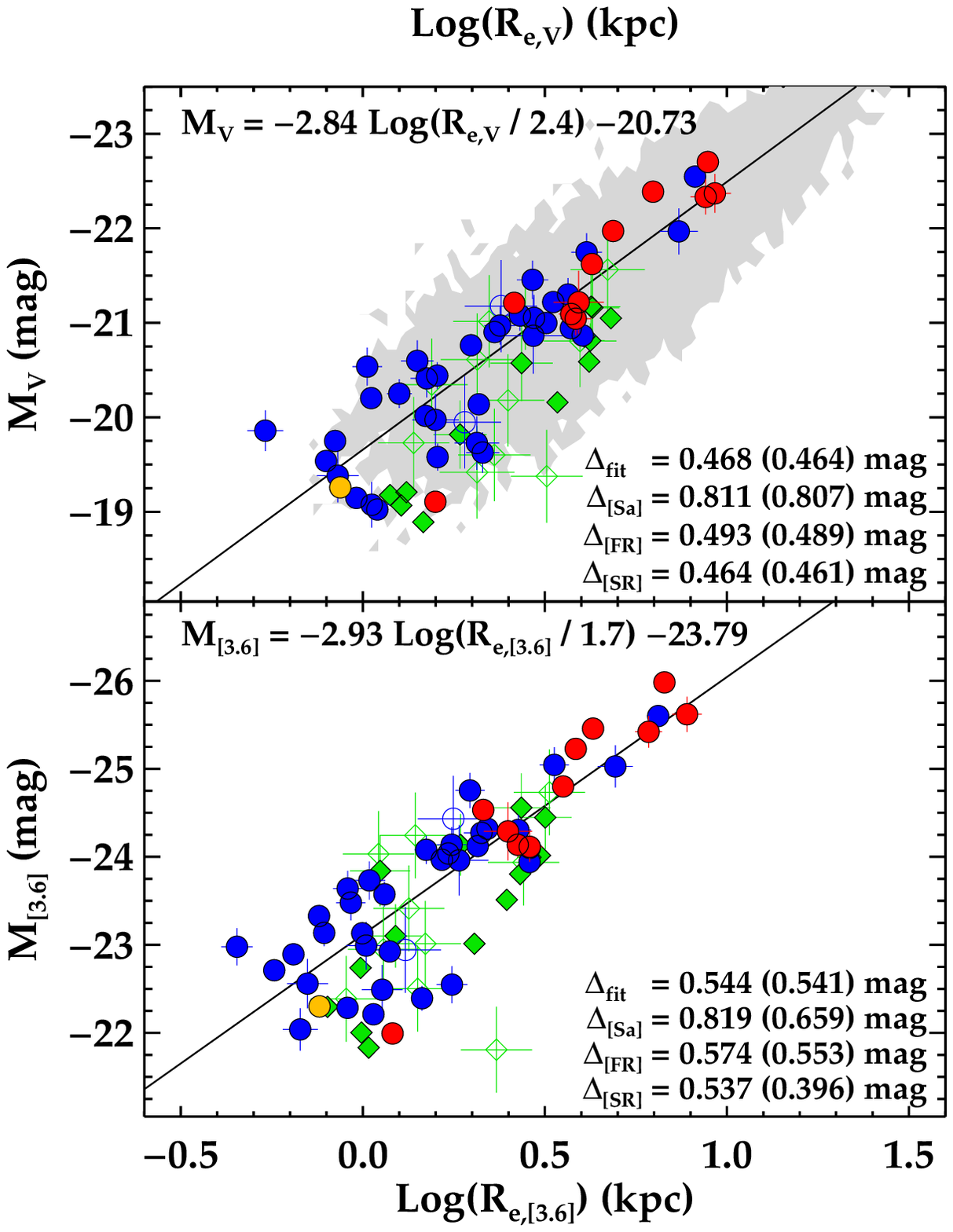}
\end{center}
\caption{Kormendy (KR) and size-luminosity (SLR) relations for the galaxies in
the \sauron\ sample in the $V$- and $3.6\mu$m-bands. Symbols as in
Fig.~\ref{fig:recomp}. The thick solid line is the best fit relation (as
indicated in the equation in each panel). $\Delta_{\rm fit}$ indicates the
observed and intrinsic scatters (the latter in parentheses) for the galaxies
used to obtain the fit, while $\Delta_{[\dots]}$ represent the measured values
for the different families around the best-fit relations. The grey area marks
the extent of the \citet{bernardi03a} sample.} 
\label{fig:Kormendy}
\end{figure*}

\subsection{Relations with colour}
\label{sec:cmd}

We presented a set of CMRs in different photometric bands in Papers
XIII and XV. In both cases the most deviant objects were galaxies displaying
widespread ongoing star formation. Here we bring a new element to these works by
investigating the location of galaxies with distinct kinematic properties (e.g.
SR/FR) in the CMR.\looseness-1 

In Figure~\ref{fig:CMR} (top panels) we plot the CMRs for our sample galaxies
using the effective colour ($V-[3.6]$)$_{\rm e}$. There are two main sources
that can change the slope and increase the intrinsic scatter in these CMRs:
young stellar populations and dust extinction. They have opposite effects. Young
populations will, in principle, shift galaxies down from the relation (galaxies
become bluer), while dust will make galaxies appear redder than their stellar
populations. Our sample contains galaxies which clearly exhibit dust (most
notably Sa galaxies). Since we have made no attempt to correct for internal
extinction, the position of these galaxies in this CMR cannot be used to infer
information about their stellar populations. The reddest objects (NGC\,1056,
NGC\,2273, NGC\,4220, NGC\,4235 and NGC\,5953) in the plotted relations are
indeed Sa galaxies with very prominent dust lanes. 

The bottom-left panel of Fig.~\ref{fig:CMR} shows the colour versus \se\
relation. Here both the FR and SR families define a much tighter sequence than
in the CMRs shown in the top panels. Particularly striking are the relations
defined by the SRs and FRs, with intrinsic scatters of only 0.019 and 0.064 mag
respectively, as opposed to the Sa galaxies with deviations of about 0.30 mag,
similar to those of the CMRs. The Sa galaxies systematically populate a region
above the best-fit relation. Given the relative insensitivity of \se\ to dust
(at least much less than absolute magnitudes in the optical), this diagram is
potentially useful to remove dusty objects in samples destined for CMR studies.
It also confirms that the average dust content is larger in spiral galaxies than
in earlier-type systems. Despite these deviant objects, it is worth noting that
there are a few Sa galaxies in the region populated by SR/FR galaxies.

As already shown in Papers XI, XIII and XV, our sample contains a number of
objects with clear signs of widespread star formation (NGC\,3032, NGC\,3156,
NGC\,4150, NGC\,3489, NGC\,4369, NGC\,4383, NGC\,4405 and NGC\,5953). Following
the arguments outlined above, one might expect these galaxies to display the
bluest colours (in the absence of dust). Surprisingly, this is not the case. In
order to understand this behaviour, we have plotted in the bottom-right panel of
Fig.~\ref{fig:CMR} the effective colour versus the \hbetao\ index measured
within the same aperture. In addition to our data points, we also include colour
predictions from \citet[][hereafter
MAR08]{marigo08}\footnote{http://stev.oapd.inaf.it/cmd, version 2.2} together
with line-strength predictions from the recently released MILES
models\footnote{http://miles.iac.es} (VAZ10). This combination of colours and
line indices is consistent, in the sense that both predictions are based on the
same set of isochrones \citep{girardi00} and were computed for a Kroupa
(\citeyear{kroupa01}) initial mass function. These isochrones take into account
the latest stages of stellar evolution through the thermally pulsing asymptotic
giant branch (AGB) regime to the point of complete envelope ejection. These
models show that the dependence of the ($V-[3.6]$)$_{\rm e}$ colour with age is
rather subtle even for young stellar populations, being much more sensitive to
metallicity than age. This is the reason behind the lack of a strong signature
of the youngest objects in the CMRs of Figure~\ref{fig:CMR}. 

Now that we have assessed that metallicity is the main driver of the
($V-[3.6]$)$_{\rm e}$ colour, we can deduce from the colour-\se\ relation that 
there must be a strong correlation between metallicity and \se. We have
made an attempt to determine this relation by computing the best linear
relation between metallicity ($[M/H]$) and $V-[3.6]$ colour in the MAR08
models ($V-[3.6] = 0.73 [M/H] + 3.28$, see Appendix~\ref{app:models}), and then
substituting that in the colour - \se\ relation found here. The selected 
galaxies are predominantly old and thus are well reproduced by single stellar 
population models. This yields a relation of the form
\begin{equation}
	[M/H] =  0.88 \, (\pm 0.07) \log(\sigma_{\rm e}) - 1.92 \, (\pm 0.16).
\end{equation}
%
%
%
%
%
%
A similarly strong correlation was presented in Paper XVII ($[Z/H]($\re$) = 0.32
\log($\se$)-0.82$) from an independent set of model predictions
\citep[][]{schiavon07} and methodology (i.e. using line-strength indices rather
than colours). Note, however, that the two relations cannot be directly compared
as the one presented in Paper XVII is based on non-solar scaled stellar
population models, while the one derived here is not. In order to compare them
we have determined the metallicity of our models using the relation from the
MAR08 models and then adjusted it assuming $[Z/H]=[M/H]+0.75[\alpha/$Fe$]$,
where the factor 0.75 is a constant that depends on the element partition used
for the models (R. Schiavon, private communication) and $[\alpha/$Fe$]$ is taken
from Paper XVII. The slope of the resulting relation still appears to be a
factor of 2 steeper than that of Paper XVII. The apparent inconsistency between
the two determinations might indicate some issues in the modelling of the always
complex AGB phase. The comparison with other relations in the literature
\citep[e.g.][]{jorgensen99,kuntschner00,trager00,thomas05,sanchez06b,proctor08,
allanson09,graves09} is not straightforward either since, in addition to the
stellar population models used, they are mostly based on central aperture
measurements (both metallicity and velocity dispersion). 

%
%

\begin{figure}
\begin{center}
\includegraphics[angle=0,width=0.99\linewidth]{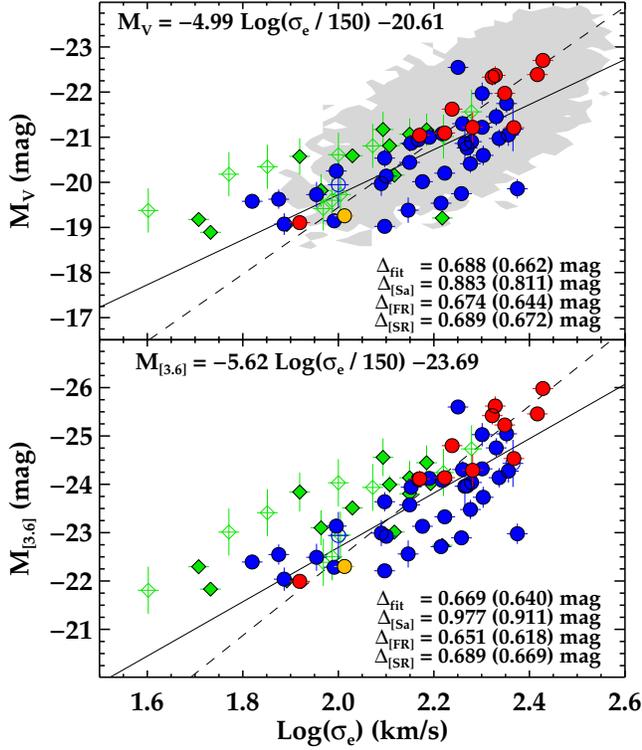}
\caption{Faber-Jackson relation for the galaxies in the \sauron\ sample in the
$V$- and $3.6\mu$m-band. Symbols as in Fig.~\ref{fig:recomp}. The solid line is
the best fit relation (as indicated in the equation on the top left in each
panel). The dashed line shows the best fit to the SR family (e.g. predominantly
old systems). $\Delta_{\rm fit}$ indicates the observed and intrinsic scatters
(the latter in parentheses) for the galaxies used to obtain the fit, while
$\Delta_{[\dots]}$ represent the measured values for the different families
around the best-fit relations. The grey area marks the extent of the
\citet{bernardi03a} sample.}
\label{fig:faberJ}
\end{center}
\end{figure}

\subsection{Kormendy and size-luminosity relations}
\label{sec:kormendy}

The KR and SLR for the \sauron\ sample of galaxies are shown together in
Fig.~\ref{fig:Kormendy}. As already noticed by other authors, the value one
obtains for these relations critically depends on the adopted selection
criteria. In particular, the Malmquist bias caused by selecting galaxies
according to their luminosities can have a large impact on the resulting
best-fit relation \citep{nigoche08}. The volume and luminosity range of our
survey means we lack small, high surface brightness, faint objects \citep[as
estimated from a comparison with the parent sample of galaxies in the
ATLAS$^{\rm 3D}$ survey,][]{cappellari10}. Nonetheless, limiting our fit to
galaxies with distances below 25 Mpc (where our sample does not suffer those
limitations) results in best-fit relations that are identical within the
uncertainties. The comparison of our data in the KR and SLR with that of B03
shows that the strongest bias introduced in our analysis by the sample selection
is in mean surface brightness. With the exception of NGC\,5845 (the galaxy with
the smallest \re), our galaxies seem to populate rather homogeneously the area
defined by the B03 sample for \muev$\lesssim21$ mag arcsec$^{-2}$. This implies
a shallower KR and steeper SLR relation than in the more complete B03 sample. 

The observed scatter around the best-fit relations is consistent with that
found in others studies for galaxies within the same magnitude range
\citep[e.g.][]{nigoche08}. It appears that there is still a significant amount
of intrinsic scatter in both relations, with the SR family displaying the
smallest values. SR galaxies tend to populate the high luminosity end (with the
exception of NGC\,4458), while FR extend across the whole luminosity range
displayed by our sample. In general, Sa galaxies appear to be fainter (by
$\approx$0.8 mag) than SR/FR galaxies of the same size. This result is
consistent with the observed strong dependence of these relations on
morphological type \citep[e.g.][]{courteau07}.

One interesting feature is the lack of large, high surface brightness (or small
and very luminous) galaxies. This \textit{zone of avoidance} (ZOA) was already
noted by \citet{bender92} and has been later better defined by
\citet{donofrio06} and \citet{cappellari10}.\looseness-2

\subsection{Faber-Jackson relation}
\label{sec:faber}

In Figure~\ref{fig:faberJ}, we show FJRs for our sample. As in previous
relations, the SR family tends to occupy the high luminosity end of the
relations. Sa galaxies, however, deviate from the best-fit relations in the
sense that they systematically populate high luminosities for a given \se\ (for
\se$\lesssim$125 \kms). Disc galaxies are often characterised by the
Tully-Fisher relation \citep{tf77} instead of the FJR. This is because in low
mass systems the maximum rotational velocity is a much better tracer of total
mass than the traditionally measured central velocity dispersions. The use of a
large aperture for the velocity dispersion measurement (\se) presented here
significantly helps to improve matters by using a parameter close to the
integrated second moment (see Paper IV). We believe that the observed offset is
mainly a stellar population effect, whereby Sa galaxies are on average younger
(and thus have in general lower stellar mass-to-light ratios) than earlier
types, as already discussed in the literature (e.g. Paper XI). This effect is
further supported by the best-fitting relation for the SRs (e.g. predominantly
old systems). As shown in the figure, the slope of the relation for the SRs is
larger (dashed line) than the original fit and in better agreement with B03.
Nevertheless, the number of SRs in our sample is rather limited and thus
complete samples (e.g. ATLAS$^{\rm 3D}$ survey) are required to confirm the
observed trend.\looseness-2

The FJR for our sample appears to 'bend'  toward low \se\ values at low
luminosities. There is a hint of this feature in the relations for the Coma
Cluster by \citet{bower92b} and \citet{jfk96}, but it is somehow not seen in
other relations in the literature based on much larger numbers of galaxies (e.g.
B03, \citealt{labarbera10}). This might be because galaxies with low velocity
dispersions are often removed from the samples. Irrespective of selection
effects, the slopes derived here are slightly shallower than previously found
\citep[e.g.][]{pahre99}. In part this is because it is common to use central
velocity dispersions, which will increasingly deviate from the values reported
here (measured in a larger aperture) for larger galaxies, as the velocity
dispersion gradients are steeper in the inner parts of larger galaxies. 

\begin{figure*}
\begin{center}
\includegraphics[angle=90,width=0.99\linewidth]{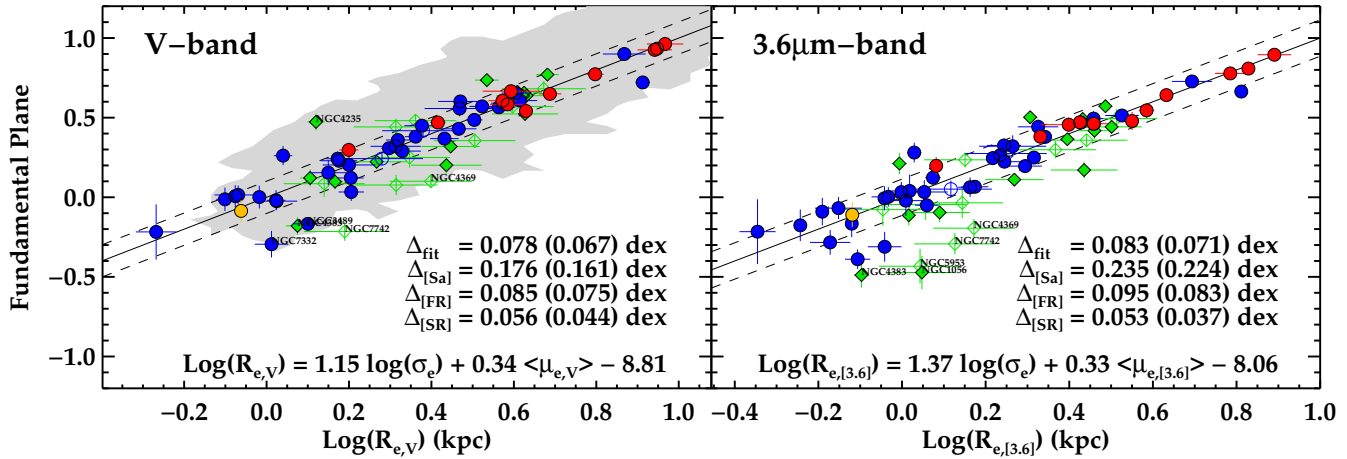}
\caption{Edge-on views of the Fundamental Plane relation for the galaxies in the
\sauron\ sample of galaxies in $V$- and $3.6\mu$m-bands. Symbols as in
Fig.~\ref{fig:recomp}. The dashed lines mark the $1\sigma$ uncertainty (measured
in log(\re)) considering all objects. The solid line is the best fit relation
(as indicated in the equation in each panel). $\Delta_{\rm fit}$ indicates the
observed and intrinsic scatters (the latter in parentheses) measured along the
log(\re) axis for the galaxies used to obtain the fit, while $\Delta_{[\dots]}$
represent the measured values for the different families around the best-fit
relations. The grey area marks the extent of the \citet{bernardi03a} sample.\looseness-2}
\label{fig:FP}
\end{center}
\end{figure*}

\section{FUNDAMENTAL PLANE}
\label{sec:FP}

In this paper we adopt the following notation for the FP:
\begin{equation}
\label{eq:FP}
\log(R_{\rm e}) = \alpha\ \log(\sigma_{\rm e}) + \beta\ \langle\mu_{\rm e}\rangle + \gamma.
\end{equation}
The multivariate nature of the FP makes common least-squares minimisation
algorithms not suitable to determine the main parameters ($\alpha, \beta,
\gamma$). 
The literature is vast on alternative methods
\citep[e.g.][]{graham97,labarbera00,saglia01}, several of which we tested to
yield consistent best-fit parameters within the corresponding estimated errors.
The FP results presented in this paper are determined via an orthogonal fit by
minimizing the sum of the absolute residuals perpendicular to the plane 
\citep[e.g.][]{jfk96}. As in \S~\ref{sec:scaling}, we include the uncertainties in
all contributing quantities and take into account correlations in the
photometric quantities via our Monte Carlo realizations.

In addition to the fitting scheme, sample selection biases in any of the
quantities of Eq.~\ref{eq:FP} can have an important impact on the resulting
coefficients \citep{nigoche09}. We have tested the sensitivity of our best-fit
parameters to magnitude and distance (the two main selection criteria in our
sample). Our sample is a good representation, in terms of luminosity, of the
complete early-type galaxy population up to a distance of 25\,Mpc, beyond which
we lack small, faint, high surface brightness objects. Nevertheless, the
inclusion of galaxies at larger distances hardly alters the best-fit parameters.
In terms of luminosity alone, the best-fit parameters remain within the
uncertainties as long as the faintest level in absolute magnitude is brighter
than M$_{\rm V}\approx-19$ and M$_{\rm [3.6]}\approx-22$ magnitudes. Therefore,
while distance has no major impact on the best-fit parameters, the removal of
the fainter objects in our sample does.
As shown in Fig.~7 of \citet{hb09}, the expected bias in the $\alpha$ 
parameter due to our sample selection in terms of absolute magnitude is below 
10\%.
As in other relations we restrict our fit to
E/S0 galaxies with good distance determinations (i.e. filled SR/FR galaxies in
the relations). From the comparison of our data with those of B03, we have
established that the lack of objects with \muev$\lesssim21.0$ mag arcsec$^{-2}$
in our sample effectively means that we miss galaxies with \re$\gtrsim$12 kpc.
We have checked, using the B03 sample, that this effect does not seem to 
bias our results in any particular way.\looseness-1

\subsection{Classic FP}
\label{sec:FPfit}
In Figure~\ref{fig:FP} we present the FP for our sample galaxies in both the
$V$-band and $3.6\mu$m-band. Sa galaxies share the same relation defined by
early-type systems and are not clearly displaced below the relation as they
usually are. We believe that the main reason for this behaviour is the way we
have derived the quantities involved. When including spiral galaxies in the FP,
it is customary to measure the photometric and spectroscopic quantities only
within the bulge dominated region \citep[e.g.][]{fb02}. Here we have measured
the half-light radius, mean surface brightness and velocity dispersion in a
consistent and homogeneous way for all galaxies regardless of morphological
type. The importance of also computing the velocity dispersion within the same
half-light radius was already shown in Figure~\ref{fig:litcomp}: even though the
photometric quantities taken from different literature sources for galaxies in
our sample individually might vary substantially, when combined with a
consistently measured integrated velocity dispersion, they all fall on our
best-fit classic FP. In \S~\ref{sec:FPscatter} we investigate the effects in the
best-fitting parameters if non-consistent velocity dispersion measurements are
used.

The best-fit coefficients in both bands are somewhat inconsistent with some of
the most recent works in the literature \citep[e.g.][]{pahre99, bernardi03b,
proctor08, hb09, labarbera09}. Although difficult to assess accurately, we
believe that the main differences are due to a combination of photometric bands
employed, fitting methods, and specially sample selection (since we include the
Sa galaxies, unlike previous works). There is nevertheless a large disparity in
the literature regarding the value of the FP coefficients. There seems to be,
however, a consensus on the fact that $\alpha$ changes gradually with
photometric band (increasing with wavelength) while $\beta$ remains almost
constant \citep[e.g.][]{hb09}. The relations derived here share that property.
The parameters also deviate from the virial theorem predictions (i.e. $\alpha=2$
and $\beta=0.4$ in the notation used here), an effect known as the
'\textit{tilt}' of the FP. At first sight it might seem surprising that the
$\beta$ coefficient in the $V$-band, more sensitive to young stellar populations
than infrared bands, has a similar value to that in the $3.6\mu$m-band. It is
important to remember, however, that the $\gamma$ parameter, while assumed
constant, is in fact a function of the total mass-to-light ratio (\ml), which is
not necessarily constant among galaxies. In fact this \ml\ depends on the
stellar populations, i.e., one can express the total \ml\ as a function of the
stellar mass-to-light ratio \steml\ and the relative fraction of luminous to
dark matter. It seems that during the fitting process stellar population effects
conspire to keep $\beta$ constant while affecting $\alpha$ and
$\gamma$.\looseness-2

\subsection{Outliers and residuals}
\label{sec:FPres}
Not all the galaxies in our sample appear to follow the main FP relation.
NGC\,4382 is a well-known shell galaxy \citep[e.g.][]{kormendy09} which must
have suffered some recent interaction event. It contains a rather young nuclei
(3.7 Gyr, see Paper XVII), it is the youngest of all the non-CO detected
\sauron\ galaxies \citep{combes07} and it also displays a prominent central
velocity dispersion dip (see Paper III). NGC\,1056, NGC\,4369, NGC\,4383, NGC\,5953 and
NGC\,7742 are Sa galaxies that contain a significant fraction of young stars
within \re\ (see Paper XI). NGC\,3489 is one of the few early-type galaxies with
a strong starburst in the last 2 Gyr (see Paper XVII). NGC\,7332 was already
recognised as an outlier of the FP in the optical and near-IR bands by
\cite{fb02}. Although it shows signs of widespread young populations, it is not
at the level of other early-type galaxies. Nevertheless, this object is peculiar
in that it is rich in kinematic substructure \citep{fb04} and might have
suffered from a recent interaction with NGC\,7339. Moreover, the  presence of
two gaseous counter-rotating discs \citep{pb96} suggests that this galaxy might
not be in dynamical equilibrium. It is interesting to note that the object in
our sample with the strongest presence of widespread young populations,
NGC\,3032, does not deviate at all from the best fits.\looseness-2

\begin{figure}
\begin{center}
\includegraphics[angle=0,width=0.99\linewidth]{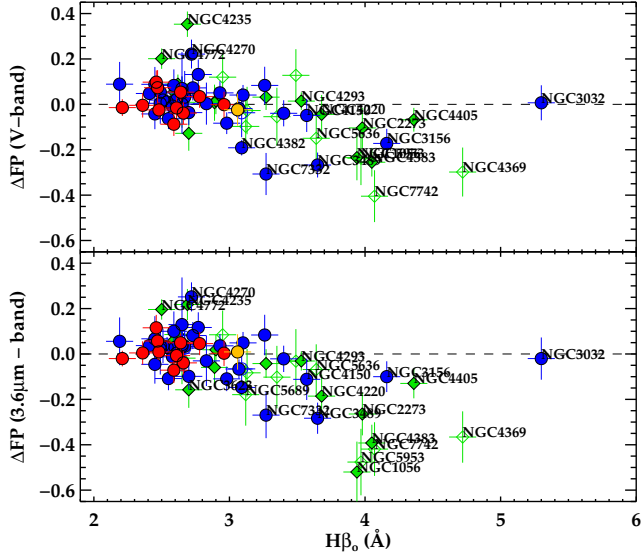}
\caption{Residuals (along the log(\re) axis) of the Fundamental Planes in
Fig.~\ref{fig:FP} as a function of the \hbetao\ line-strength index. Symbols as
in Fig.~\ref{fig:recomp}.}
\label{fig:FPres}
\end{center}
\end{figure}

Individual inspection of the outliers to understand the reasons for their
unusual location with respect to the main FP strongly suggests young stellar
populations as a common factor. We investigate this further by exploring whether
the residuals correlate with stellar population quantities, such as \hbetao\
shown in Figure~\ref{fig:FPres}. A correlation with the \hbetao\ index indeed
exists in both bands. With the exception of NGC\,3032 noted above, galaxies with
young stellar populations are systematically located below the main relation.
These residuals suggest that, while metallicity gradients may contribute to the
scatter in the FP, young populations are the dominant factor.\looseness-2

\subsection{Influence of discs}
\label{sec:fakegal}

\begin{figure}
\begin{center}
\includegraphics[angle=0,width=0.99\linewidth]{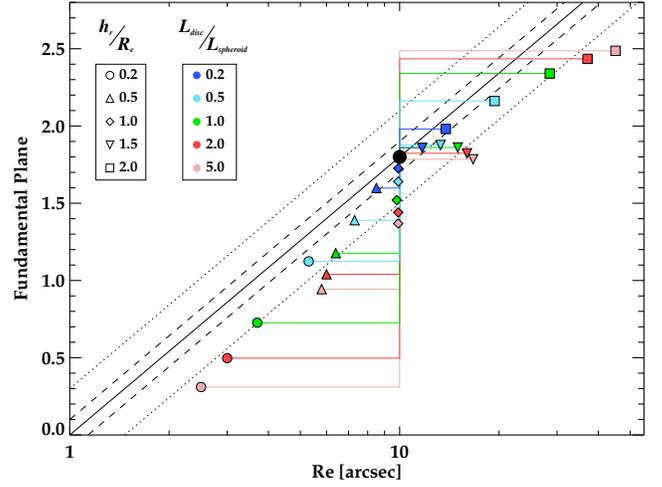}
\caption{Simple experiment to show the effects of adding a disc to a spheroid 
on the location of galaxies in the FP (see \S~\ref{sec:fakegal}). 
The black dot marks the location of a pure spheroid, by construction on the FP. 
The black solid line marks the FP, and for reference dashed and dotted 
lines denote the 1$\sigma$ and 3$\sigma$ scatter in the relation shown in 
Figure~\ref{fig:FP}. Different symbols indicate the ratio between the scale 
length of the added disc ($h_r$) and the effective radius (\re) of the 
underlying spheroid. Colours correspond to different fractions of disc light 
($L_{\rm disc}$) relative to the spheroid ($L_{\rm spheroid}$).} 
\label{fig:fake}
\end{center}
\end{figure}

In order to disentangle the possible effects on the FP due to structural
non-homology in galaxies from those due to young stellar populations, we have
carried out a simple experiment illustrated in Figure~\ref{fig:fake}.
We have defined a mock
early-type galaxy as a spheroid with a light profile described by a de
Vaucouleurs' law. By construction, we put this spheroid on a generic FP (black
filled circle and solid line). In order to assess the impact of an additional
disc component, we added exponential discs to the spheroid, with scale lengths
of respectively 0.2, 0.5, 1.0, 1.5 and 2.0 times the \re\ of the spheroid
(circles, upward triangles, diamonds, downward triangles and squares,
respectively). For reference, we note that the typical values are $\gtrsim5$ for
late-type galaxies \citep{balcells07}. Additionally, we set the light of the
disc to be 0.2, 0.5, 1.0, 2.0 and 5.0 times the light of the spheroid (dark
blue, cyan, green, red, and pink, respectively). We have analysed the growth
curves of all these mock galaxies in the same way as the real galaxies presented
in this paper. We have assumed the same \se\ for all mock galaxies as its value
does not seem to depend on the S\'ersic $n$ used in the growth curves of our
real galaxies.

The figure shows that with an increase of the disc light fraction, the mock
galaxies tend to deviate more and more from the original FP. This is
expected as the increase of light will lead to higher surface brightnesses,
which in turn will push the objects below the relation. It is interesting,
however, that the deviations only seem significant when the light fraction of
the disc relative to the bulge is above a factor 2, regardless of the size of
the disc relative to the spheroid. We note the large excursions in the \re\ and
vertical directions for large disc light fractions.
Figure~\ref{fig:fake} also shows the effect of a compact or extended disc
relative to the underlying spheroid. Compact bright discs tend to shift the
galaxies towards the left of the relation, while extended ones shift towards the
right. It is important to realise that the presence of a disc has little
influence on the end location of a galaxy, unless the disc is rather bright
relative to the spheroid. 

There are, in principle, other sources for the scatter and tilt of the FP that
we are not accounting for here in detail: projection effects, rotation,
structural homology, etc. All those, besides stellar populations, have been
studied in detail in the literature \citep[e.g.][]{guzman93, saglia93, ps94,
ciotti96, jfk96, graham97, ps97, pahre98, mobasher99, trujillo04b}, although
with different and sometimes conflicting conclusions. Recent studies, however,
point to stellar populations and dark matter as the main drivers \citep{treu06,
cappellari06, bolton08, graves10}. In this paper we focus on the effect of
stellar populations only.

Results on the importance of young stellar populations in determining the
location of galaxies in the FP were already highlighted in Paper XIII, based on
\textit{GALEX} observations in the ultraviolet regime. Although the analysis in
that paper was limited to fewer galaxies than the sample presented here, it
appears that an important fraction of the tilt and scatter of the UV FPs is due
to the presence of young stars in preferentially low-mass early-type galaxies.
Triggered by those results, we have attempted here to take a step further and
correct the FP for stellar population effects in order to bring \textit{all}
galaxies into a common FP relation. The results of this exercise are shown in
the following section.

\subsection{FP corrected for stellar populations}
\label{sec:FPml}
  
A first test to assess the importance of young stellar populations on the tilt
of the FP is to derive the best-fit relation after restricting the sample to
predominantly old galaxies, i.e. SRs and FRs with \hbetao$\leq 3.0$\,\AA, and
good distance determinations (35 objects). This simple experiment yields the
best-fit parameters $\alpha = 1.39$, $\beta = 0.35$ and $\gamma = -9.61$ in the
$V$-band and $\alpha = 1.63$, $\beta = 0.34$ and $\gamma = -8.85$ in the
$3.6\mu$m-band. The dramatic change compared to Fig.~\ref{fig:FP} is the sudden
increase of the $\alpha$ parameter, while $\beta$ remains almost unaltered. This
object selection reduces the tilt of the FP and brings it much closer to the
virial prediction. Nevertheless it is worth noting that the two $\alpha$ values
(i.e. one for each band) are not close to each other, as one might expect if we
had fully corrected for stellar population effects and hence were left with
effects, in particular the luminous-to-dark matter fraction, that are
insensitive to wavelength. 

\begin{figure}
\begin{center}
\includegraphics[angle=0,width=0.99\linewidth]{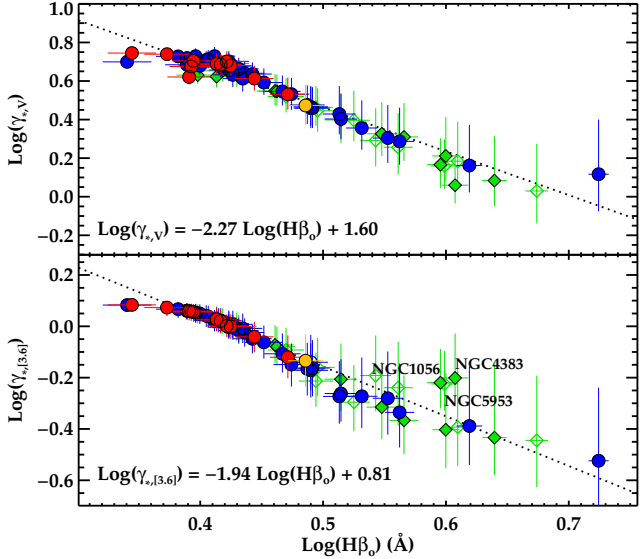}
\end{center}
\caption{\steml\ vs \hbetao\ relations for the \sauron\ sample galaxies in the
$V$ and 3.6$\mu$m-band. Best-fitted (log-log) linear relations are indicated in
each panel and plotted with a dotted line. \steml\ values have been determined
as explained in Appendix~\ref{app:ML}.}
\label{fig:mlhb}
\end{figure}

Considering the residuals in Fig.~\ref{fig:FPres}, one can take this experiment
a step further and try to fully compensate for stellar populations (young and
old) simultaneously. \hbetao\ is not a quantity that directly enters the FP
equation as defined in Eq.~\ref{eq:FP}. However, as stated in
Sect.~\ref{sec:FPfit}, the $\gamma$ coefficient does depend on stellar
populations via \steml. We have made an attempt to estimate the effective
\steml\ for each galaxy based on the combined MILES+MAR08 models, assuming a
combination of two single stellar populations as the baseline star formation
history (see Appendices~\ref{app:models} and \ref{app:ML}). The relations
between our computed \steml\ and observed \hbetao\ index are shown in
Fig.~\ref{fig:mlhb}, one for each photometric band. It appears that \hbetao\ is
strongly correlated with our measured \steml\ and thus could be used as a rough
surrogate for \steml. This behaviour was already shown in Paper IV for the
\hbeta\ index. The fitted relations, despite being strong, have two important
shortcomings: (1) they over-predict \steml\ at the low \hbetao\ end and (2) they
are rather uncertain for \hbetao\ values above 3.0\,\AA. The first effect is an
inherent limitation of the models used, i.e. any given set of models predicts a
maximum \steml\ for a given initial mass function. The second effect is related
to the wide range of allowed \steml\ for a given \hbetao\ value, as parametrised
by the set of two single stellar population models used here.

There are a few notable exceptions to the strong correlation at 3.6$\mu$m. These
are objects located very close to the peaks of the $V-[3.6]$ colour in the
MAR08 models for solar metallicities or above (i.e. NGC\,1056, NGC\,4383 and
NGC\,5953; see Fig.~\ref{fig:CMR}). Their \hbetao\ values range from 3.8 to
4.2\,\AA. Since our \stemlIR\ predictions are made after correcting for the colour
difference with \stemlV\ (see Appendix~\ref{app:models}), it is not completely
surprising that the \stemlIR\ values in that region somehow deviate from the
trend defined by neighbouring objects. However, sudden jumps in \steml\ of that
magnitude seem unrealistic for these galaxies, and thus we opted to correct their
values using the linear \steml\ vs \hbetao\ relation in that band presented in
Figure.\ref{fig:mlhb}.

\begin{figure*}
\begin{center}
\includegraphics[angle=90,width=0.99\linewidth]{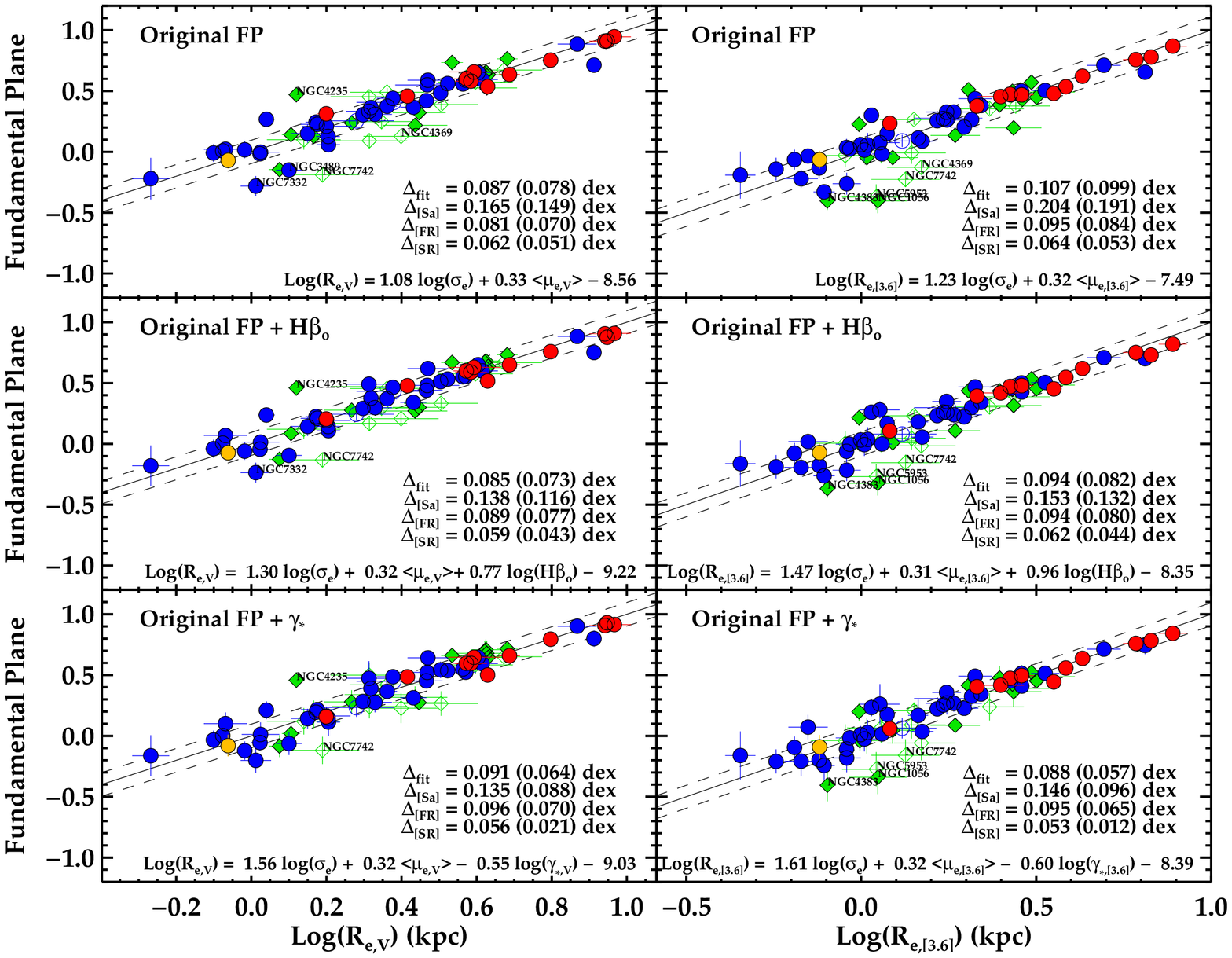}
\end{center}
\caption{Edge-on views of the Fundamental Plane relations for the galaxies in
the \sauron\ sample in the $V$- and $3.6\mu$m-band. Symbols as in
Fig.~\ref{fig:recomp}. (Top row) Original FP relation. (Middle row) Original FP
relation with a \hbetao\ term. (Bottom row) Original FP relation with a \steml\
term. The dashed lines mark the $1\sigma$ uncertainty (measured along the 
log(\re) axis) considering all objects. The solid line is the best fit relation 
(as indicated in the equation in each panel). $\Delta_{\rm fit}$ indicates the
observed and intrinsic scatters (the latter in parentheses) measured along the
log(\re) axis for the galaxies used to obtain the fit, while $\Delta_{[\dots]}$
represent the measured values for the different families around the best-fit
relations.}
\label{fig:FPML}
\end{figure*}

\begin{figure*}
\begin{center}
\includegraphics[angle=90,width=0.99\linewidth]{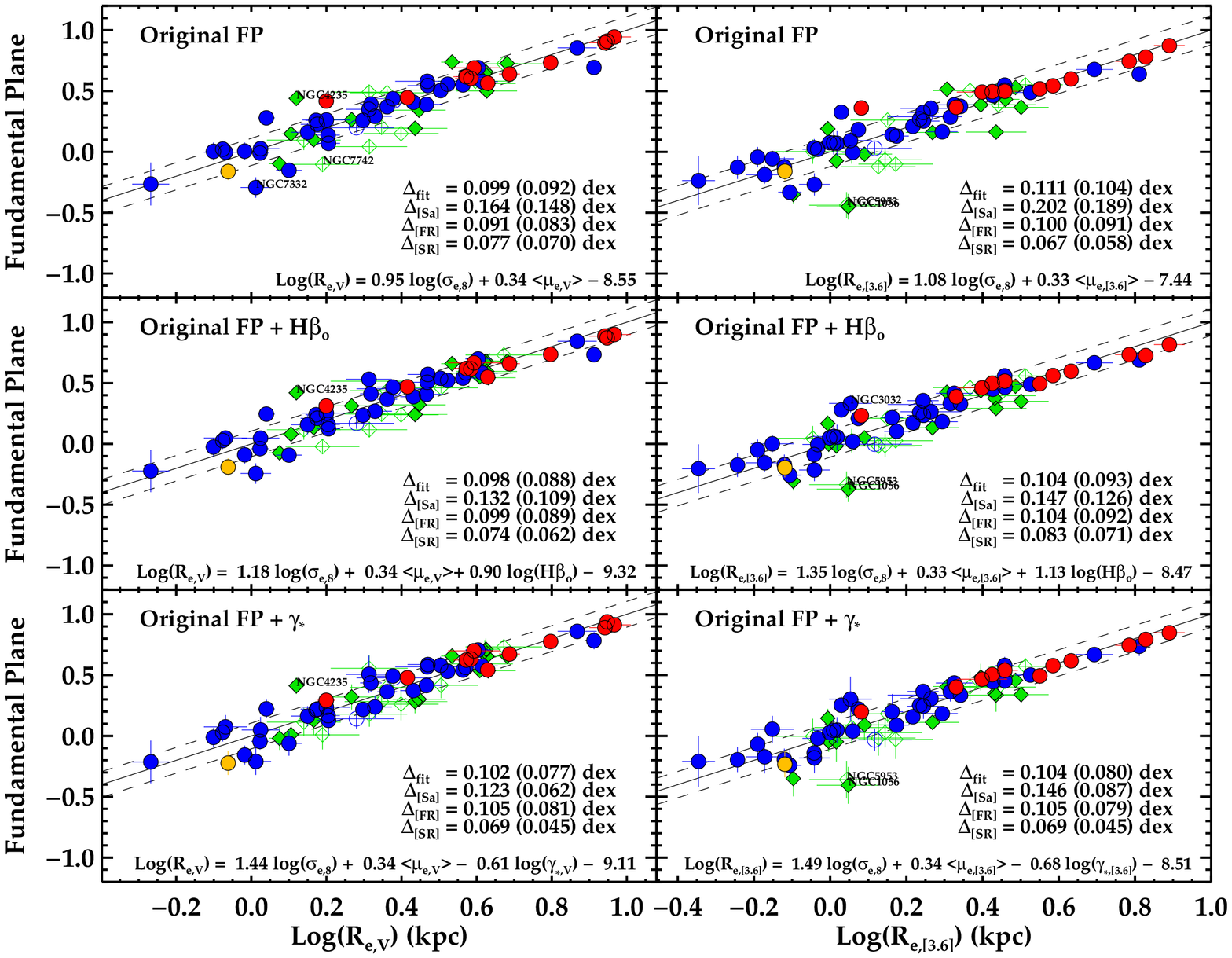}
\end{center}
\caption{Edge-on views of the Fundamental Plane relations for the galaxies in
the \sauron\ sample in the $V$- and $3.6\mu$m-band. Here \se/8 ($\sigma_{\rm
e,8}$) is used instead of \se. Symbols as in Fig.~\ref{fig:recomp}. (Top row)
Original FP relation. (Middle row) Original FP relation with a \hbetao\ term.
(Bottom row) Original FP relation with a \steml\ term. The dashed lines mark the
$1\sigma$ uncertainty (measured in log(\re)) considering all objects. The solid
line is the best fit relation (as indicated in the equation in each panel).
$\Delta_{\rm fit}$ indicates the observed and intrinsic scatters (the latter in
parentheses) measured along the log(\re) axis for the galaxies used to obtain
the fit, while $\Delta_{[\dots]}$ represent the measured values for the
different families around the best-fit relations.}
\label{fig:FPML_Re8}
\end{figure*}

Considering the nearly linear log-log relations between \steml\ and \hbetao, we
can re-write the original FP equation in terms of these variables:
\begin{equation}
\label{eq:FPml1}
\log(R_{\rm e}) = \alpha^\prime \log(\sigma_{\rm e}) + 
                 \beta^\prime \langle\mu_{\rm e}\rangle + 
                 \gamma^\prime +
                 \delta^\prime \log({\rm H}\beta_o),
\end{equation}
\begin{equation}
\label{eq:FPml2}
\log(R_{\rm e}) = \alpha^\prime \log(\sigma_{\rm e}) + 
                 \beta^\prime \langle\mu_{\rm e}\rangle + 
                 \gamma^\prime +
                 \delta^\prime \log(\gamma_\ast),
\end{equation}
\noindent where $\gamma^\prime$ now depends (mainly) on the luminous-to-dark 
matter ratio.

In Figure~\ref{fig:FPML} we plot the results of this experiment. In order to
make a meaningful estimate of the effects of the stellar populations, we
consider \textit{all} objects with good distance estimates in the fit (60
galaxies, as opposed to the fits in Fig.~\ref{fig:FP} where only 46 E and S0
with good distances were considered). We perform this test in both bands. The
top row in Fig.~\ref{fig:FPML} shows the resulting fits using the classic FP
definition (eq.~\ref{eq:FP}), whereas the middle and bottom panels present the
fits including the \hbetao\ and \steml\ terms (equations~\ref{eq:FPml1} and
\ref{eq:FPml2}). It seems that the inclusion of the Sa galaxies in the fits of
the original FP relation (top panels) results in smaller slopes, which
translates into larger departures from the virial expectations. 

Our first attempt to bring deviant objects closer to the relation is to add the
\hbetao\ term in the FP relation. Unfortunately, leaving all the input
parameters unconstrained results in very poor fits that, while successful in
bringing back the most deviant galaxies, display much larger scatters than
previous relations derived in this paper and, more importantly, have
unreasonable coefficients. In order to get meaningful fits, we have instead
fixed the \hbetao\ terms to the coefficients expected from the fits to the
original FP residuals versus \hbetao\ shown in Fig.~\ref{fig:FPres}. The results
of fixing this term are shown in the middle panels. These fits are able to bring
the young objects much closer to the relations and at the same time to slightly
reduce the intrinsic scatters. The $\alpha^\prime$ coefficients have increased
with respect to the original relation, getting them a step closer to the virial
prediction, while the $\beta^\prime$ values have only changed within the
uncertainties ($\approx0.01$). The main disadvantage of \hbetao\ as a substitute
to \steml\ can be seen in the case of NGC\,3032. As already shown in
Fig.~\ref{fig:mlhb}, this galaxy is a deviant object in the \steml\ vs \hbetao\
relation, and this translates in it being pushed out of the relation (its
measured \hbetao\ value over-predicts its true \steml).

Finally, in the bottom panels of Figure~\ref{fig:FPML}, we add the \steml\ term
in the FP relation. In this case the unconstrained fits result in physically
meaningful coefficients, while at the same time bringing the most deviant points
closer to the relation. The best-fit parameters reduce the FP tilt compared to
the original relations by almost 50\% in each band. It is important to remark
that best-fit coefficients in both bands are now consistent within the
errors, as expected if we had corrected for the effects of stellar populations.
The coefficients, at least in the 3.6$\mu$m-band, are also within the
uncertainties of those obtained by fitting the old galaxies only (see above).
While in the $V$-band most of our young objects get closer to the relation, in
the 3.6$\mu$m-band this is not the case (even though we have corrected their
\steml\ estimates based on Fig.~\ref{fig:mlhb} as explained above). In fact,
these objects cannot be brought back to the 3.6$\mu$m-band relation even with
the most extreme \steml\ values allowed by our stellar population fitting
procedure. 

As an additional exercise (not shown here) we have studied the effect of
using the ($V-[3.6]$)$_{\rm e}$ colour, instead of \hbetao\ or \steml, in
reducing the tilt of the FP. As expected, given the poor sensitivity of this
colour to age (as shown in Fig.~\ref{fig:CMR}), this quantity does not help
reducing the tilt. At the same time it also demonstrates that metallicity cannot
be a major player in producing the tilt (since this colour correlates strongly
with metallicity, see Appendix~\ref{app:models}).

It is important to remind the reader that we have only corrected for the effects
produced by stellar populations, and partly by rotation (by using \se). Still
there are a number of other factors that influence the final location of
galaxies in the FP. The complex manner in which star formation and dust
compensate each other might be at the heart of the location discrepancies of
some of the objects in the $V$- and 3.6$\mu$m-band FPs. Nevertheless, the
results of our exercise demonstrate that it is possible to correct for stellar
population effects and bring most galaxies to a common relation, regardless of
their morphological type and photometric band employed by including sensible
estimates of the stellar mass-to-light ratios. This approach was previously
adopted by \citet{ps96} in a sample of elliptical galaxies and has been more
recently exploited by other groups on much larger samples, though still
restricting the analysis to early-type systems
\citep[e.g.][]{allanson09,hb09,graves10}. Our results are largely consistent
with those.

\subsection{Scatter in the FP}
\label{sec:FPscatter}
 
One of the most striking features observed in our fits of the FP is the very
tight relation defined by the SR galaxies in both bands. This is not a totally
unexpected result as SR galaxies are uniformly old, but it is remarkable how the
trend is kept even for smaller and fainter galaxies (as SR galaxies extend over
the whole range in luminosity of our sample). The FR family displays slightly
larger rms ($\Delta_{\rm [FR]}$) values. The scatter of the Sa galaxies appears
to be the largest of the cases we have studied here. The different panels in
Fig.~\ref{fig:FP} and \ref{fig:FPML} show the observed as well as the intrinsic
scatter (within brackets) for each family. For easily comparison with other
works in the literature we choose to provide them measured along the log(\re)
direction.\looseness-2

The first thing to notice for the classic fits in Fig.~\ref{fig:FP} and
\ref{fig:FPML} is the differences between the values reported. The scatters for
the Sa galaxies in the first figure are larger than in the second. Conversely
the rms for the SR/FR families is smaller in the first figure than in the second
one. This is simply due to the fact that for determining the FP we fit only the
SR/FR galaxies in Fig.~\ref{fig:FP}, while Fig.~\ref{fig:FPML} includes the Sa
galaxies as well. If we focus on this last figure we observe, apart from a
change in the slopes, an improvement (i.e. a decrease) of the scatter for
\textit{all} families when we include the \hbetao\ and \steml\ terms to correct
for the effects of young populations. It is also interesting to note that once
corrected for this effect, the intrinsic values found for the different families
indicate that while FR and Sa galaxies are consistent within the uncertainties,
they are clearly different from those of the SRs. This finding emphasises the
results presented in Paper IX and subsequent papers in the SAURON series
suggesting, based on our kinematic classification, that these may be
intrinsically a different kind of galaxies. This clear distinction is not found
between morphologically classified elliptical and lenticular galaxies.

The comparison of the scatter presented here with others in the literature has
to be treated with caution as, while most works concentrate in E/S0 galaxies,
the different samples were selected with very different criteria in terms of
luminosity or velocity dispersion. In addition, the methods used to fit the
relation can affect this comparison, although the minimisation of the residuals
via an orthogonal fit is the most widely adopted. There is a general agreement
that the observed scatter is consistent among photometric bands, a result we
share. The typical observed scatter for a general population of E/S0 galaxies is
around 0.10 dex in log(\re) \citep[e.g.][]{jfk96,pahre98,zibetti02,labarbera10},
with decreasing values ($\sim$0.07 dex) as soon as low velocity dispersion
($\sigma\lesssim70$\kms) or fainter galaxies (M$_{r}\gtrsim-23$ mag) are removed
from their samples \citep[e.g.][]{jfk96,bernardi03b,labarbera08,gargiulo09}. The
values measured here for the SR/FR are in good agreement with those in the
literature for general samples of E/S0 galaxies. In the specific case of
\citet[][]{jfk96}, the lowest observed scatter, 0.047 dex, is measured for
galaxies with no presence of discs and M$_{r}\lesssim-23.1$ mag. They report
that this value is consistent with the measurement errors. The closest set of
galaxies in our sample matching those constraints is the SR family. 
For this group, we find a very small amount of intrinsic scatter left,
$0.021\pm0.029$ dex and $0.012\pm0.050$ dex in the V-band and 3.6mu-band
respectively. The uncertainties in the our estimates suggest that the scatter 
around the best-fit FP are fully explained by the observational errors.

Another interesting point to discuss is whether the use of a large aperture
velocity dispersion, instead of a central value, helps to reduce the intrinsic
scatter in the relation. This was nicely discussed in \citet{graham97} and
\citet{bus97}, using samples of elliptical and lenticular galaxies and dynamical
models with some assumptions on the shape of the velocity ellipsoid. The
\sauron\ spectrograph gives us now the chance to repeat this exercise using real
data. We have measured the velocity dispersion within \rev/8 ($\sigma_{\rm
e,8}$) for our sample of galaxies and then fit the FP in the same manner done in
Fig.~\ref{fig:FPML}. This is presented in Fig.~\ref{fig:FPML_Re8}. The different
panels show that the use of a small aperture velocity dispersion affects the
resulting fits in two important ways: (1) increasing the intrinsic scatter of
the relations for all families of objects, (2) the measured parameter $\alpha$
has lower values than those obtained when \se\ is used. The first point is
mainly due to the fact that with the smaller aperture we are probing regions of
the galaxies where intense star formation is taking place or inner discs exist,
and thus the measured velocity dispersion is more representative of areas around
the nuclei than the galaxy as a whole. The second effect is likely produced by
the fact that $\sigma_{\rm e,8}$ systematically departs from \se\ (i.e. it is
larger) for increasing \se\ values. This, together with the expected relation
between \re\ and \se\ (given the SLR and FJR), helps to reduce the measured
slope of the FP relation. Overall, we estimate an increase of around 15\% in the
intrinsic scatter and an average 10\% decrease in $\alpha$ as a result of using
a central value for the velocity dispersion term. It is also worth noting that
the scatter of the Sa galaxies has not increased as much as for the SR and FR
families. This supports the idea that star formation in early-type systems takes
place predominantly in their central regions, while for Sa galaxies this can
happen over a much larger region (see also Paper XI), and therefore it does not
necessarily make much difference to measure the velocity dispersion in a small
or large aperture. Alternatively, this result could be due to the shallower
nature of the stellar velocity dispersion profiles of the Sa galaxies compared
to those of the SR/FR galaxies.  

\section{Conclusions}
\label{sec:conclusions}

In this paper we report results from photometric follow-up conducted in the
context of the \sauron\ project. We use ground-based MDM
$V$-band and \textit{Spitzer}/IRAC 3.6$\mu$m-band imaging to characterise our
sample of E, S0 and Sa galaxies. We perform aperture photometry to derive
homogeneous half-light radii, mean effective surface brightnesses and total
magnitudes. Combined with the \sauron\ integral-field spectroscopic
observations, this allows us to explore and understand the location of the
galaxies in the main scaling relations as a function of the level of rotation,
kinematic substructure, stellar populations and environment.

A number of conclusions can be derived from the work presented and are 
summarised in the following points:

\begin{itemize}

\item The level of kinematic substructure (as determined from
our stellar kinematic maps) or the environment do not lead to a preferred
location in any of the scaling relations investigated. However, this is 
subject to the potential biases introduced by our sample selection. Our 
\textit{cluster} environment is defined by the Virgo Cluster and Leo group.

\item The Slow Rotator (SR) galaxies define tighter relations than Fast Rotator
(FR) galaxies. While this is not totally unexpected (i.e. SR are uniformly old
systems), it is remarkable how the trend is kept both for small and large \re. 
We find that the SR and FR  galaxies do not populate distinct locations in the 
scaling relations. The Sa family is the main contributor to the scatters.

\item Sa galaxies deviate from the colour-magnitude and colour-\se\ 
relations due to the presence of dust. SR/FR galaxies instead define very tight 
relations.

\item Surprisingly, extremely young objects do not display the bluest
$(V-[3.6])$ colours of our sample, as is usually the case for optical colours.
This can be understood in the context of the large contribution of TP-AGB stars
to the infrared even for young populations, which makes the $(V-[3.6])$ colour
almost insensitive to age for populations above $\approx1$~Gyr. This effect
results in a very tight $(V-[3.6])-$\se\ relation that allows us to define a
strong correlation between metallicity and \se.

\item  A large number of Sa galaxies appear to follow the main relations defined
by earlier-type systems, and are not necessarily offset as previously found.  We
believe this is due to the use of $\sigma_e$ (the stellar velocity dispersion
integrated within \re), which is not as much influenced by the presence of young
inner discs and accounts to a large extent for rotation in the galaxies. 

\item FR and Sa galaxies that appear offset from the relations correspond mostly
to objects with extremely young populations and signs of on-going, extended star
formation (as already highlighted in Papers XIII and XV). For the specific case
of the Fundamental Plane, we made an attempt to correct for this effect so that
all galaxies are part of a tight, single relation. Once this is done, the FP
coefficients in the $V$- and 3.6$\mu$m-bands are the same within the
uncertainties. The new estimated coefficients suggest that differences in
stellar populations account for about 50\% of the observed tilt with respect to
the virial prediction.

\item The observed scatter of the SR family around the Fundamental Plane is
smaller than that of the FR or Sa galaxies. After correcting for stellar 
populations, the SR family shows almost no intrinsic scatter around the 
best-fit Fundamental Plane.

\item The use of a velocity dispersion within a small aperture (e.g. \re/8) in
the Fundamental Plane results in an increase of around 15\% in the intrinsic 
scatter and an average 10\% decrease of the tilt away from the virial relation 
(as a result of the correlation between \re\ and \se, given by the 
size-luminosity and Faber-Jackson relations).\looseness-1 

\end{itemize}

We have deliberately focused on the role of stellar populations on the main
scaling relations of early-type galaxies. The effects other galaxy properties
exert on them (e.g. non-homology, anisotropy and dark matter variations) will be
studied in ongoing and upcoming integral-field spectroscopic surveys.

\section*{Acknowledgments}
The authors want to thank Maaike Damen, Kambiz Fathi, Katia Ganda, Bettina
Gerken and Peter Kamphuis for their contribution to this work and assistance
during the observing campaigns at the MDM Observatory. We also thank the
referee for very useful comments that have helped improving the presentation
of our results. JFB would like to thank A. Vazdekis and J.A.L. Aguerri for
insightful discussions on stellar population models and scaling relations. JFB
and KLS acknowledge the repeated hospitality of the Institute for Advanced
Study, to which collaborative visits contributed greatly to the quality of this
work. We are grateful to the Isaac Newton Group and MDM staff for their
excellent support and assistance during and after the observations. JFB
acknowledges support from the Ram\'on y Cajal Program as well as grant
AYA2010-21322-C03-02 by the Spanish Ministry of Science and Innovation. HJ
acknowledges support from the International Research Internship Program of the
Korea Science and Engineering Foundation. MB and RLD are grateful for
postdoctoral support through STFC rolling grant PP/E001114/1. MC acknowledges
support from a STFC Advanced Fellowship PP/D005574/1 and a Royal Society
University Research Fellowship. RMcD is supported by the Gemini Observatory,
which is operated by the Association of Universities for Research in Astronomy,
Inc., on behalf of the international Gemini partnership of Argentina, Australia,
Brazil, Canada, Chile, the United Kingdom, and the United States of America. KLS
acknowledge support from Spitzer Research Award 1359449. SKY acknowledges
support from the National Research Foundation of Korea to the Center for Galaxy
Evolution Research and through Doyak grant (No. 20090078756).. The SAURON
project is made possible through grants from NWO and financial contributions
from the Institut National des Sciences de l'Univers, the Universit\'e Lyon I,
the Universities of Durham, Leiden and Oxford, the Programme National Galaxies,
the British Council, STFC grant 'Observational Astrophysics at Oxford' and
support from Christ Church Oxford and the Netherlands Research School for
Astronomy NOVA. We acknowledge the usage in pPXF of the MPFIT routine by
\citet{mpfit}. The SAURON observations were obtained at the William Herschel
Telescope, operated by the Isaac Newton Group in the Spanish Observatorio del
Roque de los Muchachos of the Instituto de Astrof\'isica de Canarias. This work
is based in part on observations made with the Spitzer Space Telescope, which is
operated by the Jet Propulsion Laboratory, California Institute of Technology
under a contract with NASA. This research has made use of the NASA/IPAC
Extragalactic Database (NED) which is operated by the Jet Propulsion Laboratory,
California Institute of Technology, under contract with the National Aeronautics
and Space Administration. We acknowledge the usage of the HyperLeda database
(http://leda.univ-lyon1.fr).

\bibliographystyle{mn2e} 

\begin{landscape}
\begin{table}
\begin{minipage}{230mm}
\caption{Scaling relations best-fit parameters}
\label{tab:params}
\begin{center}
\begin{tabular}{lcrrrrcccccl}
\hline
\hline
Relation & Band & $\alpha$ & $\beta$ & $\gamma$ & $\delta$ &  $\Delta_{\rm fit}$ & $\Delta_{\rm [SR]}$ &  $\Delta_{\rm [FR]}$ &  $\Delta_{\rm [Sa]}$ & N & Sample \\ 
(1) & (2) & (3) & (4) & (5) & (6) & (7) & (8) & (9) & (10) & (11) & (12)\\
\hline
Colour-magnitude      & $V$   & $ -0.06 \pm 0.02$ & $  3.30 \pm 0.02$ &  -                & -                 & $0.092 \pm 0.010$ & $0.067 \pm 0.009$ & $0.101 \pm 0.011$ & $0.245 \pm 0.010$ & 46 & E/S0 \\
Colour-\se            &   -   & $  0.64 \pm 0.07$ & $  3.27 \pm 0.01$ &  -                & -                 & $0.057 \pm 0.011$ & $0.019 \pm 0.017$ & $0.064 \pm 0.010$ & $0.295 \pm 0.010$ & 46 & E/S0 \\
Kormendy              & $V$   & $  2.05 \pm 0.22$ & $ 19.75 \pm 0.07$ &  -                & -                 & $0.447 \pm 0.061$ & $0.400 \pm 0.029$ & $0.464 \pm 0.072$ & $0.743 \pm 0.079$ & 46 & E/S0 \\
Size-luminosity       & $V$   & $ -2.84 \pm 0.22$ & $-20.73 \pm 0.07$ &  -                & -                 & $0.464 \pm 0.028$ & $0.461 \pm 0.026$ & $0.489 \pm 0.028$ & $0.807 \pm 0.040$ & 46 & E/S0 \\
Faber-Jackson         & $V$   & $ -4.99 \pm 0.71$ & $-20.61 \pm 0.11$ &  -                & -                 & $0.662 \pm 0.071$ & $0.672 \pm 0.057$ & $0.644 \pm 0.088$ & $0.811 \pm 0.193$ & 46 & E/S0 \\
Fund. Plane           & $V$   & $  1.15 \pm 0.05$ & $  0.34 \pm 0.01$ & $ -8.81 \pm 0.19$ & -                 & $0.067 \pm 0.015$ & $0.044 \pm 0.016$ & $0.075 \pm 0.017$ & $0.161 \pm 0.037$ & 46 & E/S0 \\
Fund. Plane           & $V$   & $  1.08 \pm 0.04$ & $  0.33 \pm 0.01$ & $ -8.56 \pm 0.18$ & -                 & $0.078 \pm 0.018$ & $0.051 \pm 0.015$ & $0.070 \pm 0.018$ & $0.149 \pm 0.038$ & 60 & E/S0/Sa \\
Fund. Plane + \hbetao & $V$   & $  1.30 \pm 0.04$ & $  0.32 \pm 0.01$ & $ -9.22 \pm 0.19$ & $  0.77 $         & $0.073 \pm 0.016$ & $0.043 \pm 0.014$ & $0.077 \pm 0.016$ & $0.116 \pm 0.038$ & 60 & E/S0/Sa \\
Fund. Plane + \steml  & $V$   & $  1.56 \pm 0.13$ & $  0.32 \pm 0.01$ & $ -9.03 \pm 0.31$ & $ -0.55 \pm 0.13$ & $0.064 \pm 0.031$ & $0.021 \pm 0.029$ & $0.070 \pm 0.029$ & $0.088 \pm 0.051$ & 60 & E/S0/Sa \\
\hline                                                       
Colour-magnitude      & [3.6] & $ -0.06 \pm 0.01$ & $  3.29 \pm 0.01$ &  -                & -                 & $0.087 \pm 0.010$ & $0.060 \pm 0.010$ & $0.095 \pm 0.011$ & $0.234 \pm 0.010$ & 46 & E/S0 \\
Kormendy              & [3.6] & $  1.98 \pm 0.27$ & $ 15.94 \pm 0.08$ &  -                & -                 & $0.519 \pm 0.082$ & $0.464 \pm 0.042$ & $0.541 \pm 0.085$ & $0.745 \pm 0.097$ & 46 & E/S0 \\
Size-luminosity       & [3.6] & $ -2.93 \pm 0.26$ & $-23.79 \pm 0.08$ &  -                & -                 & $0.541 \pm 0.029$ & $0.396 \pm 0.024$ & $0.553 \pm 0.120$ & $0.659 \pm 0.128$ & 46 & E/S0 \\
Faber-Jackson         & [3.6] & $ -5.62 \pm 0.69$ & $-23.69 \pm 0.10$ &  -                & -                 & $0.640 \pm 0.069$ & $0.669 \pm 0.054$ & $0.618 \pm 0.085$ & $0.911 \pm 0.187$ & 46 & E/S0 \\
Fund. Plane           & [3.6] & $  1.37 \pm 0.05$ & $  0.33 \pm 0.01$ & $ -8.06 \pm 0.19$ & -                 & $0.071 \pm 0.014$ & $0.037 \pm 0.012$ & $0.083 \pm 0.017$ & $0.224 \pm 0.035$ & 46 & E/S0 \\
Fund. Plane           & [3.6] & $  1.23 \pm 0.04$ & $  0.32 \pm 0.01$ & $ -7.49 \pm 0.16$ & -                 & $0.099 \pm 0.016$ & $0.053 \pm 0.012$ & $0.084 \pm 0.017$ & $0.191 \pm 0.036$ & 60 & E/S0/Sa \\
Fund. Plane + \hbetao & [3.6] & $  1.47 \pm 0.04$ & $  0.31 \pm 0.01$ & $ -8.35 \pm 0.18$ & $  0.96 $         & $0.082 \pm 0.014$ & $0.044 \pm 0.011$ & $0.080 \pm 0.015$ & $0.132 \pm 0.036$ & 60 & E/S0/Sa \\
Fund. Plane + \steml  & [3.6] & $  1.61 \pm 0.12$ & $  0.32 \pm 0.01$ & $ -8.39 \pm 0.37$ & $ -0.60 \pm 0.15$ & $0.057 \pm 0.041$ & $0.012 \pm 0.050$ & $0.065 \pm 0.036$ & $0.096 \pm 0.057$ & 60 & E/S0/Sa \\     
\hline                                                                                                                                   
\hline
\end{tabular}
\end{center}
\begin{flushleft}
\small{NOTES: \\
(1) We use the following notation for each of the scaling relations:\\
Colour-magnitude: ($V-[3.6]$)$_{\rm e}=\alpha($M$+24) + \beta$ for the $V$-band and ($V-[3.6]$)$_{\rm e}=\alpha($M$+21) + \beta$ for the 3.6$\mu$m-band\\ 
Colour-\se: ($V-[3.6]$)$_{\rm e}=\alpha\log($\se$/\sigma_{\rm ref}) + \beta$, with $\sigma_{\rm ref}=150$.\\ 
Kormendy relation: \mue $=\alpha\log($\re$/R_{\rm ref}) + \beta$, with $R_{\rm ref}$ equal to $2.4$ and $1.7$ for the $V$- and 3.6$\mu$m-band respectively.\\ 
Size-luminosity: M $=\alpha\log($\re$/R_{\rm ref}) + \beta$, with $R_{\rm ref}$ equal to $2.4$ and $1.7$ for the $V$- and 3.6$\mu$m-band respectively.\\ 
Faber-Jackson: M $=\alpha\log($\se$/\sigma_{\rm ref}) + \beta$, with $\sigma_{\rm ref}=150$.\\ 
Fundamental Plane: $\log($\re$) = \alpha\log($\se$)+\beta$\mue$+ \gamma$\\ 
Fundamental Plane + \hbetao: $\log($\re$) = \alpha\log($\se$)+\beta$\mue$+ \gamma + \delta\log($\hbetao$)$\\ 
Fundamental Plane + \steml: $\log($\re$) = \alpha\log($\se$)+\beta$\mue$+ \gamma + \delta\log($\steml$)$\\
Note that the $\delta$ parameter for the Fundamental Plane + \hbetao\ term has been held fixed (see \S\ref{sec:FPml}).\\
(2) Photometric band of the fit. (3-6) Best-fit parameters and uncertainties.\\ 
(7-10) Intrinsic scatter and uncertainties for the fit and SR/FR/Sa families. Values for the colour-magnitude, colour-\se, size-luminosity and Faber-Jackson 
are expressed in magnitudes, while in mag arcsec$^{-2}$ for the Kormendy relation. For the Fundamental Plane, the intrinsic scatter is given in dex along the 
log(\re) direction.\\ 
(11-12) Number of objects in the fit and type of galaxies fitted. Only galaxies with good distance estimations have been included in the fit.}
\end{flushleft}
\end{minipage}
\end{table}
\end{landscape}

\appendix

\section{Stellar Population models}
\label{app:models}

As described in Sect.~\ref{sec:cmd}, we use a combination of the new MILES
models (VAZ10) together with those of MAR08 to interpret our results. The choice
of these models is deliberate. The MILES models currently provide one of the
best libraries for the study of stellar populations at optical wavelengths. They
are based on the MILES stellar library \citep{sanchez06} which consists of
$\approx1000$ stars spanning a large range in atmospheric parameters and
covering the wavelength range 3525-7500\AA\ at 2.3\AA\ (FWHM) spectral
resolution. The spectral resolution, spectral type coverage, flux calibration
accuracy and number of stars represent a substantial improvement over previous
libraries used in population synthesis models. The MAR08 models on the other
hand provide colour predictions at infrared wavelengths, and in particular for
the \textit{Spitzer}/IRAC 3.6$\mu$m-band data used here. These models differ
from MILES in that they have incorporated an improved treatment of the
thermally-pulsing asymptotic giant branch (TP-AGB) phase. The main advantage is
that both libraries make use of the same set of theoretical isochrones
\citep{girardi00} and therefore allow us to generate photometric and
spectroscopic predictions on the same age and metallicity scales, as well as for
the same type of initial mass function (IMF). For the purpose of our work we
focus on predictions for a Kroupa IMF (not corrected for binaries). The
spectroscopic measurements have been performed after transforming the MILES
models to the LIS-14.0\AA\ system (see VAZ10).

\begin{figure}
\begin{center}
\includegraphics[angle=0,width=0.95\linewidth]{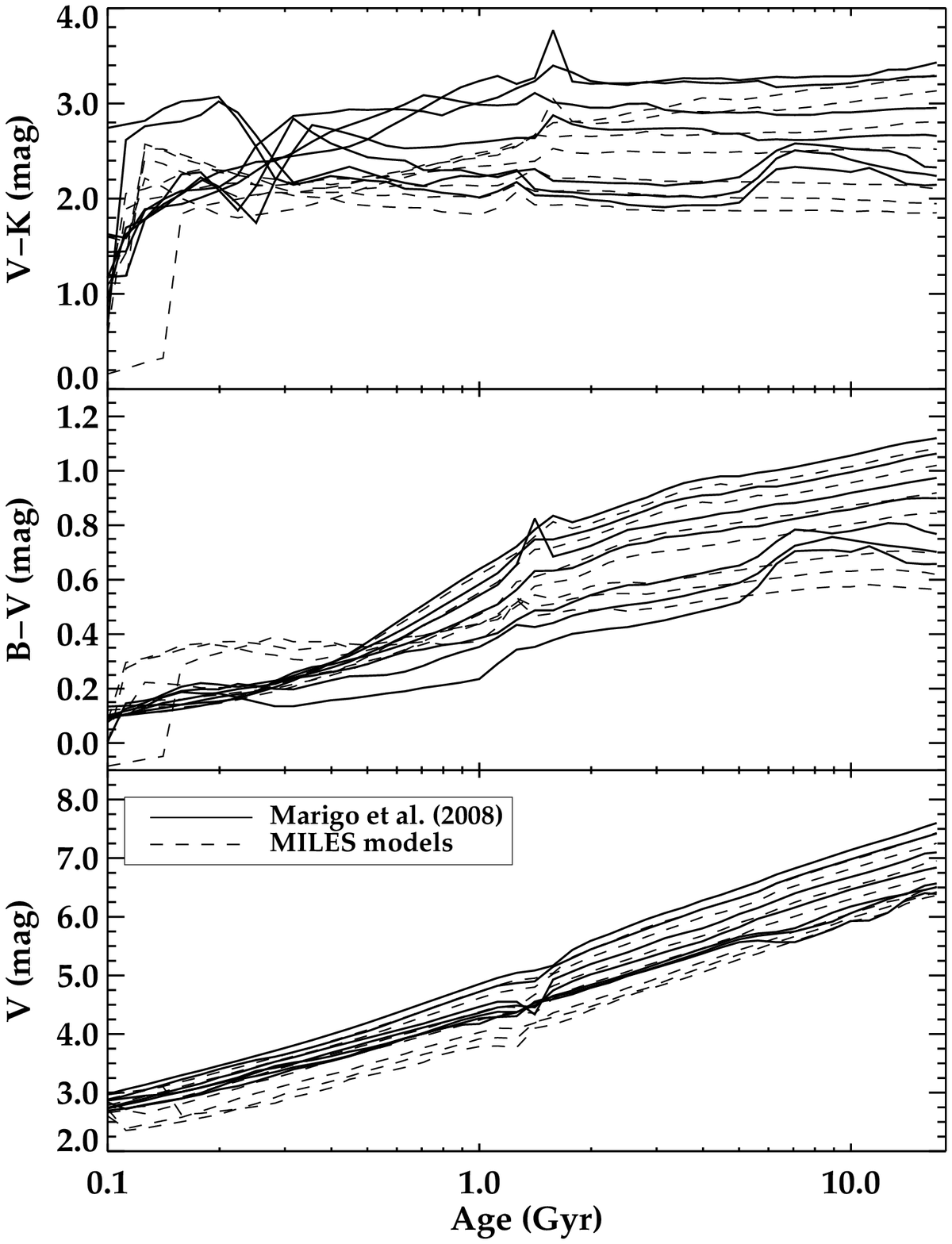}
\end{center}
\caption{Comparison of the colour and luminosity predictions of the
\citet{marigo08} and MILES \citep{vaz10} models as a function of age (in Gyr)
and metallicity (increasing from $-2.32$ to $+0.22$, from bottom to
top of the figures, with increasing colour or luminosity), for a Kroupa initial
mass function.}
\label{fig:models}
\end{figure}

Although still somewhat controversial in the details, the importance of the
TP-AGB phase on observed colours has been recognised \citep[e.g.][]{maraston05}.
In order to understand the impact the TP-AGB phase has on the optical and
infrared colours used here, we show in Figure~\ref{fig:models} the photometric
predictions of both sets of models up to the reddest band predicted by the MILES
models ($K$-band). Given the small difference observed between the $K$- and
3.6$\mu$m colours, any conclusion based on $K$-band can be extended to the
\textit{Spitzer}/IRAC 3.6$\mu$m data. The top row shows the $V-K$ colour as a
function of age for a range of metallicities. The difference between the MILES
and MAR08 models, at large metallicities in particular, is significant. While
the MILES models predict a strong correlation between colour and age, this is
essentially non-existent for ages above 1 Gyr in the MAR08 models (which show a
constant behaviour). This result contrasts with the good agreement, albeit with
a small offset, between the two sets of models for the $B-V$ colour (as shown in
the middle panel) over the range of ages and metallicities displayed by our
data. This is confirmed in the bottom panel, showing that the predicted $V$-band
magnitudes in both set of models are consistent except for an offset.

\begin{figure}
\begin{center}
\includegraphics[angle=0,width=1\linewidth]{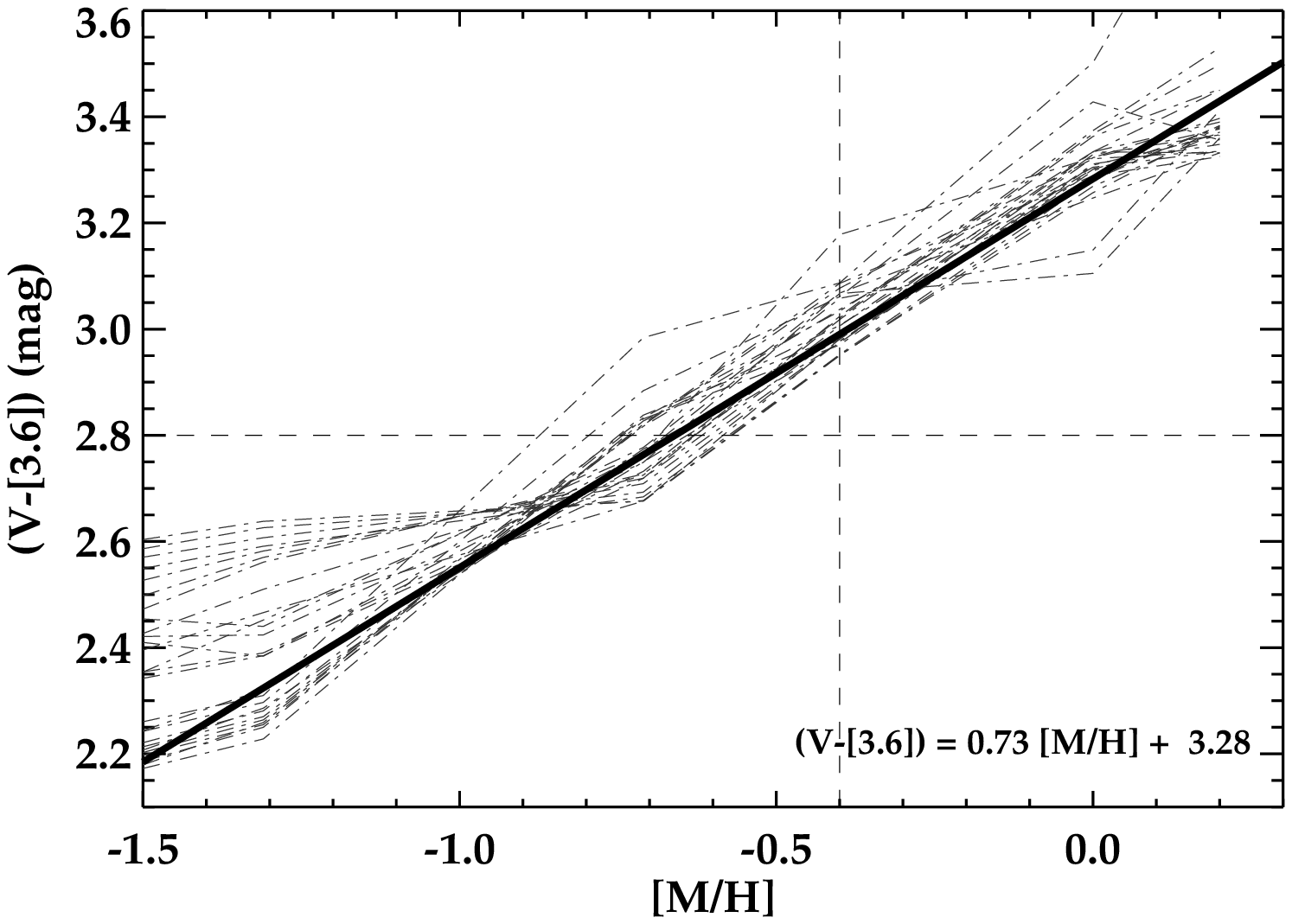}
\end{center}
\caption{($V-[3.6]$) vs metallicity relation from the single stellar population 
models of \citet{marigo08}. Dashed lines mark the predictions for different ages 
and metallicities. The thick solid line shows the best fit for the range of colours 
and \hbetao\ values defined by our sample (top-right quadrant delimited by the 
horizontal and vertical dashed lines). The fitted relation is also indicated.\looseness-2}
\label{fig:colormet}
\end{figure}

The large similarities between the MILES and MAR08 models at optical wavelengths
allow us to safely combine the spectroscopic prediction from MILES and the 
photometric ones from MAR08 (as shown in Fig.~\ref{fig:CMR}).

As shown in \S\ref{sec:cmd}, our sample galaxies display a very tight
correlation between the ($V-[3.6]$)$_e$ colour and \se. Transforming that
relation into a mass-metallicity relation requires to derive a relation between
the ($V-[3.6]$) and metallicity ($[M/H]$) first. This is shown in
Fig.~\ref{fig:colormet}, based on the MAR08 models for the range of colours and
\hbetao\ values exhibited by our sample (($V-[3.6]$)$_{\rm e}\ge2.8$ mag and
\hbetao$\leq$3.0\AA).

\section{Stellar Mass-to-light ratios}
\label{app:ML}
The effective stellar mass-to-light ratio (\steml) is one of the most sought
after quantities when it comes to determine the stellar mass content or to
derive the fraction of dark matter present in a galaxy (e.g. after comparison
with dynamical estimates). Pioneering work relied on optical and infrared
colours \citep[e.g.][]{bdj01}. This has been extended over the years to provide
new calibrations based on different sets of filters \citep{ls09,zibetti09}. The
use of full spectral fitting methods has also been extensive
\citep[e.g.][]{panter07}, but as a drawback it requires some prescription for
the amount of dust present in the fitted galaxies. An alternative not much
explored is the use of line-strength indices \citep[but
see][]{allanson09,graves10}. The advantage is that line-strengths are largely
insensitive to the presence of dust \citep{mac05} and only a small number of
indices is required to get accurate \steml\ estimates \citep{gallazzi09}.

Ideally one would like to use \steml\ from models which include a proper
treatment of the TP-AGB phase. The MAR08 models however do not provide such a
quantity. It is possible though to estimate \steml\ based on the MILES
predictions corrected for the colour differences of the two model libraries
(given their consistency in the optical range). For the $V$-band, for instance,
we scale the MILES \steml\ for the small difference between the models' $V$-band
predictions (bottom panel in Fig.~\ref{fig:models}). Once this is done, it is
possible to determine \steml\ at 3.6$\mu$m using the MAR08 ($V-[3.6]$)$_{\rm e}$
colour.

Detailed studies on the star formation histories (SFHs) of early-type galaxies
show a wide range of scenarios. It does seem however that the SFH of a galaxy is
largely dependent on its stellar mass \citep{thomas05}, with massive galaxies
showing single stellar populations while lower mass galaxies display more
extended star formation. The sample considered here consists of early-type
galaxies as well as Sa disc galaxies. A priori the expected range of SFHs is
quite large given the range of single stellar population quantities exhibited
(see Papers XI and XVII). While full, non-parametric spectral fitting is
desirable in principle, it is not clear whether and how the short \sauron\
spectral range will provide more constraints than those from the \hbeta,
\hbetao, Fe5015 and Mg$b$ line-strength indices (i.e. the main features in our
wavelength range).\looseness-1

\begin{figure}
\begin{center}
\includegraphics[angle=0,width=0.95\linewidth]{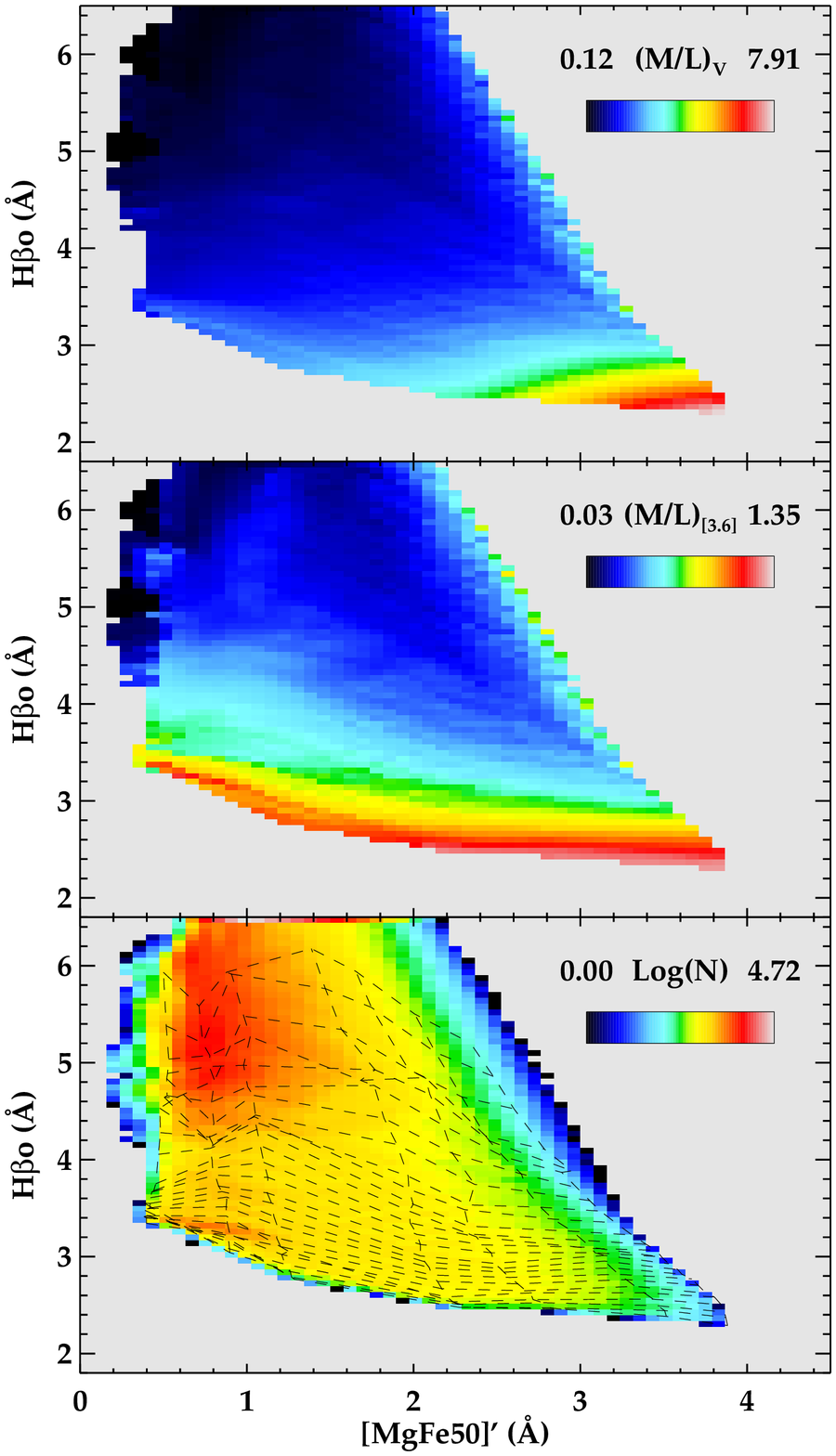}
\end{center}
\caption{Stellar mass-to-light ratios (\steml) predictions for the set of models
produced by the combination of two single stellar populations. Top and middle
panel show the median $M/L$ ratio in the $V-$ and $3.6\mu$m-band. The bottom
panel shows a two-dimensional diagram with the number of models (in logarithmic
scale) falling in a given bin in the \hbetao\ vs [MgFe50]$^\prime$ space. For
reference, dashed lines mark locations of equal age (along the vertical
direction) and metallicity (along the abcissae) for a single stellar
population.}
\label{fig:mlpred}
\end{figure}

We choose here to use as baseline SFHs different combinations of two stellar
populations. While a library with different combinations might not accurately
represent the SFH of any given galaxy, it will at least show the maximum range
of \steml\ allowed for a given pair of \hbetao\ and [MgFe50]$^\prime$ indices.
We thus built a library of $\sim5\times10^6$ models by combining two single
stellar populations in the age range from 0.1 to 17.28 Gyr (sampled
logarithmically) and $[M/H]=$[-2.32,0.22] (sampled uniformly) for different mass
fractions of the young component from 0 (i.e. an SSP) to 0.25. We set no
restriction on the metallicities of the two components, and simply impose the
age of the young one to be lower than the old one. Figure~\ref{fig:mlpred} shows
the median \steml, for the $V$- and $3.6\mu$m-bands, in the \hbetao\ vs
[MgFe50]$^\prime$ space. It also shows the number of models per bin. On average
our library of models contains at least 1000 predictions per bin.
Figure\ref{fig:mlpred} shows that \steml\ variations depend mostly on \hbetao\
and that this dependency is stronger in the $3.6\mu$m-band than in the $V-$band.

We determine the best effective \steml\ via a Bayesian approach following
\citet{kauffmann03} (see Appendix~A in that paper for details). This method
results into a probability distribution for a given set of plausible models. We
take as effective \steml\ the weighted mean value of the distribution and the
68\% confidence interval as its uncertainty. Measured values are listed in
Tables~\ref{tab:saupars_ESO} and ~\ref{tab:saupars_Sa}.

\section{Dependence of kinematic substructure and environment in scaling relations}
\label{app:relations}
In this Appendix we show the main scaling relations in the $V$- and $3.6\mu$m-band,
colour-coded according to kinematic substructure and environment. They are shown
here, rather than in the main text, for completeness as they do not reveal any
particular trend.

\begin{figure*}
\begin{center}
\includegraphics[angle=0,width=0.85\linewidth]{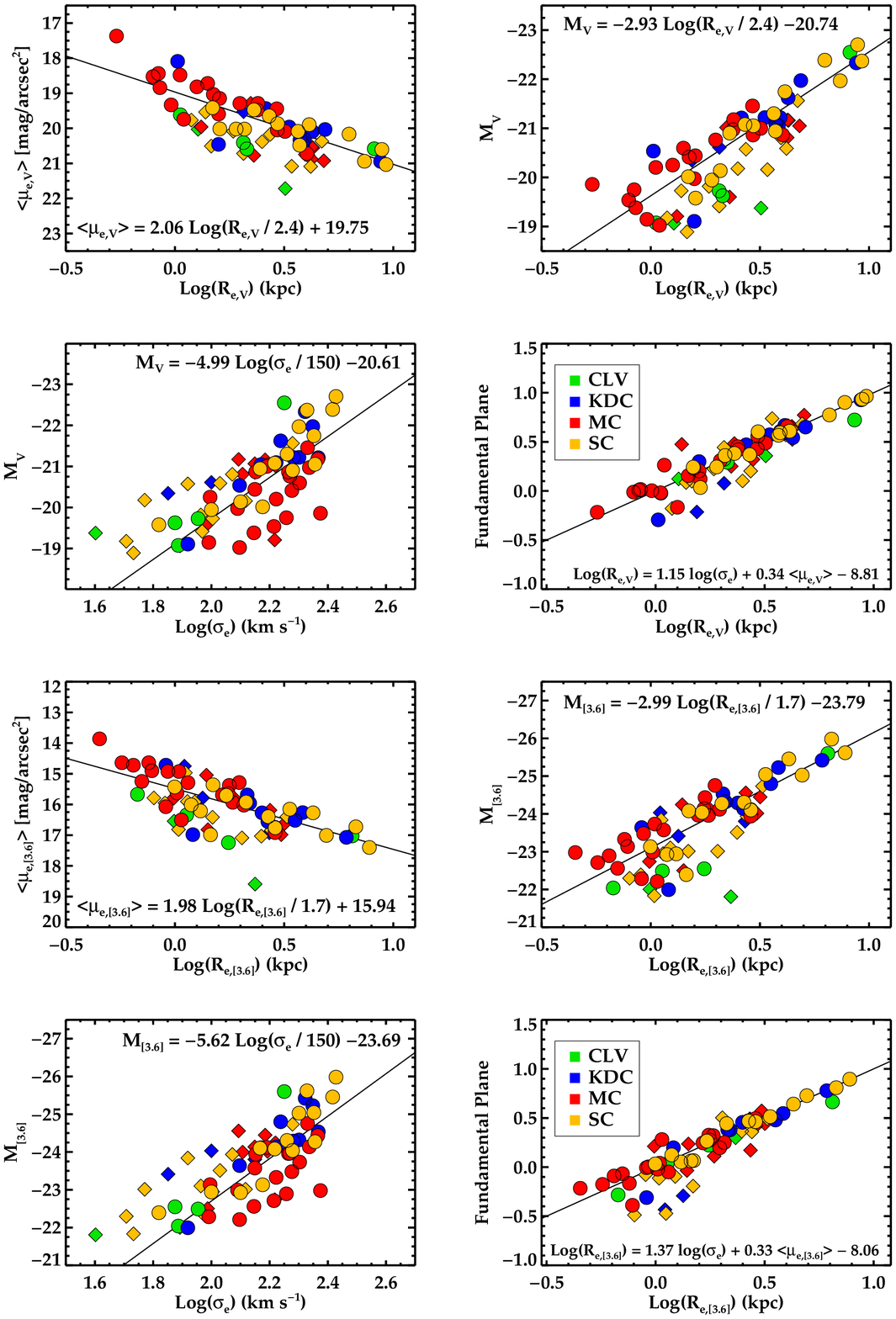}
\end{center}
\caption{Scaling relations in the $V$- and $3.6\mu$m-band for the \sauron\ sample. Galaxies are
colour-coded according to their level of kinematic substructure: (CLV) central
low velocity, (KDC) kinematically decoupled core, (MC) multiple component, (SC)
single component. Elliptical and lenticular galaxies are marked with circles,
while Sa galaxies are denoted with diamonds. Best-fit relations as
shown in \S\ref{sec:scaling} and \S\ref{sec:FP}.}
\label{fig:scaling_kinclass}
\end{figure*}

\begin{figure*}
\begin{center}
\includegraphics[angle=0,width=0.85\linewidth]{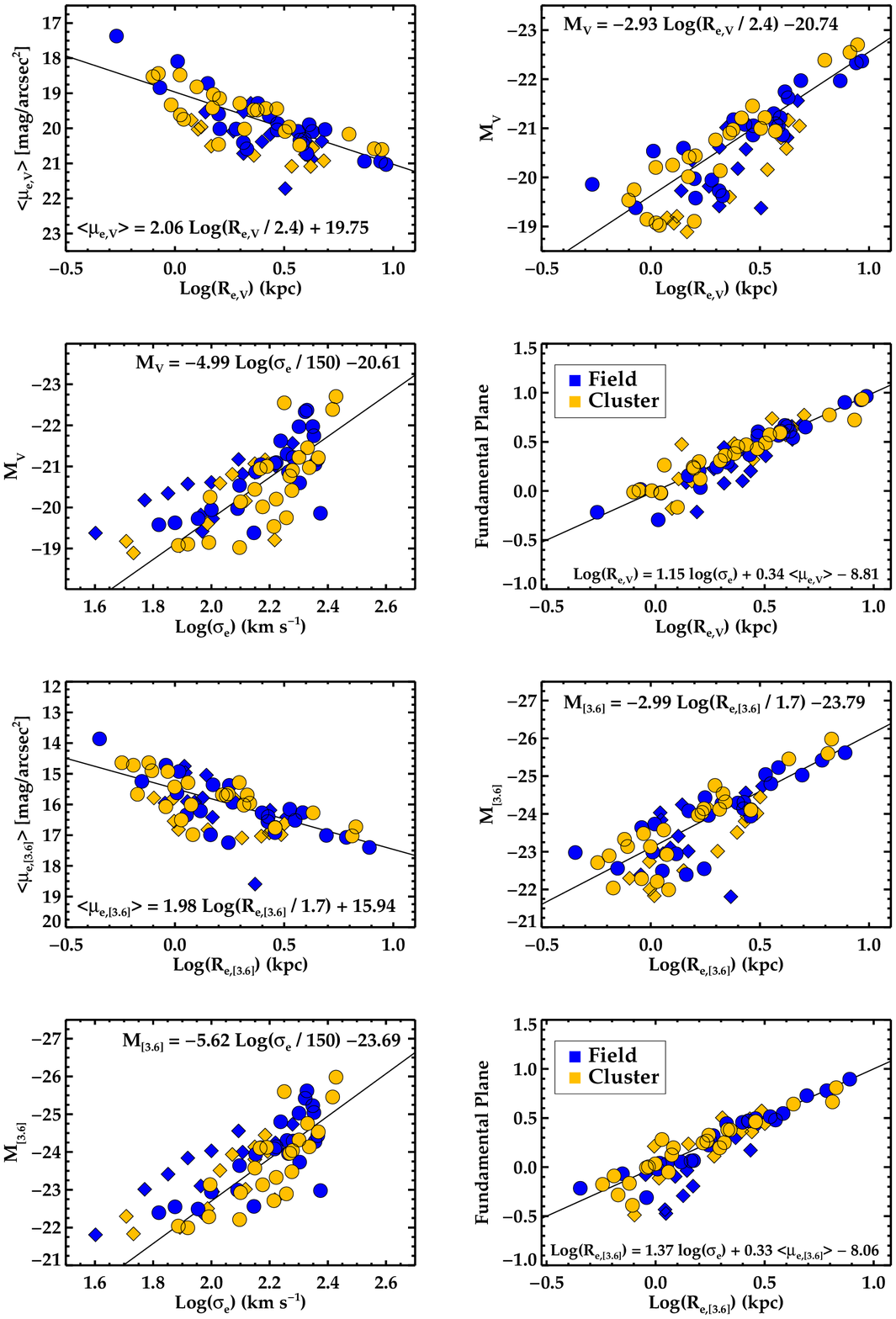}
\end{center}
\caption{Scaling relations in the $V$- and $3.6\mu$m-band for the \sauron\ sample. Galaxies are
colour-coded according to their environment as indicated. Elliptical and 
lenticular galaxies are marked with circles, while Sa galaxies are denoted with 
diamonds. Best-fit relations as shown in \S\ref{sec:scaling} and 
\S\ref{sec:FP}.}
\label{fig:scaling_environ}
\end{figure*}

\section{Photometric \& \sauron\ quantities}
\label{app:photab}
Here we present tables with the aperture photometry derived for the \sauron\ 
sample galaxies in both the $V$- and 3.6$\mu$m-bands, as described in the text 
(see \S~\ref{sec:aper_phot}). We also give tables with the spectroscopic 
quantities derived from the \sauron\ data, as described in \S~\ref{sec:sauron} 
and Appendix~\ref{app:ML}. 

\newpage

\begin{table*}
%
\caption{Photometric quantities for the \sauron\ sample of E/S0 galaxies.}
%
\label{tab:photpars_ESO}
\begin{center}
\renewcommand\tabcolsep{2.00000pt}
\begin{tabular}{ccccccccccccccc}
\hline
\hline
Galaxy & Env. & D & Ref. & R$_{\rm e,V}$ & $\langle\mu_{\rm e,V}\rangle$ & $M_{\rm V}$ & A$_{\rm V}$ & R$_{\rm e,[3.6]}$ & $\langle\mu_{\rm e,[3.6]}\rangle$ & $M_{\rm [3.6]}$ & A$_{\rm [3.6]}$ & $(V-[3.6])_{\rm e}$ & n$_{\rm V}$ & n$_{\rm [3.6]}$ \\
 &  & (Mpc) &  & (arcsec) & (mag arcsec$^{-2}$) & (mag) & (mag) & (arcsec) & (mag arcsec$^{-2}$) & (mag) & (mag) & (mag) &  &  \\
(1) & (2) & (3) & (4) & (5) & (6) & (7) & (8) & (9) & (10) & (11) & (12) & (13) & (14) & (15) \\
\hline
$   0474$ & F & $29.51$ & $ 1$ & $ 28.08\pm  1.65$ & $20.72\pm 0.09$ & $-20.87\pm  0.12$ & $0.11$ & $ 20.07\pm  0.37$ & $20.72\pm 0.11$ & $-23.94\pm  0.11$ & $0.01$ & $3.26\pm0.05$ & $ 4.0$ & $ 4.0$ \\
$   0524$ & F & $23.99$ & $ 3$ & $ 35.41\pm  1.02$ & $19.90\pm 0.07$ & $-21.75\pm  0.20$ & $0.27$ & $ 28.86\pm  0.26$ & $19.90\pm 0.08$ & $-25.04\pm  0.20$ & $0.02$ & $3.43\pm0.03$ & $ 2.8$ & $ 2.8$ \\
$   0821$ & F & $24.10$ & $ 3$ & $ 31.33\pm  1.49$ & $20.09\pm 0.08$ & $-21.30\pm  0.17$ & $0.36$ & $ 22.88\pm  0.46$ & $20.09\pm 0.10$ & $-24.31\pm  0.17$ & $0.02$ & $3.19\pm0.04$ & $ 4.0$ & $ 4.0$ \\
$   1023$ & F & $11.43$ & $ 3$ & $ 48.75\pm  1.46$ & $19.65\pm 0.05$ & $-21.08\pm  0.16$ & $0.20$ & $ 26.95\pm  0.19$ & $19.65\pm 0.08$ & $-24.08\pm  0.16$ & $0.01$ & $3.35\pm0.02$ & $ 4.0$ & $ 3.0$ \\
$   2549$ & F & $12.65$ & $ 3$ & $ 13.92\pm  0.43$ & $18.84\pm 0.16$ & $-19.38\pm  0.28$ & $0.21$ & $ 11.49\pm  0.08$ & $18.84\pm 0.19$ & $-22.56\pm  0.28$ & $0.01$ & $3.30\pm0.02$ & $ 2.9$ & $ 2.9$ \\
$   2685$ & F & $16.60$ & $ 0$ & $ 23.68\pm  1.16$ & $20.02\pm 0.10$ & $-19.95\pm  0.49$ & $0.20$ & $ 16.27\pm  0.41$ & $20.02\pm 0.14$ & $-22.94\pm  0.49$ & $0.01$ & $3.21\pm0.04$ & $ 4.0$ & $ 3.5$ \\
$   2695$ & F & $32.36$ & $ 3$ & $ 18.74\pm  1.26$ & $20.05\pm 0.12$ & $-20.86\pm  0.40$ & $0.06$ & $ 11.72\pm  0.13$ & $20.05\pm 0.19$ & $-23.96\pm  0.40$ & $0.00$ & $3.37\pm0.04$ & $ 4.0$ & $ 3.1$ \\
$   2699$ & F & $26.92$ & $ 3$ & $ 12.14\pm  0.95$ & $19.59\pm 0.18$ & $-19.97\pm  0.27$ & $0.07$ & $  7.83\pm  0.08$ & $19.59\pm 0.28$ & $-22.99\pm  0.27$ & $0.00$ & $3.29\pm0.05$ & $ 4.0$ & $ 2.9$ \\
$   2768$ & F & $22.39$ & $ 3$ & $ 67.97\pm  3.11$ & $20.94\pm 0.05$ & $-21.97\pm  0.24$ & $0.14$ & $ 45.49\pm  0.28$ & $20.94\pm 0.05$ & $-25.03\pm  0.24$ & $0.01$ & $3.29\pm0.04$ & $ 4.0$ & $ 3.5$ \\
$   2974$ & F & $21.48$ & $ 3$ & $ 28.35\pm  1.61$ & $19.87\pm 0.09$ & $-21.05\pm  0.24$ & $0.18$ & $ 20.33\pm  0.11$ & $19.87\pm 0.11$ & $-24.27\pm  0.24$ & $0.01$ & $3.42\pm0.04$ & $ 4.0$ & $ 3.4$ \\
$   3032$ & F & $21.98$ & $ 3$ & $ 19.30\pm  1.07$ & $20.41\pm 0.12$ & $-19.73\pm  0.28$ & $0.06$ & $ 10.61\pm  0.09$ & $20.41\pm 0.20$ & $-22.49\pm  0.28$ & $0.00$ & $3.08\pm0.04$ & $ 3.6$ & $ 3.6$ \\
$   3156$ & F & $22.39$ & $ 3$ & $ 14.78\pm  0.44$ & $20.02\pm 0.15$ & $-19.58\pm  0.14$ & $0.11$ & $ 13.40\pm  0.14$ & $20.02\pm 0.16$ & $-22.39\pm  0.14$ & $0.01$ & $2.89\pm0.03$ & $ 2.4$ & $ 2.4$ \\
$   3377$ & C & $11.22$ & $ 3$ & $ 38.28\pm  1.99$ & $20.02\pm 0.07$ & $-20.14\pm  0.10$ & $0.11$ & $ 21.80\pm  0.22$ & $20.02\pm 0.10$ & $-22.92\pm  0.09$ & $0.01$ & $3.08\pm0.04$ & $ 4.0$ & $ 4.0$ \\
$   3379$ & C & $10.57$ & $ 3$ & $ 44.89\pm  1.47$ & $19.47\pm 0.05$ & $-20.90\pm  0.11$ & $0.08$ & $ 33.53\pm  0.40$ & $19.47\pm 0.07$ & $-24.04\pm  0.11$ & $0.00$ & $3.31\pm0.03$ & $ 4.0$ & $ 4.0$ \\
$   3384$ & C & $11.59$ & $ 3$ & $ 28.50\pm  0.91$ & $19.15\pm 0.08$ & $-20.44\pm  0.14$ & $0.09$ & $ 20.41\pm  0.09$ & $19.15\pm 0.11$ & $-23.57\pm  0.14$ & $0.00$ & $3.28\pm0.02$ & $ 4.0$ & $ 4.0$ \\
$   3414$ & F & $25.23$ & $ 3$ & $ 32.02\pm  1.45$ & $20.31\pm 0.08$ & $-21.22\pm  0.33$ & $0.08$ & $ 20.47\pm  0.23$ & $20.31\pm 0.11$ & $-24.29\pm  0.33$ & $0.00$ & $3.31\pm0.03$ & $ 4.0$ & $ 4.0$ \\
$   3489$ & C & $12.08$ & $ 3$ & $ 21.51\pm  0.81$ & $18.82\pm 0.10$ & $-20.25\pm  0.15$ & $0.05$ & $ 13.38\pm  0.14$ & $18.82\pm 0.16$ & $-23.14\pm  0.15$ & $0.00$ & $3.15\pm0.02$ & $ 4.0$ & $ 3.0$ \\
$   3608$ & F & $22.91$ & $ 3$ & $ 33.64\pm  1.52$ & $20.34\pm 0.07$ & $-21.09\pm  0.14$ & $0.07$ & $ 23.98\pm  0.15$ & $20.34\pm 0.09$ & $-24.14\pm  0.14$ & $0.00$ & $3.24\pm0.04$ & $ 4.0$ & $ 4.0$ \\
$   4150$ & C & $13.74$ & $ 3$ & $ 15.88\pm  0.52$ & $19.62\pm 0.14$ & $-19.07\pm  0.24$ & $0.06$ & $ 10.10\pm  0.09$ & $19.62\pm 0.21$ & $-22.04\pm  0.24$ & $0.00$ & $3.24\pm0.02$ & $ 2.6$ & $ 2.6$ \\
$   4262$ & C & $15.42$ & $ 2$ & $ 10.60\pm  0.43$ & $18.53\pm 0.21$ & $-19.54\pm  0.06$ & $0.11$ & $  7.64\pm  0.07$ & $18.53\pm 0.28$ & $-22.71\pm  0.06$ & $0.01$ & $3.34\pm0.02$ & $ 4.0$ & $ 4.0$ \\
$   4270$ & C & $16.53$ & $ 5$ & $ 13.68\pm  0.47$ & $19.74\pm 0.16$ & $-19.02\pm  0.09$ & $0.07$ & $ 13.34\pm  0.14$ & $19.74\pm 0.16$ & $-22.21\pm  0.08$ & $0.00$ & $3.21\pm0.03$ & $ 2.2$ & $ 2.2$ \\
$   4278$ & C & $16.07$ & $ 3$ & $ 30.57\pm  1.07$ & $19.48\pm 0.07$ & $-20.97\pm  0.20$ & $0.09$ & $ 22.51\pm  0.39$ & $19.48\pm 0.10$ & $-24.14\pm  0.20$ & $0.00$ & $3.32\pm0.02$ & $ 4.0$ & $ 4.0$ \\
$   4374$ & C & $18.45$ & $ 2$ & $ 70.08\pm  2.95$ & $20.17\pm 0.04$ & $-22.39\pm  0.06$ & $0.13$ & $ 47.95\pm  0.60$ & $20.17\pm 0.05$ & $-25.46\pm  0.05$ & $0.01$ & $3.28\pm0.03$ & $ 4.0$ & $ 4.0$ \\
$   4382$ & C & $17.86$ & $ 2$ & $ 94.40\pm  3.17$ & $20.58\pm 0.04$ & $-22.55\pm  0.06$ & $0.10$ & $ 74.77\pm  0.70$ & $20.58\pm 0.03$ & $-25.60\pm  0.05$ & $0.01$ & $3.21\pm0.04$ & $ 4.0$ & $ 4.0$ \\
$   4387$ & C & $17.95$ & $ 2$ & $ 11.04\pm  0.22$ & $19.33\pm 0.20$ & $-19.15\pm  0.06$ & $0.11$ & $ 10.43\pm  0.08$ & $19.33\pm 0.21$ & $-22.28\pm  0.06$ & $0.01$ & $3.18\pm0.02$ & $ 2.1$ & $ 2.1$ \\
$   4458$ & C & $16.37$ & $ 2$ & $ 19.94\pm  0.96$ & $20.46\pm 0.12$ & $-19.11\pm  0.06$ & $0.08$ & $ 15.21\pm  0.14$ & $20.46\pm 0.14$ & $-21.99\pm  0.05$ & $0.00$ & $3.07\pm0.04$ & $ 3.1$ & $ 2.9$ \\
$   4459$ & C & $16.07$ & $ 2$ & $ 40.97\pm  1.48$ & $20.09\pm 0.06$ & $-21.00\pm  0.07$ & $0.15$ & $ 26.54\pm  0.36$ & $20.09\pm 0.08$ & $-24.12\pm  0.06$ & $0.01$ & $3.35\pm0.03$ & $ 4.0$ & $ 4.0$ \\
$   4473$ & C & $15.28$ & $ 2$ & $ 26.76\pm  0.89$ & $19.29\pm 0.08$ & $-20.76\pm  0.06$ & $0.09$ & $ 22.26\pm  0.31$ & $19.29\pm 0.10$ & $-23.97\pm  0.06$ & $0.00$ & $3.30\pm0.02$ & $ 4.0$ & $ 4.4$ \\
$   4477$ & C & $16.53$ & $ 5$ & $ 46.48\pm  2.22$ & $20.48\pm 0.06$ & $-20.94\pm  0.09$ & $0.10$ & $ 35.87\pm  0.69$ & $20.48\pm 0.06$ & $-24.10\pm  0.08$ & $0.01$ & $3.32\pm0.04$ & $ 4.0$ & $ 4.0$ \\
$   4486$ & C & $17.22$ & $ 2$ & $106.18\pm  4.23$ & $20.60\pm 0.05$ & $-22.70\pm  0.08$ & $0.07$ & $ 80.69\pm  0.71$ & $20.60\pm 0.03$ & $-25.98\pm  0.07$ & $0.00$ & $3.44\pm0.04$ & $ 4.0$ & $ 4.0$ \\
$   4526$ & C & $16.90$ & $ 3$ & $ 35.70\pm  1.29$ & $19.44\pm 0.07$ & $-21.46\pm  0.20$ & $0.07$ & $ 24.03\pm  0.22$ & $19.44\pm 0.09$ & $-24.75\pm  0.20$ & $0.00$ & $3.53\pm0.03$ & $ 3.3$ & $ 3.2$ \\
$   4546$ & C & $14.06$ & $ 3$ & $ 21.98\pm  0.94$ & $19.03\pm 0.10$ & $-20.41\pm  0.20$ & $0.11$ & $ 13.60\pm  0.10$ & $19.03\pm 0.16$ & $-23.48\pm  0.20$ & $0.01$ & $3.34\pm0.03$ & $ 4.0$ & $ 3.4$ \\
$   4550$ & C & $15.49$ & $ 2$ & $ 11.55\pm  0.27$ & $19.00\pm 0.19$ & $-19.26\pm  0.05$ & $0.13$ & $ 10.13\pm  0.06$ & $19.00\pm 0.21$ & $-22.30\pm  0.05$ & $0.01$ & $3.15\pm0.02$ & $ 1.8$ & $ 1.8$ \\
$   4552$ & C & $15.85$ & $ 2$ & $ 33.89\pm  1.09$ & $19.43\pm 0.07$ & $-21.21\pm  0.06$ & $0.13$ & $ 27.88\pm  0.21$ & $19.43\pm 0.08$ & $-24.53\pm  0.06$ & $0.01$ & $3.43\pm0.02$ & $ 4.0$ & $ 4.0$ \\
$   4564$ & C & $15.85$ & $ 2$ & $ 19.35\pm  1.11$ & $19.41\pm 0.12$ & $-20.02\pm  0.07$ & $0.11$ & $ 12.97\pm  0.21$ & $19.41\pm 0.17$ & $-23.13\pm  0.06$ & $0.01$ & $3.34\pm0.04$ & $ 4.0$ & $ 3.4$ \\
$   4570$ & C & $17.06$ & $ 2$ & $ 12.75\pm  0.56$ & $18.48\pm 0.17$ & $-20.20\pm  0.07$ & $0.07$ & $  9.17\pm  0.08$ & $18.48\pm 0.24$ & $-23.33\pm  0.06$ & $0.00$ & $3.34\pm0.03$ & $ 4.0$ & $ 2.4$ \\
$   4621$ & C & $14.93$ & $ 2$ & $ 46.06\pm  2.99$ & $19.96\pm 0.06$ & $-21.22\pm  0.07$ & $0.11$ & $ 30.38\pm  0.39$ & $19.96\pm 0.07$ & $-24.32\pm  0.06$ & $0.01$ & $3.33\pm0.04$ & $ 4.0$ & $ 4.0$ \\
$   4660$ & C & $15.00$ & $ 2$ & $ 11.54\pm  0.41$ & $18.44\pm 0.19$ & $-19.75\pm  0.05$ & $0.11$ & $  8.87\pm  0.11$ & $18.44\pm 0.25$ & $-22.89\pm  0.05$ & $0.01$ & $3.31\pm0.02$ & $ 4.0$ & $ 4.0$ \\
$   5198$ & F & $48.78$ & $ 8$ & $ 18.01\pm  0.69$ & $20.09\pm 0.12$ & $-21.62\pm  0.03$ & $0.08$ & $ 15.01\pm  0.14$ & $20.09\pm 0.14$ & $-24.80\pm  0.01$ & $0.00$ & $3.31\pm0.03$ & $ 2.5$ & $ 2.5$ \\
$   5308$ & F & $29.43$ & $ 4$ & $  9.89\pm  0.25$ & $18.72\pm 0.22$ & $-20.60\pm  0.22$ & $0.06$ & $  7.31\pm  0.04$ & $18.72\pm 0.30$ & $-23.73\pm  0.22$ & $0.00$ & $3.38\pm0.02$ & $ 4.0$ & $ 1.7$ \\
$   5813$ & F & $32.21$ & $ 3$ & $ 55.99\pm  2.72$ & $20.94\pm 0.06$ & $-22.33\pm  0.19$ & $0.19$ & $ 39.03\pm  0.62$ & $20.94\pm 0.06$ & $-25.42\pm  0.18$ & $0.01$ & $3.33\pm0.05$ & $ 4.0$ & $ 4.0$ \\
$   5831$ & F & $27.16$ & $ 3$ & $ 29.19\pm  1.70$ & $20.45\pm 0.08$ & $-21.04\pm  0.17$ & $0.19$ & $ 21.82\pm  0.27$ & $20.45\pm 0.10$ & $-24.11\pm  0.17$ & $0.01$ & $3.23\pm0.04$ & $ 4.0$ & $ 4.0$ \\
$   5838$ & F & $23.99$ & $ 0$ & $ 20.61\pm  0.87$ & $19.29\pm 0.11$ & $-21.18\pm  0.49$ & $0.17$ & $ 15.25\pm  0.12$ & $19.29\pm 0.14$ & $-24.43\pm  0.49$ & $0.01$ & $3.41\pm0.03$ & $ 4.0$ & $ 3.8$ \\
$   5845$ & F & $25.94$ & $ 3$ & $  4.29\pm  0.26$ & $17.37\pm 0.51$ & $-19.86\pm  0.21$ & $0.17$ & $  3.59\pm  0.08$ & $17.37\pm 0.61$ & $-22.98\pm  0.21$ & $0.01$ & $3.29\pm0.05$ & $ 4.0$ & $ 4.0$ \\
$   5846$ & F & $24.89$ & $ 3$ & $ 76.79\pm  3.32$ & $21.03\pm 0.05$ & $-22.37\pm  0.21$ & $0.18$ & $ 64.39\pm  0.61$ & $21.03\pm 0.04$ & $-25.62\pm  0.20$ & $0.01$ & $3.35\pm0.05$ & $ 4.0$ & $ 4.0$ \\
$   5982$ & F & $40.36$ & $ 1$ & $ 24.89\pm  1.19$ & $20.03\pm 0.09$ & $-21.97\pm  0.09$ & $0.06$ & $ 19.65\pm  0.19$ & $20.03\pm 0.11$ & $-25.23\pm  0.09$ & $0.00$ & $3.38\pm0.03$ & $ 4.0$ & $ 4.0$ \\
$   7332$ & F & $23.01$ & $ 3$ & $  9.21\pm  0.15$ & $18.09\pm 0.24$ & $-20.54\pm  0.20$ & $0.12$ & $  8.14\pm  0.04$ & $18.09\pm 0.27$ & $-23.64\pm  0.20$ & $0.01$ & $3.20\pm0.01$ & $ 2.0$ & $ 2.0$ \\
$   7457$ & F & $13.24$ & $ 3$ & $ 33.25\pm  1.28$ & $20.59\pm 0.08$ & $-19.63\pm  0.21$ & $0.17$ & $ 27.37\pm  0.19$ & $20.59\pm 0.08$ & $-22.55\pm  0.21$ & $0.01$ & $3.08\pm0.04$ & $ 2.5$ & $ 2.5$ \\
\hline
\hline
\end{tabular}
\end{center}
\begin{flushleft}
\small{NOTES: (1) NGC galaxy number. (2) Environment: Field (F) or Cluster (C). (3) Distance (Mpc). (4) Original reference for the distance estimate (0: redshift distance from NED; 1: \citealt{Cantiello2007}; 2: \citealt{Mei2007};  3: \citealt{Tonry2001};  4: \citealt{Reindl2005};  5: \citealt{Mei2007}, but see (iii) in \S\ref{sec:distances};  6: \citealt{Terry2002};  7: \citealt{Tully2008};  8: \citealt{Willick1997}). See \S\ref{sec:distances} for uncertainties in the measurements, (5-12) Photometric quantities: effective radii (\re), mean effective surface brightness (\mue), absolute magnitude (M) and Galactic extinction in the $V$- and $3.6\mu$m bands. (13) Effective $(V-[3.6])$ colour. (14-15) S\'ersic indices used in our growth curve analysis (see \S\ref{sec:aper_phot}).}
\end{flushleft}
\end{table*}

\newpage

\begin{table*}
%
\caption{Photometric quantities for the \sauron\ sample of Sa galaxies.}
%
\label{tab:photpars_Sa}
\begin{center}
\renewcommand\tabcolsep{2.00000pt}
\begin{tabular}{ccccccccccccccc}
\hline
\hline
Galaxy & Env. & D & Ref. & R$_{\rm e,V}$ & $\langle\mu_{\rm e,V}\rangle$ & $M_{\rm V}$ & A$_{\rm V}$ & R$_{\rm e,[3.6]}$ & $\langle\mu_{\rm e,[3.6]}\rangle$ & $M_{\rm [3.6]}$ & A$_{\rm [3.6]}$ & $(V-[3.6])_{\rm e}$ & n$_{\rm V}$ & n$_{\rm [3.6]}$ \\
 &  & (Mpc) &  & (arcsec) & (mag arcsec$^{-2}$) & (mag) & (mag) & (arcsec) & (mag arcsec$^{-2}$) & (mag) & (mag) & (mag) &  &  \\
(1) & (2) & (3) & (4) & (5) & (6) & (7) & (8) & (9) & (10) & (11) & (12) & (13) & (14) & (15) \\
\hline
$   1056$ & F & $32.21$ & $ 7$ & $ 17.48\pm  1.24$ & $20.17\pm 0.13$ & $-20.57\pm  0.40$ & $0.48$ & $  7.14\pm  0.32$ & $20.17\pm 0.31$ & $-23.84\pm  0.40$ & $0.03$ & $3.70\pm0.06$ & $ 4.0$ & $ 4.0$ \\
$   2273$ & F & $31.62$ & $ 6$ & $ 27.64\pm  1.22$ & $20.53\pm 0.09$ & $-21.17\pm  0.39$ & $0.23$ & $ 17.78\pm  0.65$ & $20.53\pm 0.12$ & $-24.56\pm  0.39$ & $0.01$ & $3.64\pm0.04$ & $ 2.9$ & $ 6.6$ \\
$   2844$ & F & $25.12$ & $ 0$ & $ 16.91\pm  1.23$ & $20.72\pm 0.14$ & $-19.42\pm  0.49$ & $0.06$ & $  7.39\pm  0.11$ & $20.72\pm 0.29$ & $-22.39\pm  0.49$ & $0.00$ & $3.41\pm0.05$ & $ 4.0$ & $ 2.9$ \\
$   3623$ & C & $11.97$ & $ 7$ & $ 48.21\pm  0.95$ & $19.73\pm 0.05$ & $-21.07\pm  0.35$ & $0.08$ & $ 31.99\pm  0.32$ & $19.73\pm 0.07$ & $-24.14\pm  0.35$ & $0.00$ & $3.45\pm0.03$ & $ 2.1$ & $ 1.5$ \\
$   4220$ & F & $18.79$ & $ 7$ & $ 20.30\pm  0.50$ & $20.08\pm 0.11$ & $-19.82\pm  0.36$ & $0.06$ & $ 13.50\pm  0.08$ & $20.08\pm 0.16$ & $-23.10\pm  0.36$ & $0.00$ & $3.62\pm0.03$ & $ 1.6$ & $ 2.1$ \\
$   4235$ & C & $16.53$ & $ 5$ & $ 16.44\pm  0.49$ & $19.96\pm 0.13$ & $-19.21\pm  0.09$ & $0.06$ & $ 12.30\pm  0.24$ & $19.96\pm 0.18$ & $-22.74\pm  0.08$ & $0.00$ & $3.71\pm0.03$ & $ 1.9$ & $ 1.9$ \\
$   4245$ & C & $13.18$ & $ 0$ & $ 35.95\pm  1.63$ & $20.77\pm 0.07$ & $-19.60\pm  0.49$ & $0.07$ & $ 22.14\pm  0.23$ & $20.77\pm 0.10$ & $-22.50\pm  0.49$ & $0.00$ & $3.24\pm0.04$ & $ 3.5$ & $ 3.5$ \\
$   4274$ & C & $19.41$ & $ 7$ & $ 45.43\pm  1.26$ & $20.56\pm 0.06$ & $-21.16\pm  0.36$ & $0.07$ & $ 33.70\pm  0.16$ & $20.56\pm 0.06$ & $-24.45\pm  0.36$ & $0.00$ & $3.53\pm0.03$ & $ 2.3$ & $ 1.6$ \\
$   4293$ & C & $16.53$ & $ 5$ & $ 52.20\pm  1.63$ & $21.09\pm 0.06$ & $-20.59\pm  0.09$ & $0.13$ & $ 31.00\pm  0.48$ & $21.09\pm 0.07$ & $-23.51\pm  0.08$ & $0.01$ & $3.36\pm0.05$ & $ 1.8$ & $ 2.4$ \\
$   4314$ & C & $17.38$ & $ 0$ & $ 46.90\pm  2.24$ & $20.74\pm 0.06$ & $-20.81\pm  0.49$ & $0.08$ & $ 32.80\pm  0.59$ & $20.74\pm 0.07$ & $-23.94\pm  0.49$ & $0.00$ & $3.31\pm0.03$ & $ 4.0$ & $ 3.0$ \\
$   4369$ & F & $20.89$ & $ 0$ & $ 24.73\pm  1.27$ & $20.38\pm 0.09$ & $-20.18\pm  0.49$ & $0.08$ & $ 14.67\pm  0.15$ & $20.38\pm 0.15$ & $-23.01\pm  0.49$ & $0.00$ & $3.13\pm0.04$ & $ 4.0$ & $ 2.7$ \\
$   4383$ & C & $16.53$ & $ 5$ & $ 14.83\pm  0.79$ & $19.77\pm 0.15$ & $-19.18\pm  0.09$ & $0.08$ & $  9.98\pm  0.19$ & $19.77\pm 0.22$ & $-22.30\pm  0.08$ & $0.00$ & $3.32\pm0.03$ & $ 4.0$ & $ 4.0$ \\
$   4405$ & C & $16.53$ & $ 5$ & $ 18.29\pm  0.48$ & $20.51\pm 0.12$ & $-18.89\pm  0.09$ & $0.08$ & $ 12.95\pm  0.17$ & $20.51\pm 0.17$ & $-21.83\pm  0.08$ & $0.00$ & $3.25\pm0.04$ & $ 1.2$ & $ 1.2$ \\
$   4425$ & C & $16.53$ & $ 5$ & $ 15.91\pm  0.65$ & $20.03\pm 0.14$ & $-19.07\pm  0.09$ & $0.09$ & $ 12.36\pm  0.09$ & $20.03\pm 0.18$ & $-22.00\pm  0.08$ & $0.00$ & $3.14\pm0.04$ & $ 2.0$ & $ 2.0$ \\
$   4596$ & C & $16.53$ & $ 5$ & $ 59.91\pm  3.05$ & $20.93\pm 0.06$ & $-21.05\pm  0.09$ & $0.07$ & $ 38.24\pm  0.45$ & $20.93\pm 0.06$ & $-24.02\pm  0.08$ & $0.00$ & $3.25\pm0.05$ & $ 4.0$ & $ 4.0$ \\
$   4698$ & C & $16.53$ & $ 5$ & $ 50.64\pm  2.40$ & $20.76\pm 0.06$ & $-20.85\pm  0.09$ & $0.08$ & $ 33.74\pm  0.22$ & $20.76\pm 0.06$ & $-23.80\pm  0.08$ & $0.00$ & $3.22\pm0.04$ & $ 4.0$ & $ 3.0$ \\
$   4772$ & C & $16.53$ & $ 5$ & $ 42.71\pm  2.34$ & $21.08\pm 0.07$ & $-20.16\pm  0.09$ & $0.09$ & $ 25.26\pm  0.31$ & $21.08\pm 0.09$ & $-23.01\pm  0.08$ & $0.00$ & $3.14\pm0.05$ & $ 4.0$ & $ 4.0$ \\
$   5448$ & F & $34.83$ & $ 6$ & $ 24.97\pm  0.87$ & $20.88\pm 0.10$ & $-20.81\pm  0.18$ & $0.05$ & $ 17.09\pm  0.26$ & $20.88\pm 0.13$ & $-24.00\pm  0.18$ & $0.00$ & $3.48\pm0.04$ & $ 1.8$ & $ 3.0$ \\
$   5475$ & F & $30.20$ & $ 0$ & $  9.42\pm  0.23$ & $19.54\pm 0.23$ & $-19.73\pm  0.49$ & $0.04$ & $  7.78\pm  0.07$ & $19.54\pm 0.28$ & $-22.95\pm  0.49$ & $0.00$ & $3.36\pm0.02$ & $ 1.3$ & $ 2.4$ \\
$   5636$ & F & $28.84$ & $ 0$ & $ 22.89\pm  0.89$ & $21.72\pm 0.11$ & $-19.38\pm  0.49$ & $0.11$ & $ 16.65\pm  0.24$ & $21.72\pm 0.13$ & $-21.81\pm  0.49$ & $0.01$ & $2.85\pm0.07$ & $ 1.0$ & $ 1.0$ \\
$   5689$ & F & $36.31$ & $ 0$ & $ 12.63\pm  0.35$ & $19.29\pm 0.17$ & $-21.02\pm  0.49$ & $0.12$ & $  7.91\pm  0.04$ & $19.29\pm 0.27$ & $-24.24\pm  0.49$ & $0.01$ & $3.56\pm0.03$ & $ 2.1$ & $ 2.1$ \\
$   5953$ & F & $33.11$ & $ 0$ & $ 12.84\pm  0.72$ & $19.53\pm 0.17$ & $-20.61\pm  0.49$ & $0.16$ & $  6.88\pm  0.14$ & $19.53\pm 0.32$ & $-24.03\pm  0.49$ & $0.01$ & $3.71\pm0.04$ & $ 4.0$ & $ 4.0$ \\
$   6501$ & F & $47.86$ & $ 0$ & $ 20.27\pm  1.40$ & $20.37\pm 0.12$ & $-21.56\pm  0.49$ & $0.29$ & $ 14.02\pm  0.33$ & $20.37\pm 0.16$ & $-24.73\pm  0.49$ & $0.02$ & $3.35\pm0.05$ & $ 4.0$ & $ 5.9$ \\
$   7742$ & F & $22.91$ & $ 0$ & $ 13.93\pm  0.36$ & $19.17\pm 0.16$ & $-20.35\pm  0.49$ & $0.18$ & $ 12.05\pm  0.07$ & $19.17\pm 0.18$ & $-23.41\pm  0.49$ & $0.01$ & $3.18\pm0.02$ & $ 2.1$ & $ 2.1$ \\
\hline
\hline
\end{tabular}
\end{center}
\begin{flushleft}
\small{NOTES: (1) NGC galaxy number. (2) Environment: Field (F) or Cluster (C). (3) Distance (Mpc). (4) Original reference for the distance estimate (0: redshift distance from NED; 1: \citealt{Cantiello2007}; 2: \citealt{Mei2007};  3: \citealt{Tonry2001};  4: \citealt{Reindl2005};  5: \citealt{Mei2007}, but see (iii) in \S\ref{sec:distances};  6: \citealt{Terry2002};  7: \citealt{Tully2008};  8: \citealt{Willick1997}). See \S\ref{sec:distances} for uncertainties in the measurements, (5-12) Photometric quantities: effective radii (\re), mean effective surface brightness (\mue), absolute magnitude (M) and Galactic extinction in the $V$- and $3.6\mu$m bands. (13) Effective $(V-[3.6])$ colour. (14-15) S\'ersic indices used in our growth curve analysis (see \S\ref{sec:aper_phot}).}
\end{flushleft}
\end{table*}

\newpage

\begin{table*}
%
\caption{Spectroscopic quantities for the \sauron\ sample of E/S0 galaxies.}
%
\label{tab:saupars_ESO}
\begin{center}
\renewcommand\tabcolsep{3.50000pt}
\begin{tabular}{cccccccccc}
\hline
\hline
Galaxy & SR/FR & Kinem. & $\sigma_{\rm e}$ & H$\beta$ & H$\beta_o$ & Fe5015 & Mg$b$ & \stemlV & \stemlIR \\
 &  &  & (km/s) & (\AA) & (\AA) & (\AA) & (\AA) & \MLsunV & \MLsun36 \\
(1) & (2) & (3) & (4) & (5) & (6) & (7) & (8) & (9) & (10) \\
\hline
\vspace{0.05cm}
$   0474$ & FR &  MC & $142$ & $1.85$ & $2.93$ & $3.84$ & $2.96$ & $3.52_{- 0.59}^{+ 0.67}$ & $0.78_{- 0.13}^{+ 0.14}$ \\
\vspace{0.05cm}
$   0524$ & FR &  SC & $225$ & $1.52$ & $2.58$ & $4.26$ & $3.63$ & $5.36_{- 0.60}^{+ 0.58}$ & $1.05_{- 0.10}^{+ 0.10}$ \\
\vspace{0.05cm}
$   0821$ & FR &  SC & $182$ & $1.65$ & $2.65$ & $3.69$ & $3.22$ & $4.54_{- 0.60}^{+ 0.63}$ & $1.02_{- 0.12}^{+ 0.12}$ \\
\vspace{0.05cm}
$   1023$ & FR &  SC & $165$ & $1.57$ & $2.55$ & $3.85$ & $3.55$ & $5.17_{- 0.57}^{+ 0.56}$ & $1.09_{- 0.11}^{+ 0.11}$ \\
\vspace{0.05cm}
$   2549$ & FR &  MC & $140$ & $2.01$ & $3.26$ & $4.14$ & $3.10$ & $2.68_{- 0.65}^{+ 0.90}$ & $0.53_{- 0.13}^{+ 0.17}$ \\
\vspace{0.05cm}
$   2685$ & FR &  SC & $100$ & $2.05$ & $3.09$ & $3.33$ & $2.59$ & $2.88_{- 0.69}^{+ 0.74}$ & $0.72_{- 0.18}^{+ 0.18}$ \\
\vspace{0.05cm}
$   2695$ & FR &  MC & $184$ & $1.36$ & $2.19$ & $3.39$ & $3.44$ & $5.01_{- 0.28}^{+ 0.28}$ & $1.21_{- 0.05}^{+ 0.03}$ \\
\vspace{0.05cm}
$   2699$ & FR &  MC & $123$ & $1.78$ & $2.83$ & $3.81$ & $3.11$ & $3.91_{- 0.57}^{+ 0.65}$ & $0.87_{- 0.13}^{+ 0.13}$ \\
\vspace{0.05cm}
$   2768$ & FR &  SC & $200$ & $1.67$ & $2.67$ & $3.47$ & $3.12$ & $4.28_{- 0.62}^{+ 0.65}$ & $1.02_{- 0.13}^{+ 0.12}$ \\
\vspace{0.05cm}
$   2974$ & FR &  SC & $227$ & $1.73$ & $2.77$ & $4.01$ & $3.50$ & $4.33_{- 0.56}^{+ 0.60}$ & $0.89_{- 0.12}^{+ 0.11}$ \\
\vspace{0.05cm}
$   3032$ & FR & CLV & $ 90$ & $3.85$ & $5.30$ & $3.19$ & $1.74$ & $1.31_{- 0.58}^{+ 0.85}$ & $0.30_{- 0.14}^{+ 0.20}$ \\
\vspace{0.05cm}
$   3156$ & FR &  SC & $ 66$ & $2.93$ & $4.16$ & $2.90$ & $1.55$ & $1.45_{- 0.47}^{+ 0.70}$ & $0.41_{- 0.14}^{+ 0.22}$ \\
\vspace{0.05cm}
$   3377$ & FR &  SC & $126$ & $1.95$ & $3.10$ & $3.48$ & $2.74$ & $2.88_{- 0.68}^{+ 0.72}$ & $0.69_{- 0.17}^{+ 0.17}$ \\
\vspace{0.05cm}
$   3379$ & FR &  SC & $190$ & $1.54$ & $2.60$ & $3.74$ & $3.53$ & $4.90_{- 0.59}^{+ 0.58}$ & $1.06_{- 0.11}^{+ 0.11}$ \\
\vspace{0.05cm}
$   3384$ & FR &  MC & $141$ & $1.87$ & $2.98$ & $4.05$ & $3.17$ & $3.40_{- 0.64}^{+ 0.66}$ & $0.71_{- 0.14}^{+ 0.13}$ \\
\vspace{0.05cm}
$   3414$ & SR & KDC & $191$ & $1.58$ & $2.47$ & $3.49$ & $3.17$ & $4.74_{- 0.43}^{+ 0.45}$ & $1.14_{- 0.09}^{+ 0.08}$ \\
\vspace{0.05cm}
$   3489$ & FR &  MC & $ 99$ & $2.53$ & $3.65$ & $3.47$ & $2.21$ & $1.94_{- 0.54}^{+ 0.78}$ & $0.46_{- 0.15}^{+ 0.20}$ \\
\vspace{0.05cm}
$   3608$ & SR & KDC & $167$ & $1.71$ & $2.78$ & $3.73$ & $3.24$ & $4.10_{- 0.59}^{+ 0.64}$ & $0.91_{- 0.13}^{+ 0.12}$ \\
\vspace{0.05cm}
$   4150$ & FR & CLV & $ 77$ & $2.69$ & $3.57$ & $3.27$ & $2.00$ & $2.02_{- 0.60}^{+ 0.79}$ & $0.52_{- 0.17}^{+ 0.22}$ \\
\vspace{0.05cm}
$   4262$ & FR &  MC & $164$ & $1.52$ & $2.45$ & $3.45$ & $3.33$ & $4.84_{- 0.43}^{+ 0.42}$ & $1.15_{- 0.09}^{+ 0.07}$ \\
\vspace{0.05cm}
$   4270$ & FR &  MC & $125$ & $1.77$ & $2.72$ & $3.64$ & $2.77$ & $4.10_{- 0.63}^{+ 0.68}$ & $0.98_{- 0.13}^{+ 0.13}$ \\
\vspace{0.05cm}
$   4278$ & FR &  MC & $217$ & $1.65$ & $2.73$ & $3.62$ & $3.63$ & $4.37_{- 0.57}^{+ 0.63}$ & $0.95_{- 0.12}^{+ 0.12}$ \\
\vspace{0.05cm}
$   4374$ & SR &  SC & $261$ & $1.49$ & $2.48$ & $3.61$ & $3.47$ & $5.06_{- 0.48}^{+ 0.48}$ & $1.14_{- 0.10}^{+ 0.08}$ \\
\vspace{0.05cm}
$   4382$ & FR & CLV & $178$ & $1.99$ & $3.09$ & $3.63$ & $2.80$ & $2.93_{- 0.66}^{+ 0.71}$ & $0.67_{- 0.16}^{+ 0.16}$ \\
\vspace{0.05cm}
$   4387$ & FR &  MC & $ 98$ & $1.60$ & $2.51$ & $3.58$ & $3.18$ & $4.78_{- 0.49}^{+ 0.49}$ & $1.12_{- 0.11}^{+ 0.09}$ \\
\vspace{0.05cm}
$   4458$ & SR & KDC & $ 83$ & $1.64$ & $2.46$ & $3.17$ & $2.88$ & $4.18_{- 0.37}^{+ 0.36}$ & $1.15_{- 0.10}^{+ 0.08}$ \\
\vspace{0.05cm}
$   4459$ & FR &  MC & $155$ & $1.92$ & $3.07$ & $3.62$ & $2.91$ & $2.99_{- 0.68}^{+ 0.70}$ & $0.68_{- 0.16}^{+ 0.16}$ \\
\vspace{0.05cm}
$   4473$ & FR &  MC & $186$ & $1.56$ & $2.54$ & $3.82$ & $3.45$ & $5.11_{- 0.57}^{+ 0.54}$ & $1.10_{- 0.11}^{+ 0.10}$ \\
\vspace{0.05cm}
$   4477$ & FR &  SC & $147$ & $1.60$ & $2.61$ & $3.70$ & $3.29$ & $4.71_{- 0.59}^{+ 0.59}$ & $1.05_{- 0.12}^{+ 0.12}$ \\
\vspace{0.05cm}
$   4486$ & SR &  SC & $268$ & $1.25$ & $2.21$ & $3.52$ & $3.87$ & $5.56_{- 0.30}^{+ 0.30}$ & $1.21_{- 0.05}^{+ 0.03}$ \\
\vspace{0.05cm}
$   4526$ & FR &  MC & $214$ & $1.68$ & $2.70$ & $3.89$ & $3.58$ & $4.61_{- 0.56}^{+ 0.61}$ & $0.96_{- 0.11}^{+ 0.11}$ \\
\vspace{0.05cm}
$   4546$ & FR &  MC & $189$ & $1.60$ & $2.49$ & $3.64$ & $3.39$ & $5.02_{- 0.49}^{+ 0.48}$ & $1.13_{- 0.10}^{+ 0.09}$ \\
\vspace{0.05cm}
$   4550$ & SR &  SC & $103$ & $2.01$ & $3.06$ & $3.46$ & $2.55$ & $2.97_{- 0.69}^{+ 0.73}$ & $0.73_{- 0.17}^{+ 0.17}$ \\
\vspace{0.05cm}
$   4552$ & SR & KDC & $233$ & $1.55$ & $2.64$ & $4.06$ & $3.79$ & $5.03_{- 0.58}^{+ 0.60}$ & $1.00_{- 0.10}^{+ 0.10}$ \\
\vspace{0.05cm}
$   4564$ & FR &  SC & $150$ & $1.62$ & $2.65$ & $3.84$ & $3.47$ & $4.75_{- 0.59}^{+ 0.62}$ & $1.01_{- 0.11}^{+ 0.11}$ \\
\vspace{0.05cm}
$   4570$ & FR &  MC & $167$ & $1.51$ & $2.45$ & $3.75$ & $3.43$ & $5.24_{- 0.47}^{+ 0.45}$ & $1.15_{- 0.09}^{+ 0.08}$ \\
\vspace{0.05cm}
$   4621$ & FR & KDC & $200$ & $1.50$ & $2.41$ & $3.70$ & $3.56$ & $5.34_{- 0.44}^{+ 0.41}$ & $1.17_{- 0.09}^{+ 0.06}$ \\
\vspace{0.05cm}
$   4660$ & FR &  MC & $181$ & $1.55$ & $2.59$ & $3.72$ & $3.48$ & $4.89_{- 0.57}^{+ 0.59}$ & $1.07_{- 0.11}^{+ 0.11}$ \\
\vspace{0.05cm}
$   5198$ & SR & KDC & $173$ & $1.60$ & $2.59$ & $3.75$ & $3.43$ & $4.89_{- 0.57}^{+ 0.58}$ & $1.07_{- 0.11}^{+ 0.11}$ \\
\vspace{0.05cm}
$   5308$ & FR &  MC & $201$ & $1.52$ & $2.49$ & $3.83$ & $3.60$ & $5.36_{- 0.53}^{+ 0.50}$ & $1.13_{- 0.10}^{+ 0.09}$ \\
\vspace{0.05cm}
$   5813$ & SR & KDC & $210$ & $1.54$ & $2.61$ & $3.75$ & $3.47$ & $4.84_{- 0.59}^{+ 0.59}$ & $1.05_{- 0.11}^{+ 0.12}$ \\
\vspace{0.05cm}
$   5831$ & SR & KDC & $148$ & $1.87$ & $2.96$ & $3.86$ & $2.87$ & $3.40_{- 0.62}^{+ 0.68}$ & $0.76_{- 0.14}^{+ 0.15}$ \\
\vspace{0.05cm}
$   5838$ & FR &  MC & $232$ & $1.66$ & $2.67$ & $3.94$ & $3.58$ & $4.76_{- 0.57}^{+ 0.62}$ & $0.99_{- 0.11}^{+ 0.11}$ \\
\vspace{0.05cm}
$   5845$ & FR &  MC & $237$ & $1.57$ & $2.65$ & $4.19$ & $3.76$ & $5.03_{- 0.57}^{+ 0.61}$ & $0.99_{- 0.10}^{+ 0.10}$ \\
\vspace{0.05cm}
$   5846$ & SR &  SC & $213$ & $1.35$ & $2.36$ & $3.74$ & $3.60$ & $5.48_{- 0.42}^{+ 0.37}$ & $1.18_{- 0.09}^{+ 0.05}$ \\
\vspace{0.05cm}
$   5982$ & SR & KDC & $223$ & $1.63$ & $2.66$ & $4.05$ & $3.38$ & $4.79_{- 0.57}^{+ 0.62}$ & $1.00_{- 0.11}^{+ 0.11}$ \\
\vspace{0.05cm}
$   7332$ & FR & KDC & $125$ & $2.10$ & $3.27$ & $3.87$ & $2.75$ & $2.53_{- 0.61}^{+ 0.76}$ & $0.55_{- 0.14}^{+ 0.17}$ \\
$   7457$ & FR & CLV & $ 75$ & $2.26$ & $3.40$ & $3.60$ & $2.40$ & $2.27_{- 0.59}^{+ 0.75}$ & $0.53_{- 0.15}^{+ 0.19}$ \\
\hline
\hline
\end{tabular}
\end{center}
\begin{flushleft}
\small{NOTES: (1) NGC galaxy number. (2) Slow/Fast Rotator class. (3) Kinematic substructure as defined in Paper XII. (4) Stellar velocity dispersion within \rev. (5-8) Line-strength indices within \rev\ measured in the LIS-14\AA\ system. (9-10) Stellar mass-to-light ratios (\steml) in the $V$- and $3.6\mu$m bands.}
\end{flushleft}
\end{table*}

\newpage

\begin{table*}
%
\caption{Spectroscopic quantities for the \sauron\ sample of Sa galaxies.}
%
\label{tab:saupars_Sa}
\begin{center}
\renewcommand\tabcolsep{3.50000pt}
\begin{tabular}{cccccccccc}
\hline
\hline
Galaxy & SR/FR & Kinem. & $\sigma_{\rm e}$ & H$\beta$ & H$\beta_o$ & Fe5015 & Mg$b$ & \stemlV & \stemlIR \\
 &  &  & (km/s) & (\AA) & (\AA) & (\AA) & (\AA) & \MLsunV & \MLsun36 \\
(1) & (2) & (3) & (4) & (5) & (6) & (7) & (8) & (9) & (10) \\
\hline
\vspace{0.05cm}
$   1056$ & FR &  SC & $ 83$ & $2.85$ & $3.94$ & $1.76$ & $1.24$ & $1.47_{- 0.41}^{+ 0.46}$ & $0.60_{- 0.14}^{+ 0.18}$ \\
\vspace{0.05cm}
$   2273$ & FR &  MC & $124$ & $2.72$ & $3.98$ & $3.13$ & $2.27$ & $1.63_{- 0.50}^{+ 0.75}$ & $0.40_{- 0.14}^{+ 0.20}$ \\
\vspace{0.05cm}
$   2844$ & FR &  SC & $ 93$ & $2.45$ & $3.49$ & $2.56$ & $1.70$ & $1.96_{- 0.61}^{+ 0.75}$ & $0.64_{- 0.21}^{+ 0.23}$ \\
\vspace{0.05cm}
$   3623$ & FR &  MC & $141$ & $1.69$ & $2.70$ & $3.70$ & $3.27$ & $4.39_{- 0.59}^{+ 0.65}$ & $0.98_{- 0.12}^{+ 0.12}$ \\
\vspace{0.05cm}
$   4220$ & FR &  SC & $ 92$ & $2.61$ & $3.68$ & $3.89$ & $2.40$ & $2.04_{- 0.59}^{+ 0.83}$ & $0.43_{- 0.13}^{+ 0.17}$ \\
\vspace{0.05cm}
$   4235$ & FR &  MC & $165$ & $1.84$ & $2.69$ & $3.67$ & $2.97$ & $4.29_{- 0.61}^{+ 0.66}$ & $1.00_{- 0.13}^{+ 0.13}$ \\
\vspace{0.05cm}
$   4245$ & FR &  MC & $ 97$ & $1.94$ & $2.95$ & $3.43$ & $2.76$ & $3.31_{- 0.63}^{+ 0.73}$ & $0.81_{- 0.14}^{+ 0.16}$ \\
\vspace{0.05cm}
$   4274$ & FR &  MC & $153$ & $1.88$ & $2.89$ & $3.51$ & $2.85$ & $3.53_{- 0.62}^{+ 0.71}$ & $0.85_{- 0.14}^{+ 0.15}$ \\
\vspace{0.05cm}
$   4293$ & FR &  SC & $107$ & $2.44$ & $3.53$ & $3.71$ & $2.30$ & $2.12_{- 0.56}^{+ 0.79}$ & $0.48_{- 0.14}^{+ 0.20}$ \\
\vspace{0.05cm}
$   4314$ & FR &  SC & $118$ & $2.13$ & $3.13$ & $3.37$ & $2.60$ & $2.79_{- 0.70}^{+ 0.74}$ & $0.69_{- 0.18}^{+ 0.18}$ \\
\vspace{0.05cm}
$   4369$ & FR &  SC & $ 59$ & $3.39$ & $4.72$ & $1.99$ & $1.30$ & $1.07_{- 0.41}^{+ 0.60}$ & $0.36_{- 0.15}^{+ 0.21}$ \\
\vspace{0.05cm}
$   4383$ & FR &  SC & $ 51$ & $3.05$ & $4.05$ & $0.40$ & $0.89$ & $1.15_{- 0.25}^{+ 0.40}$ & $0.63_{- 0.10}^{+ 0.25}$ \\
\vspace{0.05cm}
$   4405$ & FR &  SC & $ 54$ & $3.13$ & $4.36$ & $2.34$ & $1.53$ & $1.21_{- 0.37}^{+ 0.64}$ & $0.37_{- 0.12}^{+ 0.24}$ \\
\vspace{0.05cm}
$   4425$ & FR & CLV & $ 78$ & $1.89$ & $2.90$ & $3.62$ & $2.78$ & $3.52_{- 0.62}^{+ 0.70}$ & $0.83_{- 0.14}^{+ 0.15}$ \\
\vspace{0.05cm}
$   4596$ & FR &  MC & $156$ & $1.66$ & $2.62$ & $3.51$ & $3.21$ & $4.49_{- 0.58}^{+ 0.61}$ & $1.05_{- 0.12}^{+ 0.12}$ \\
\vspace{0.05cm}
$   4698$ & FR & KDC & $141$ & $1.67$ & $2.59$ & $3.30$ & $2.90$ & $4.20_{- 0.55}^{+ 0.53}$ & $1.08_{- 0.13}^{+ 0.11}$ \\
\vspace{0.05cm}
$   4772$ & FR &  SC & $131$ & $1.54$ & $2.50$ & $3.07$ & $3.19$ & $4.27_{- 0.43}^{+ 0.42}$ & $1.13_{- 0.11}^{+ 0.09}$ \\
\vspace{0.05cm}
$   5448$ & FR &  MC & $128$ & $2.19$ & $3.27$ & $3.23$ & $2.65$ & $2.50_{- 0.68}^{+ 0.79}$ & $0.62_{- 0.18}^{+ 0.20}$ \\
\vspace{0.05cm}
$   5475$ & FR &  SC & $101$ & $2.17$ & $3.35$ & $4.07$ & $2.85$ & $2.49_{- 0.62}^{+ 0.88}$ & $0.51_{- 0.13}^{+ 0.17}$ \\
\vspace{0.05cm}
$   5636$ & FR & CLV & $ 40$ & $2.49$ & $3.64$ & $2.53$ & $1.75$ & $1.81_{- 0.59}^{+ 0.73}$ & $0.58_{- 0.20}^{+ 0.24}$ \\
\vspace{0.05cm}
$   5689$ & FR &  MC & $166$ & $1.98$ & $3.12$ & $4.02$ & $2.93$ & $2.91_{- 0.64}^{+ 0.74}$ & $0.61_{- 0.14}^{+ 0.15}$ \\
\vspace{0.05cm}
$   5953$ & FR & KDC & $100$ & $2.86$ & $3.97$ & $1.69$ & $1.29$ & $1.44_{- 0.39}^{+ 0.46}$ & $0.60_{- 0.14}^{+ 0.18}$ \\
\vspace{0.05cm}
$   6501$ & FR &  SC & $190$ & $1.65$ & $2.65$ & $4.06$ & $3.58$ & $4.91_{- 0.57}^{+ 0.61}$ & $1.00_{- 0.11}^{+ 0.11}$ \\
$   7742$ & FR & KDC & $ 71$ & $2.81$ & $4.07$ & $2.87$ & $2.04$ & $1.53_{- 0.49}^{+ 0.72}$ & $0.40_{- 0.14}^{+ 0.21}$ \\
\hline
\hline
\end{tabular}
\end{center}
\begin{flushleft}
\small{NOTES: (1) NGC galaxy number. (2) Slow/Fast Rotator class. (3) Kinematic substructure as defined in Paper XII. (4) Stellar velocity dispersion within \rev. (5-8) Line-strength indices within \rev\ measured in the LIS-14\AA\ system. (9-10) Stellar mass-to-light ratios (\steml) in the $V$- and $3.6\mu$m bands.}
\end{flushleft}
\end{table*}

\end{document}